\documentclass{article}

\usepackage{PRIMEarxiv}

\usepackage[utf8]{inputenc} 
\usepackage[T1]{fontenc}    
\usepackage{hyperref}       
\usepackage{url}            
\usepackage{booktabs}       
\usepackage{amsfonts}       
\usepackage{nicefrac}       
\usepackage{microtype}      
\usepackage{lipsum}
\usepackage{xcolor} 
\usepackage{fancyhdr}       
\usepackage{graphicx}       
\graphicspath{{images/}} 

\usepackage{amssymb}

\usepackage{placeins}
\usepackage[normalem]{ulem}
\usepackage{multirow}
\usepackage{bbding}
\usepackage[nointegrals]{wasysym}
\usepackage{booktabs}
\usepackage{tabularx}
\usepackage{xltabular}
\usepackage{soul}
\usepackage{array}
\usepackage{adjustbox}
\usepackage{amsmath}
\usepackage{rotating}
\usepackage{longtable}
\usepackage{pdflscape}
\usepackage{threeparttable, caption}
\usepackage{pifont}
\newcommand{\cmark}{\ding{51}}%
\newcommand{\xmark}{\ding{55}}%
 
\usepackage{lscape}
\usepackage{xcolor}
\usepackage{caption}
\usepackage{subcaption}
\usepackage{url}
\usepackage{comment}
\usepackage{graphicx}
\usepackage{epstopdf}     
\usepackage{placeins} 
\usepackage{hyperref}       
\graphicspath{{images/}} 
\usepackage{url}
\usepackage{comment}
\usepackage{amsmath} 
\usepackage{multirow}
\usepackage[nointegrals]{wasysym}
\usepackage{url}
\usepackage{longtable}
\usepackage{soul}
\usepackage{caption}
\usepackage{subcaption}
\usepackage{ragged2e}
\emergencystretch=\maxdimen
\usepackage{tikz}
\usetikzlibrary{positioning,chains}
\tikzset{
    mynode/.style={
        draw, rectangle, align=center, text width=5cm, inner sep=3ex},
    mylabel/.style={
        draw, rectangle, align=center, rounded corners, font=\bf, inner sep=2ex, 
        fill=cyan!30, minimum height=3.8cm},
    arrow/.style={
        very thick,->,>=stealth}
}

\usepackage{tikz}
\usetikzlibrary{positioning,chains}
\tikzset{
    mynode/.style={
        draw, rectangle, align=center, text width=5cm, inner sep=3ex},
    mylabel/.style={
        draw, rectangle, align=center, rounded corners, font=\bf, inner sep=2ex, 
        fill=cyan!30, minimum height=3.8cm},
    arrow/.style={
        very thick,->,>=stealth}
}

\newcolumntype{R}[2]{%
    >{\adjustbox{angle=#1,lap=\width-(#2)}\bgroup}%
    l%
    <{\egroup}%
}

\title{Security through the Eyes of AI: How Visualization is Shaping Malware Detection}

\author{
Matteo Brosolo \\
Department of Mathematics,\\
University of Padova, Italy \\
\texttt{matteo.brosolo@unipd.it}\\
\And
Asmitha K. A. \\
Department of Computer Applications,\\
Cochin University of Science\\
and Technology, Kochi, India\\
\texttt{asmitha@pg.cusat.ac.in} \\
\And
Mauro Conti \\
Department of Mathematics,\\
University of Padova, Italy \\
\texttt{mauro.conti@unipd.it}\\
\And
Rafidha Rehiman K. A. \\
Department of Computer Applications,\\
Cochin University of Science\\
and Technology, Kochi, India \\
\texttt{rafidharehimanka@cusat.ac.in}
\And
Muhammed Shafi K. P. \\
Department of Computer Applications, \\
Cochin University of Science \\
and Technology, Kochi, India\\
\texttt{shafikp@cusat.ac.in} \\
\And
Serena Nicolazzo \\
Department of Computer Science, \\
University of Milan, \\
G. Celoria, 20, Milan, Italy\\
\texttt{serena.nicolazzo@unimi.it} \\
\And
Antonino Nocera \\
Department of Electrical, Computer \\
and Biomedical Engineering, \\
University of Pavia, \\
A. Ferrata, 5, Pavia, Italy \\
\texttt{antonino.nocera@unipv.it} \\
\And
Vinod P. \\
Department of Computer Applications, \\
Cochin University of Science \\
and Technology, Kochi, India \\
\texttt{vinod.p@cusat.ac.in} \\
}

\begin{document}
\maketitle

\begin{abstract}
Malware, a persistent cybersecurity threat, increasingly targets interconnected digital systems such as desktop, mobile, and IoT platforms through sophisticated attack vectors. By exploiting these vulnerabilities, attackers compromise the integrity and resilience of modern digital ecosystems. To address this risk, security experts actively employ Machine Learning or Deep Learning-based strategies, integrating static, dynamic, or hybrid approaches to categorize malware instances. Despite their advantages, these methods have inherent drawbacks and malware variants persistently evolve with increased sophistication, necessitating advancements in detection strategies. Visualization-based techniques are emerging as scalable and interpretable solutions for detecting and understanding malicious behaviors across diverse platforms including desktop, mobile, IoT, and distributed systems as well as through analysis of network packet capture files.
In this comprehensive survey of more than $100$ high-quality research articles, we evaluate existing visualization-based approaches applied to malware detection and classification. As a first contribution, we propose a new all-encompassing framework to study the landscape of visualization-based malware detection techniques. Within this framework, we systematically analyze state-of-the-art approaches across the critical stages of the malware detection pipeline. By analyzing not only the single techniques but also how they are combined to produce the final solution, we shed light on the main challenges in visualization-based approaches and provide insights into the advancements and potential future directions in this critical field.
\end{abstract}


\section{Introduction}

Malware samples are becoming increasingly sophisticated, often employing obfuscation and polymorphic techniques to evade traditional signature-based and behavior-based detection. Moreover, due to the increase in reliance on technology and connectivity, malware analysts have noted an expansion in the scope and strength of attacks against various types of infrastructure. Windows malware, in particular, has become a critical area of concern due to its prevalence and the potential for significant financial losses, as well as the threat it poses to the security of sovereign nations. 
Windows remains the operating system most targeted for malware, with ransomware attacks increasing 2.75 times in 2024, and 92\% of these attacks affecting Windows devices\cite{Microsoft}. Additionally, threat actors in 2025 are increasingly exploiting Windows vulnerabilities through AI-driven malware\cite{Kerr2025}.
The widespread use of Windows-based systems in personal, enterprise, and government environments has made them a prime target for attackers seeking to exploit vulnerabilities in the operating system and associated software \cite{chayal2024review}. 

Other platforms, such as Android, are not safe either, and an increase in Android malware has also been observed in recent years after the widespread adoption of mobile phones \cite{bhavan2024android}. The threats against Android devices differ in some aspects from those of their Windows counterparts. Firstly, the popularity and widespread adoption of Android mobile devices make them an attractive target for cyber-criminals, whereas Windows malware has historically targeted desktop computers. Additionally, the open nature of the Android platform allows for easier distribution of malicious apps. Furthermore, Android malware often uses different techniques and vulnerabilities specific to the mobile environment. Malware propagates across desktop and mobile platforms through various means. On desktops, it can infiltrate systems via malevolent email attachments, compromised websites, or tainted software downloads. Once present, malware exploits vulnerabilities, self-replicates, and extends its reach to connected devices and networks \cite{alsmadi2021asurvey}. On mobile platforms, malware commonly disguises itself as legitimate apps or leverages third-party app stores. Users unknowingly install these pernicious apps, thus granting access to sensitive data or device control. Additional dissemination routes encompass text messages, phishing links, and compromised Wi-Fi networks, accentuating the substantial risk posed to users on both PC and mobile platforms \cite{felt2011mobile}.
\par Another fertile ground for malware is the IoT landscape. Malware spreads in IoT devices through the exploitation of vulnerabilities in interconnected systems, allowing it to quickly infiltrate and compromise a multitude of smart devices. The interconnected nature of IoT ecosystems provides a fertile ground for the swift propagation of malicious software, thus fostering the need for robust security measures to mitigate these risks \cite{mahoubi2020stochastic}.

In this context, security researchers face the challenge of combating widespread malware campaigns and safeguarding against both new and familiar strains. However, traditional approaches that rely on signatures are inadequate in the face of ever-evolving threats. Attackers employ sophisticated methods, such as polymorphic and metamorphic malware, to avoid detection by altering their code. Consequently, there is an increasing demand for innovative techniques such as behavioral analysis, Machine Learning, and dynamic analysis. These methods provide deeper insights into malware behavior, enabling analysts to better detect and respond to threats.

In this paper, we focus on one particular aspect, namely the use of malware visualization for the classification of malware, which is a novel and advanced method for malware analysis. With the integration of Deep Learning and image-based analysis, visualization methods have shown great promise in recent years for creating interpretable results for a variety of machine learning tasks, including malware classification \cite{mathews2019explainable}. 
In particular, Convolutional Neural Networks (CNNs) are effective in malware classification due to their ability to extract features from binary representations of samples without the help of domain experts.

We survey the state-of-the-art in machine learning-based malware classification across various platforms, focusing on high-quality works published over the past seven years, from 2018 to 2025. We categorize the research literature according to the fundamental steps of the visualization-based malware classification, namely: Dataset Collection, Image Generation, Feature Extraction, Classification,
Evaluation, Model Robustness, and Adaptation. Furthermore, we explore the challenges associated with this approach, including the need for large and diverse datasets as well as the potential for adversarial attacks, and the most important technique for interpretability.
Despite growing interest, visualization-based malware detection remains a relatively young field. Hence, by gaining a deeper understanding of the strengths and limitations of Machine Learning-based approaches to malware classification, we can better evaluate their potential for improving the accuracy and efficiency of malware analysis. Through our survey of the literature, we provide a comprehensive overview of the current state of the art in this area, propose best practices, and help unify future research identifying areas for forthcoming development.

In summary, the main contributions of this survey are as follows.
\begin{enumerate}
\item To the best of our knowledge, we are the first to have a comprehensive and methodologically rigorous survey on existing studies that employ visualization-based techniques in malware detection, structured around the proposed unified framework that captures key characteristics of existing approaches and enables systematic comparison, trend analysis, and identification of research gaps.

\item We distill the entire pipeline of visualization-based malware detection from numerous concrete image-based approaches. This provides readers with an overview, aiding in understanding the general process of visualization-based methods. The overview covers the datasets used, image representation, feature processing techniques, model generation, performance evaluation, and considerations about robustness and model adaptation.
\item We explore the aspects of explainability in various visualization techniques applied to malware detection, providing valuable insights.
\item To gain a deeper understanding of adversarial attacks in visualization-based malware detection and of the existing defenses against them, we systematically survey various state-of-the-art approaches. This includes exploring adversarial attack methods on ML-based malware classifiers, DL classifiers, and defensive strategies to enhance the adversarial robustness of visualization-based malware classifiers.
\item We delve deeper into the gaps present in current studies and thoroughly explore future research challenges and directions within the domain of visualization-based malware analysis and detection. This discussion presents readers with potential avenues for developing innovative solutions.
\end{enumerate}

This survey is organized as follows. Section\ref{sec:relatedWork} introduces related surveys, while Section \ref{sec:methodology} outlines the approach adopted to carry out this survey. In section \ref{sec:background}, we present an overview of the main background concepts about malware detection and classification techniques. Section  \ref{sec:framework} describes the current literature classified according to the steps of visualization-based malware classification, namely
Dataset Collection, Image Generation, Feature Extraction, Classification, and Evaluation. In particular, in Section \ref{sec:dataset} we examine the most used dataset about malware. Section \ref{sec:visualization} deals with the techniques employed to convert a malware file into an image, utilizing diverse file extensions acquired through static and dynamic analysis approaches. In Section \ref{sec:featureExtraction}, we explore how researchers have approached the processing of features, which they subsequently use as input for classifiers. Section \ref{sec:classification} is devoted to the description of the various Machine Learning methods used to classify images through their extracted features.
Section \ref{sec:evaluation} provides a discussion of how authors decided to evaluate their classifiers, compare their results with those of other researchers, consider relevant parameters, and assess whether the choices made are fair. Section \ref{sec:adversarial} provides a description of model robustness and adaptation, specifically targeting adversarial attacks targeting visualization-based malware detectors. In Section \ref{sec:interpretability}, we explore the issue of interpretability in the context of malware classification, examining how authors have addressed the problem and suggesting potential improvements for the situation. Section \ref{sec:lesson} lists the lessons learned from our investigation. Section \ref{sec:challenges} examines several open challenges and offers perspectives on possible directions for future research. Finally, Section \ref{sec:conclusion} gives some general considerations on the state of research in the field of image-based malware classification and tries to advise to move forward.

\section{Related Work}
\label{sec:relatedWork}

This section offers an overview of surveys that explore visual malware classification. For each survey, we describe along with its advantages and disadvantages. It is worth noting that most of the surveys dealt with general techniques in malware detection and only a few others addressed the visualization of visual malware images despite brief mentions of the concept in surveys with broader scopes. 
Table \ref{tab:surveys} offers a summary of the reviewed surveys, in which we consider the 
year of publication, the number of analyzed papers, and the main scope of the work (i.e., image generation technique analysis, feature extractors analysis, classifiers analysis, interpretability analysis, sustainability analysis, different training methods analysis, dataset coverage, adversarial attacks analysis, few-shot analysis, metrics analysis.
\begin{table*}[!htb]
\scriptsize
\centering
\caption{Comparison with existing survey papers \label{tab:surveys}}

\renewcommand{\arraystretch}{1.5}

\begin{tabular}
{|l|c|c|c|c|c|c|c|c|c|c|c|c|c|}
\hline
    \multirow{2}*{\textbf{Ref.}} & \multirow{2}*{\textbf{Year}} & \multirow{2}*{\textbf{Literature}} & \multirow{2}*{\textbf{\#Papers}}& \multicolumn{10}{c|}{\multirow{1}{*}{\textbf{Paper Scope}}}\\
    \cline{5-14} 
 & & \textbf{timeline} & & $DC$ & $IG$ & $FE$ & $Class$ & $Inter$ & $S$ & $Diff\_TM$ & $AA$ & $FS$ & $M$\\ \hline

  \cite{gibertsurvey} & 2020 & 2008-2019 & 67 & \Circle & \CIRCLE & \LEFTcircle & \CIRCLE & \LEFTcircle & \LEFTcircle & \Circle & \LEFTcircle & \Circle & \LEFTcircle\\ \hline 
  
  \cite{survey3} & 2021 & 2009-2020 & 20 & \LEFTcircle & \LEFTcircle & \LEFTcircle & \LEFTcircle & \Circle & \Circle & \Circle & \LEFTcircle & \LEFTcircle & \LEFTcircle\\ \hline 
  
  \cite{zhao2021review} & 2021 & 2009-2019 & - & \Circle & \Circle & \LEFTcircle & \LEFTcircle & \Circle & \Circle & \Circle & \Circle & \Circle & \Circle\\ \hline 
  
  \cite{ucci2019survey} & 2019 & 2001-2017 & 64  & \LEFTcircle & \CIRCLE & \LEFTcircle & \LEFTcircle & \Circle & \Circle & \Circle & \Circle & \Circle & \Circle\\ \hline 
  
  \cite{survey1} & 2022 & 2016-2020 & 4 & \Circle & \Circle & \LEFTcircle & \Circle & \Circle & \Circle & \Circle & \Circle & \Circle & \Circle\\ \hline 
  
  \cite{survey2} & 2022 &- & 10 & \Circle & \LEFTcircle & \LEFTcircle & \LEFTcircle & \Circle & \Circle & \Circle & \Circle & \Circle & \Circle\\ \hline 
  
  \cite{gopinath2023acomprehensive} & 2023 &- & - & \Circle & \LEFTcircle & \LEFTcircle & \LEFTcircle & \Circle & \Circle & \Circle & \LEFTcircle & \Circle & \Circle \\ \hline
  
  \cite{deldar2023deep} & 2023 & -& - & \LEFTcircle & \LEFTcircle & \LEFTcircle & \LEFTcircle & \Circle & \LEFTcircle & \Circle & \CIRCLE & \LEFTcircle & \LEFTcircle \\ \hline

  \cite{brosolo2024sok} & 2024 & 2018-2023 & - & \LEFTcircle & \LEFTcircle & \LEFTcircle & \LEFTcircle & \LEFTcircle & \LEFTcircle & \LEFTcircle & \LEFTcircle & \LEFTcircle & \LEFTcircle \\ \hline
  \hline
  
  \textbf{Our} & 2025 & 2018-2025 & 103 & \CIRCLE & \CIRCLE & \CIRCLE & \CIRCLE & \CIRCLE & \CIRCLE & \CIRCLE & \CIRCLE & \CIRCLE & \CIRCLE \\
  \hline
\end{tabular}

\footnotesize \Circle - Topic not covered or minimally analyzed; \LEFTcircle - Topic addressed to a reasonable extent; \CIRCLE - Topic covered in depth.\\ ${DC}$: Dataset Coverage, $IG$: Image generation technique analysis, $FE$: Feature extractors analysis, $Class$: Classifiers analysis, $Inter$: Interpretability analysis, $S$: Sustainability analysis, $Diff\_TM$: Different training methods analysis, $AA$: Adversarial attacks analysis, $Fs$: Few-shot analysis, $M$: Metrics analysis.
\end{table*} 

\subsection{General Machine Learning and Malware Detection Surveys}
In \cite{gibertsurvey} the authors provides an extensive examination of ML models applied to malware classification and detection. The survey's main focus is not specifically on visualization methods. Emerging and popular challenges, such as interpretability and sustainability are briefly mentioned. Despite these limitations, this survey serves as a valuable starting point, providing an overview of traditional Machine Learning and Deep Learning workflows used in malware classification and detection. Ucci et al. in \cite{ucci2019survey} give a thorough analysis of the Machine Learning techniques. They provide a well-thought taxonomy to categorize the different Machine Learning methods. The section on feature extraction techniques covers the subject in great depth. This survey excels in analyzing Machine Learning techniques broadly, but it falls short in giving adequate attention to recent advancements, such as Deep Learning, and specifically, the utilization of CNNs. Additionally, the survey lacks sufficient emphasis on the contemporary challenges researchers face today. Naik et al. in \cite{survey3} provide a general overview of image-based malware analysis. The authors discuss the main steps of the workflow from the dataset to the evaluation. The inclusion of 20 analyzed papers also highlights the study's significance. However, it is essential to acknowledge the limitations of this study, such as the lack of in-depth analysis regarding datasets and feature extraction. Moreover, it is noteworthy that there is no discussion on current issues like interpretability and sustainability. Gopinath et al. \cite{gopinath2023acomprehensive} present a comprehensive survey encompassing a broad range of papers that employ Machine Learning and Deep Learning techniques. The survey primarily examines various approaches to developing classifiers rather than focusing on image-based visualization methods. The authors explain each paper with due detail, characterizing each model well. The survey divides the publications based on the approach used for the classifier, therefore each paper is named and explained just once. For this reason, it is hard to abstract the information received and categorize the different approaches to feature extraction and image generation. Furthermore, the authors do not extensively delve into contemporary challenges encountered in the field, aside from brief tangential discussions.

\subsection{Visualization-Based Malware Detection Surveys}

Ahmad et al. \cite{survey2} survey visualization-based malware detection and classification. The authors present a detailed review of existing malware detection methods, leveraging both ML and DL techniques, and outline the implementation of eleven distinct models. This survey effectively identifies the essential components of each model and employs a concise listing approach, facilitating comparisons between them. However, It is worth mentioning that we have conducted a thorough analysis of the entire pipeline of visualization approaches.
Shah et al. in \cite{survey1} analyze visualization-based malware classification models. We particularly focus on the computational performance of the model. However, it is crucial to emphasize that the number of models considered in this survey is limited. Therefore, the authors cannot provide extensive insights into the broader landscape of visualization-based malware analysis.  
Deldar et al. \cite{deldar2023deep} provide a survey specifically on the detection of zero-day malware. The taxonomy employed neatly divides the models and enables the reader to identify the possible choices at each step of the classification pipeline. However, the survey does not delve into the different possibilities in depth; instead, it merely presents or briefly discusses them. Remarkable is exploring the subject of most interest for zero-day classification, meaning adversarial attacks and few-shot learning. The authors do not directly deal with modern problems in Machine Learning malware visual classification, like sustainability and interpretability. The researchers in \cite{zhao2021review} apply computer vision to network security, broadly covering phishing detection, malware detection, and traffic anomaly detection. They highlight its broader cybersecurity implications, including applications in physical security and critical infrastructure protection, and bridge computer vision, machine learning, and network security. However, their study lacks an in-depth focus on visualization-based malware detection. In contrast, our study provides a detailed analysis of visualization techniques, starting from dataset collection, image types, feature extraction, classification methods, interpretability, robustness evaluation, and adaptation in malware detection.
\section{Methodology}
\label{sec:methodology}
In this study, we design a structured search strategy to gather relevant research on visualization-based malware detection according to the objective of our survey article. Initially, we explore academic databases such as Google Scholar and Web of Science and review reputable repositories such as SpringerLink, IEEE Xplore, ScienceDirect, and ACM Digital Library. The search focuses on publications from 2018 to 2025 to ensure a comprehensive examination of recent advancements. An initial global scan reveals that, before 2018 research on visualization-driven malware detection remains limited, with only a few notable publications. As a result, we select a seven-year time frame for analysis to capture the most relevant and recent works. Recently, research on visualization-based malware detection has expanded rapidly, driving significant innovation in the field.
 \par
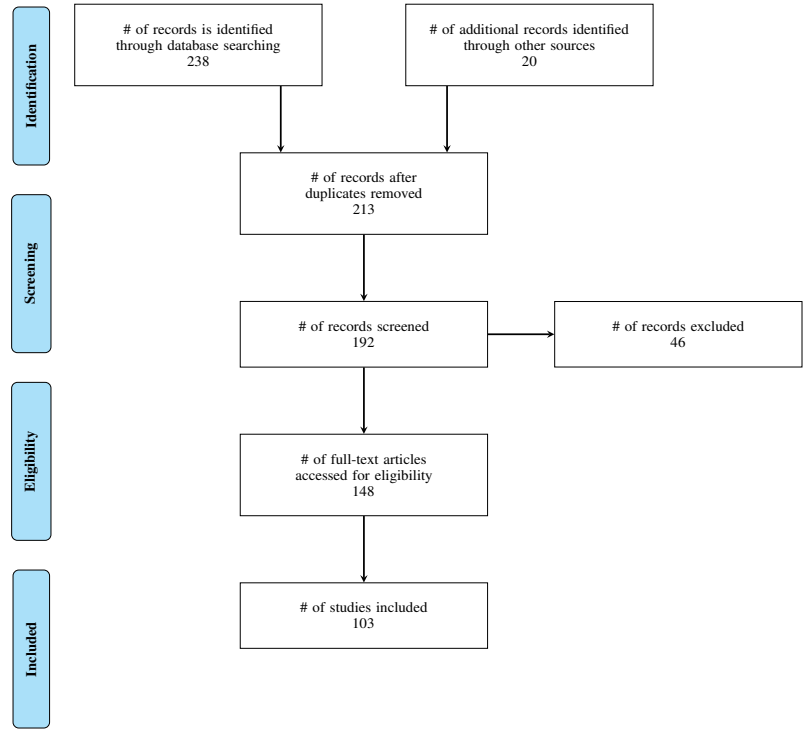
\begin{figure*}[!htb]
\centering
\scalebox{.55}{
\begin{tikzpicture}[
    node distance=1.6cm,
    start chain=1 going below,
    every join/.style=arrow,
    ]
    \coordinate[on chain=1] (tc);
    \node[mynode, on chain=1] (n2)
        {\# of records after duplicates removed \\ 213};
    \node[mynode, join, on chain=1] (n3)
        {\# of records screened \\ 192};
    \node[mynode, join, on chain=1] (n4)
        {\# of full-text articles accessed for eligibility \\ 148};
    \node[mynode, join, on chain=1] (n5)
        {\# of studies included \\ 103};

    \begin{scope}[start chain=going right]
        \chainin (n3);
        \node[mynode, join, on chain]
            {\# of records excluded \\ 46};
    \end{scope}

    \node[mynode, left=1cm of tc, anchor=south east] (n1l)
        {\# of records is identified through database searching  \\ 238};
    \node[mynode, right=1cm of tc, anchor=south west] (n1r) 
        {\# of additional records identified\\ through other sources \\ 20};

    \coordinate (n2nl) at ([xshift=-2cm]n2.north);
    \coordinate (n2nr) at ([xshift= 2cm]n2.north);
    \draw[arrow] (n1l.south -| n2nl) -- (n2nl);
    \draw[arrow] (n1r.south -| n2nr) -- (n2nr);

    \begin{scope}[start chain=going below, xshift=-8cm, node distance=.7cm]
        \node[mylabel, on chain] {\rotatebox{90}{Identification}};
        \node[mylabel, on chain] {\rotatebox{90}{Screening}};
        \node[mylabel, on chain] {\rotatebox{90}{Eligibility}};
        \node[mylabel, on chain] {\rotatebox{90}{Included}};
    \end{scope}

\end{tikzpicture}}
\caption{PRISMA Diagram} \label{fig:prisma}
\end{figure*}

 We ensure comprehensive coverage by formulating search queries using specific keywords. The primary search terms include ``malware visualization'' and ``image-based malware detection''. To refine our search, we incorporate additional terms such as ``Deep Learning'', ``feature extraction from malware images'', ``Machine Learning'', ``Vision Transformers'', ``adversarial robustness in malware images'', ``obfuscation'', ``interpretability'', and ``sustainability''. This approach captures a broad spectrum of studies related to malware visualization and classification. Figure \ref{fig:prisma} visually represents our study selection process in the PRISMA flow diagram, detailing the number of studies we identify, screen, exclude, and include in our final review.
 

\subsection{Selection and Filtering Criteria}
This section explains the methodology we use to evaluate the quality and relevance of the academic papers considered in this survey on visualization-based malware detection. Articles are selected when they fulfill at least one inclusion criterion and are not disqualified by any exclusion criteria. We evaluate the suitability of the paper through the following selection criteria.

\subsubsection{Inclusion Guidelines}
We evaluate a paper's suitability for this survey article based on the following selection criteria
\begin{itemize}
    \item The significance and credibility of the publication venue according to Scimago and Core.edu rankings.
    \item The citation impact through rankings on Google Scholar and Scopus.
    \item The publication date. In particular, we prioritize recent research, especially studies published in the last six years.
    \item Significance of the contribution to malware visualization.
\end{itemize}

\subsubsection{Exclusion Guidelines}
We also exclude studies that meet any of the following conditions.
\begin{itemize}
    \item Publication date before 2018, as the article would fall outside the seven-year review period.
    \item Lack of focus on visualization-based malware detection.
    \item Lack of peer review and/or publication in a language other than English.
\end{itemize}

Table \ref{tab:SystemSymbols} provides a summary of the acronyms used throughout the paper.

\begin{table}[!htb]
\small
\centering
\caption{List of the acronyms used in the paper\label{tab:SystemSymbols}}

\begin{tabular}{|l|l|}
\hline
    \textbf{Acronyms} & \textbf{Description}\\
    \hline \hline
    CNN & Convolutional Neural Network\\ \hline
    DL & Deep Learning \\ \hline
    DT & Decision Tree \\ \hline
    ELM & Extreme Learning Machine \\ \hline
    GAN & Generative Adversarial Network \\ \hline
    IoT & Internet of Things \\ \hline
    KNN & K-Nearest Neighbors \\ \hline
    ML & Machine Learning \\ \hline
    RF & Random Forest \\ \hline
    RNN & Recurrent Neural Network \\ \hline
    SFC & Space-Filling Curve \\ \hline
    SVM & Support Vector \\ \hline    

\end{tabular}

\end{table}
\section{Background}
\label{sec:background}
An important task in cybersecurity is the categorization of malware in families. This task is crucial for categorizing malware based on characteristics and behavior, enabling researchers to understand its functionality and develop effective countermeasures. It also reveals trends in attacker techniques, guiding proactive prevention strategies.
Malware classification differs from malware detection, which identifies the presence of malware in a system. Malware classification categorizes and organizes various types of malware. In contrast, malware detection directly identifies specific malware instances, allowing timely intervention to prevent or minimize their impact.
When analyzing malware, researchers can use different techniques to help explore the structure and behavior of the malicious code. These techniques broadly fall into two macro-categories: {\em(i)} dynamic analysis and {\em(ii)} static analysis.
Security researchers must tackle the challenge of widespread malware campaigns and identify and protect from new and old malware strains. However, traditional signature-based approaches to malware analysis are no longer enough in the face of constantly evolving malware \cite{aslan2020acomprehensive}. Attackers use increasingly sophisticated techniques to evade detection, such as polymorphic and metamorphic malware, which can alter their code to avoid signature detection. As a result, there is a growing need for novel and advanced methods to analyze malware, including behavioral analysis, and integrating {\em(i)} Machine Learning and Deep Learning. These techniques allow for a more in-depth understanding of the behavior of malware and its impact on systems, enabling analysts to identify and respond to threats more effectively.

\subsection{Static Analysis}
Static analysis is a technique employed in malware analysis that involves examining the binary or code of a malicious program without executing it. This approach focuses on analyzing the structural and content aspects of malware samples to identify potential malicious behaviors and understand their operational mechanisms without direct execution. Widely used tools for static analysis include disassemblers, decompilers, and debuggers, which facilitate the extraction of the instructions used by malware to execute malicious actions, such as data theft, system modification, or propagation. In addition, static analysis enables the identification of indicators of compromise (IOCs), such as file names, network traffic patterns, registry keys, and other attributes associated with malware. By examining the code or binary, researchers can extract signatures or behavioral patterns that aid in detecting the presence of malware on a system \cite{quoc2020asurvey}. These signatures can detect malware in a system by identifying matching patterns~\cite{yara}. A convenient approach explored in this survey is to represent information extracted through static analysis by visualizing it in images, primarily due to visual similarity among variants of the same malware. Using images helps researchers to have a global view of the entire malware without losing local details. Combining feature extractors and Machine Learning models can use this characteristic to achieve accurate classification and detection results.

\subsection{Dynamic Analysis}
Dynamic analysis is a technique for examining malware behavior by executing it in a secure controlled environment, such as a sandbox or virtual machine. Unlike static analysis, which inspects code without running it, dynamic analysis allows direct observation of malware interactions with the operating system and network resources. Online sandbox platforms, including Any.Run~\cite{anyrun}, VirusTotal~\cite{virustotal}, Joe Sandbox~\cite{joesandbox}, Cuckoo Sandbox~\cite{cuckoo}, and Hybrid Analysis~\cite{hybrid} provide automated environments for the safe execution of malware and the generation of detailed behavioral reports.
Security analysts employ monitoring tools, such as system monitors, network analyzers, and debuggers, to detect malicious activities such as file creation, system modification, or communication with command-and-control (C2) servers. Furthermore, dynamic analysis reveals evasion techniques, such as anti-debugging or antivirtualization strategies \cite{chen2008towards}, improving the understanding of sophisticated malware.
By analyzing malware behavior within an isolated environment, analysts can more accurately assess the nature and potential threat of malicious code. Moreover, dynamic analysis can be augmented through visualization techniques, where run-time behavioral data is transformed into image-based representations, facilitating clearer interpretation and efficient handling of emerging malware variants.
\subsection{Machine Learning}
Machine learning has become a fundamental approach in malware analysis, enabling automated classification and detection of malicious software based on its inherent characteristics and behavior \cite{ucci2019survey}. By examining various features, including file attributes (such as size and type), system interactions, network activity, and other indicators, machine learning algorithms can effectively distinguish between benign and malicious software.
ML models monitor the runtime behavior of malware during dynamic analysis, detecting malicious patterns by examining its interactions with system components and network communications. In static analysis, these algorithms assess code structure and binaries to detect malicious patterns.
The adoption of deep learning, particularly Deep Neural Networks (DNNs), represents a significant advancement in this domain. Over the past decade, deep learning techniques have gained prominence due to their remarkable performance across various fields, including image recognition and natural language processing. Convolutional neural networks~(CNNs) excel in malware analysis by providing superior performance in malware visualization compared to traditional machine learning models. Despite their reliance on substantial computational resources and extensive datasets, DNNs have significant potential to improve malware detection and classification.
\subsection{Malware Visualization}
Malware visualization is an analytical technique that transforms the behavior and characteristics of malicious software into visual representations, facilitating its analysis and classification. Visualization provides a clear view of system and network activities, helping analysts identify patterns, uncover connections, and detect possible attack pathways. Visualization techniques generate graphical representations of different malware attributes, including network interactions, file operations, and system calls. By examining these visual representations, analysts can gain insights into the malware’s operational techniques, such as evasion strategies and exploited vulnerabilities. In addition to aiding malware detection and analysis, visualization can also enhance classification tasks by depicting malware samples as images, facilitating the differentiation between various strains.
Furthermore, malware visualization is instrumental in identifying minor variations among related malware samples. As malware authors frequently produce new variants with slight code modifications, visualization effectively captures these differences, aiding in the detection of related strains. By leveraging visual representations of malware behavior in combination with machine learning techniques, experts can efficiently identify similar and distinct patterns in malware samples, facilitating the classification of new or previously unknown threats.

\section{Unified Framework for Visualization-based Malware Detection}
\label{sec:framework}
\begin{figure*}[!htb]
    \centering
    \includegraphics[scale=0.2]{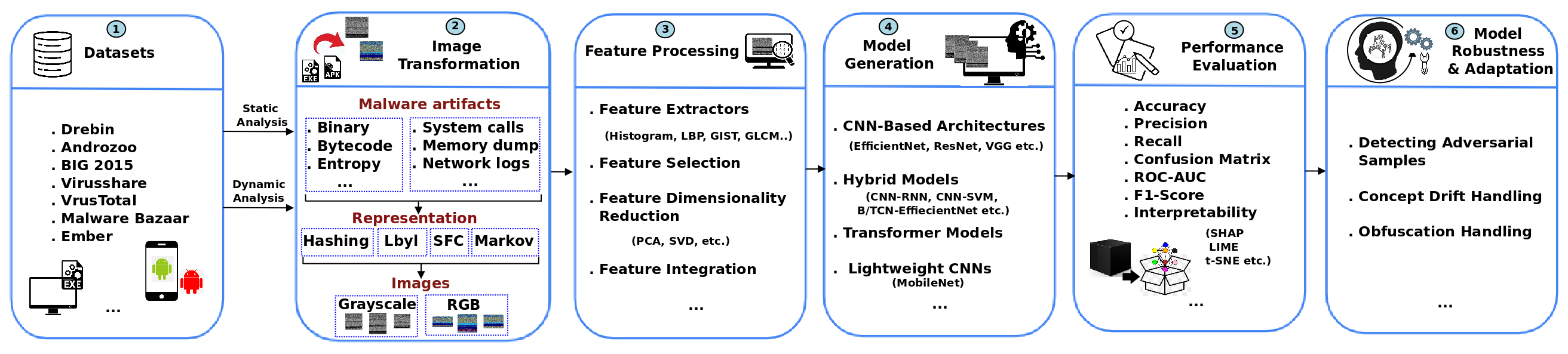}
    \caption{An End-to-End Framework for Visualization-Based Malware Detection Across Diverse and Evolving Threat Landscapes.}
    \label{workflow}
\end{figure*}
\begin{figure*}[!htbp]
   \minipage{\textwidth}
      \centering
      \includegraphics[width=\linewidth]{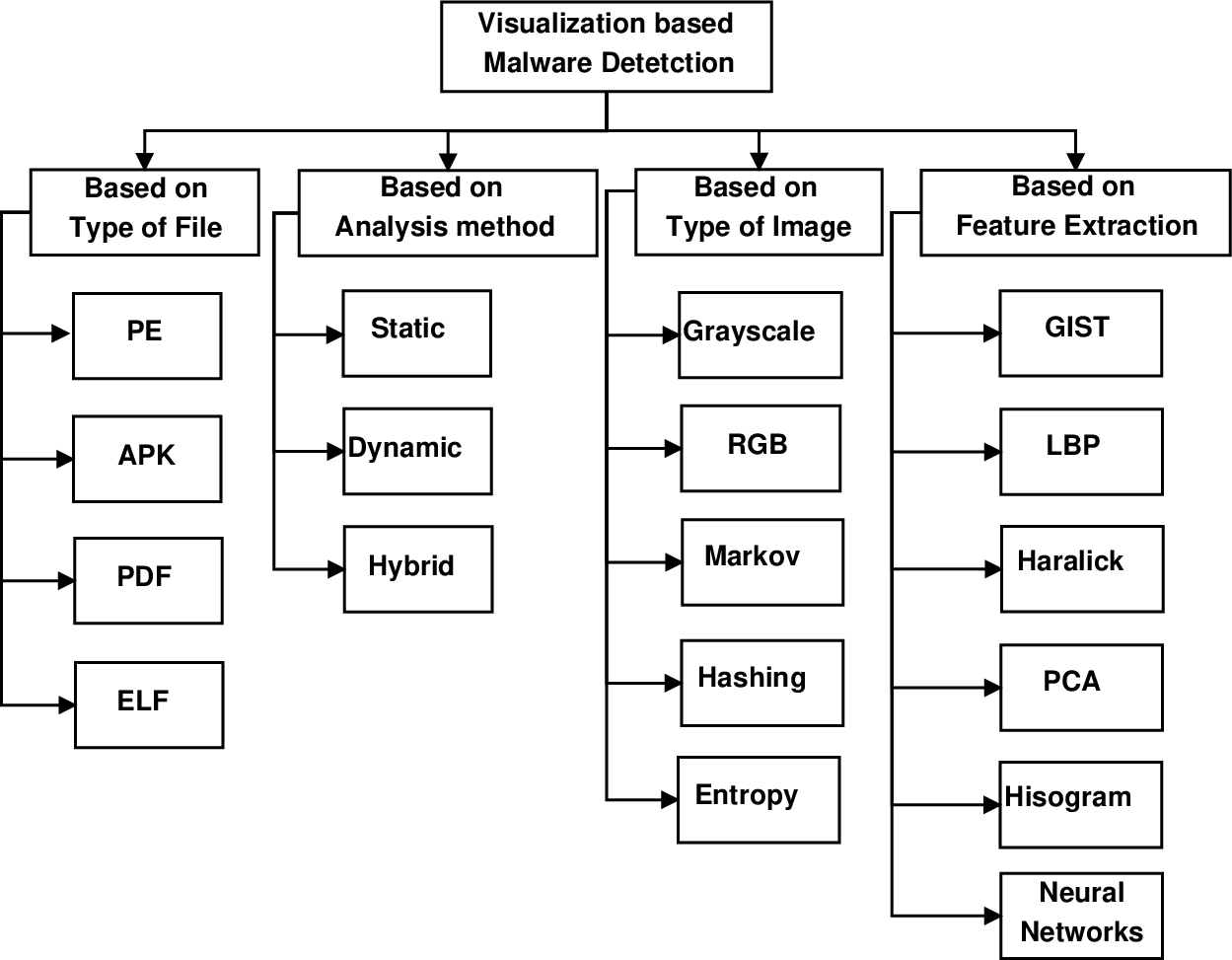}
      \caption{Taxonomy of visualization-based malware detection}
      \label{tax}
   \endminipage
  \end{figure*}

This paper presents a unified framework for visualization-based malware detection. This framework aims to comprehensively survey various state-of-the-art research in the visualization domain and to identify future research directions to improve security. We outline the fundamental process of visualization-based malware classification into five phases: Dataset Collection, Image Generation, Feature Extraction, Classification, Evaluation, and Model Robustness and Adaptation. These steps form the pipeline as seen in Figure~\ref{workflow}.
Figure \ref{tax} provides a more detailed taxonomy of the first part of the pipeline, from the selection of the file type to the feature extraction step.
\begin{enumerate}
\item \textbf{Dataset Collection}. Researchers collect datasets of malware and benign samples in row form or image format in this step. Some benchmark datasets are designed explicitly for visualization-based malware detection, such as MalImg.
\item \textbf{Image Transformation}. This phase distinguishes visualization-based malware detection approaches from other methods of malware detection. During this phase, researchers generate several images for each malware, employing various algorithms such as B2IMG and Word2Vec.
\item \textbf{Feature Processing}. In this phase, high-level features are extracted from malware images to highlight essential patterns, reduce dimensionality, and generate meaningful input for the classification model. Techniques such as Local Binary Patterns~(LBP), GIST descriptors, and dimensionality reduction methods like Principal Component Analysis~(PCA) can be employed to enhance feature quality and efficiency.
\item \textbf{Classification}. For predicting input features, we opt for Machine Learning (ML) and Deep Learning (DL) models as malware classifiers. Basic ML models like Support Vector Machine(SVM), Random Forest(RF), XGBoost, etc., and DL models such as Convolutional Neural Networks(CNN), Deep Neural Networks(DNN), etc., can serve as effective malware classifiers.
\item \textbf{Evaluation}. Visualization-based methods quantitatively evaluate the predictions of individual classifiers using accuracy measures and qualitatively assess them using explainability and interpretable terms. Binary classification methods provide probabilities for malicious or benign samples, and multi-classification methods offer probabilities for various malware families.
\item \textbf{Model Robustness and Adaptation}.  This phase involves techniques to enhance the resilience of the model against adversarial attacks, obfuscation techniques, and concept drift. Additionally, adaptive learning techniques ensure that the model remains effective against evolving malware patterns.
\end{enumerate}

\subsection{Dataset Collection}
\label{sec:dataset}

Researchers have extensively explored a wide range of datasets in their experimental investigations. The selection of appropriate datasets is paramount, as it forms the foundation for conducting rigorous and reliable research. Consequently, it is crucial to identify and prioritize datasets that are most prevalent and widely used within the domain.

When researchers seek malware samples, they can adopt two distinct approaches. The most common approach involves selecting a malware dataset explicitly designed for classification tasks\cite{yang2018aconvolutional,ni2018malware,sun2021deep,xiao2020malfcs,kumar2018malicious,cui2018detection}. These datasets provide readily available information that other researchers can easily leverage to construct image representations. However, these datasets have limitations regarding the flexibility of the available information and their overall size. An alternative approach is to create a dataset by utilizing expansive malware databases such as Malware Bazaar, Virus Share, or VX Heaven\cite{hsiao2019malware,naeem2019identification,baptista2019anovel,keyes2021entroplyzer,jiang2023apyramid}. This approach offers authors more freedom in selecting modern malware families to consider and determining the type of information to gather. However, it comes with the trade-off of increased complexity in collecting the data and the opportunity to benchmark their model against services such as VirusTotal(VT). Researchers often tailor datasets created using malware repositories(MalwareBazaar, VirusShare, VX-Underground, etc.) to the specific use cases presented in their respective research papers, resulting in significant variations in content~\cite{wang2022image,conti2022few}.

The MalImg and BIG2015 datasets have gained significant attention in studies focusing on executable malware, as highlighted in Table~\ref{tab:datasets_table}. These datasets collectively account for a substantial percentage of citations among the datasets.  Among them, the MalImg dataset~\cite{nataraj2011malware} is the most frequently cited dataset for visualization-based malware detection using PE malware. It is a publicly available image dataset comprising $9,339$ malware executables from 25 families. The dataset derives its executables from the Anubis analysis system, converting each into a grayscale image by mapping individual bytes to pixels.\ This dataset comprises a mixture of packed and unpacked malwaanging from e, with specific malware families such as Yuner.A, VB.AT, Malex.gen\!J, Autorun.K, and Rbot\!gen being packed using UPX \cite{upxwebsite}. Additionally, the dataset exhibits a high level of imbalance, spanning from $2,949$ samples belonging to the Allaple.A family to merely $80$ samples associated with the Skintrim.N family.
\par The second most cited dataset, the Microsoft Malware Classification Challenge Dataset (BIG 2015 or MMCC) \cite{ronen2018microsoft}, comprises $10,860$ malware executables belonging to 9 distinct families. Each malware executable in the dataset has a unique identifier consisting of a 20-character hash value paired with an integer denoting the malware family name. In addition, the dataset includes metadata that collects various information, such as function calls, strings, and more, from the binaries. It also provides assembly files associated with each binary. Similar to MalImg, the dataset exhibits imbalanced distribution among the malware families.
Furthermore, the Dumpware10 dataset \cite{bozkir2021catch} consists of $3,686$ malware executable samples belonging to 10 different families. The dataset also includes $608$ benign samples. The images in this dataset are in PNG format and available in dimensions $224$, $300$, and $4096$.
\par The Malevis dataset \cite{bozkir2019utilization} is an RGB-based dataset specifically designed for multi-class malware categorization. It comprises $9,100$ training samples and $5,126$ validation samples, with images distributed across 26 classes. Among these classes, 25 correspond to different types of malware, while one class represents benign samples. The dataset was constructed by extracting binary images from malware files provided by Comodo Inc. The images are available in two square resolutions, namely $224 \times 224$  and $300\times 300$ pixels. Additionally, the Malheur dataset~\cite{rieck2011automatic}, contributed by a security research team at the University of Erlangen, encompasses $3,131$ samples from 24 malware families. Among the research papers considered, samples from the VirusShare repository rank third in terms of usage. The authors of these papers generated their dataset by randomly selecting samples from VirusShare, resulting in variations in the number of families considered. 
Another frequently referenced dataset in the literature is the MALICIA dataset introduced in \cite{malicia}. This dataset comprises samples collected by analyzing servers responsible for distributing drive-by malware downloads. In addition to these datasets containing raw executable malware samples, researchers have provided datasets consisting solely of pre-extracted feature vectors from the binaries. Notable examples include the EMBER dataset \cite{ember} and the more recent BODMAS dataset \cite{bodmas}. Such datasets typically offer a more significant number of samples and malware families, but the absence of actual binary files limits them.
\par Most of the research conducted on visualization-based Android malware has mainly focused on two datasets, Drebin and MalGenome, as visible in Table~\ref{tab:datasets_table}. The Drebin dataset holds significance in Android malware analysis as it contains $5,560$ malware samples and $123,453$ benign applications collected between August 2010 and October 2012. Similarly, the MalGenome dataset \cite{zhou2012dissecting}, collected between 2010 and 2011, consists of $1,260$ malware applications. It is important to note that these datasets are relatively old and, as technology advances, the samples they contain become outdated and less representative. The authors of \cite{alam2025miracle} introduced a novel Android malware dataset for binary classification, comprising $16,868$ samples ($14,285$ malware, $2,583$ benign). This dataset, derived from the Drebin and Androzoo repositories, is publicly available, along with the accompanying code, to facilitate reproducibility and future research. Table \ref{tab:datasets_table} provides a comprehensive overview of the frequently used datasets for visualization-based malware detection. For each benchmark dataset listed, we provide accompanying details, namely the title of the dataset, the year of publication, the number of malware and benign samples, the number of families (identified by $\#Fam$), the time duration for which the samples are collected (identified as $POT$), the type of file identified as $TOF$ (e.g., an image, apk, executable, elf, or another file extensions), the availability status of each dataset (i.e., whether it is publicly accessible or not), the total number of citations for the dataset identified as $\#Cit.$, the references leveraging the dataset. All datasets listed in Table \ref{tab:datasets_table}, except AMD and Malgenome, are publicly available, facilitating further analysis and comparison. MalImg, Malheur, and Dumpware10 are image datasets, while all other datasets are available in their original .exe or .apk format. However, it is worth noting that AMD and MalGenome are no longer active projects, rendering their data inaccessible. As a result, the MalImg and Drebin datasets remain the primary choice for obtaining malware samples. Hence, researchers commonly use these datasets for visualization-based malware detection. However, specific datasets like BIG2015 have limitations, such as the unavailability of PE files and the absence of benign samples. These limitations hinder the comprehensive extraction of features and make these datasets more suitable for close-set inference tasks focused on identifying known malicious files rather than general malware detection.

Regarding Android malware, Drebin stands out as the most widely used dataset in numerous studies. However, ~\cite{irolla2018duplication} observed that the Drebin dataset may contain duplicate files, limiting access to genuine APKs and reducing the amount of usable malware data. Despite these limitations, researchers continue to extensively utilize Drebin and the aforementioned PE malware datasets, even in recent research, as the primary sources of malware data for developing visualization-based algorithms targeting modern malware detection.
The AMD, BIG2015, and Ember datasets are the exceptions to the rule, as they cover only short periods and do not account for the significant transformation and progression of malware over time. Neglecting this aspect can substantially impact the effectiveness of malware detectors, as the significance of features in effectively distinguishing malware may vary over time. Furthermore, many recent detection methods rely on outdated malware data sets to develop and evaluate solutions. This approach poses a potential risk to the models' ability to generalize to new and emerging malware.

For robust and accurate malware detection, it's essential to consider malware's dynamic nature and use up-to-date datasets reflecting the current threat landscape. Due to differing objectives and methodologies in sample collection, many datasets lack balance in the number of samples per family, with only a few authors addressing this issue. Commonly used techniques are oversampling families with fewer samples or undersampling families with the most samples \cite{cui2018detection} \cite{bouchaib2021transfer}.  Other authors group different family variations in a single family with similar content to show a coarse-grained classification \cite{gibert2019using}. An alternative to these solutions proposed in \cite{jain2020convo} employs a weighted linear system to solve the weights of the output layer of the convolution neural network~(CNN) to correct the bias of the Neural Network toward more common families.

\begin{figure*}[!htb]
    \centering
    \includegraphics[scale=0.5]{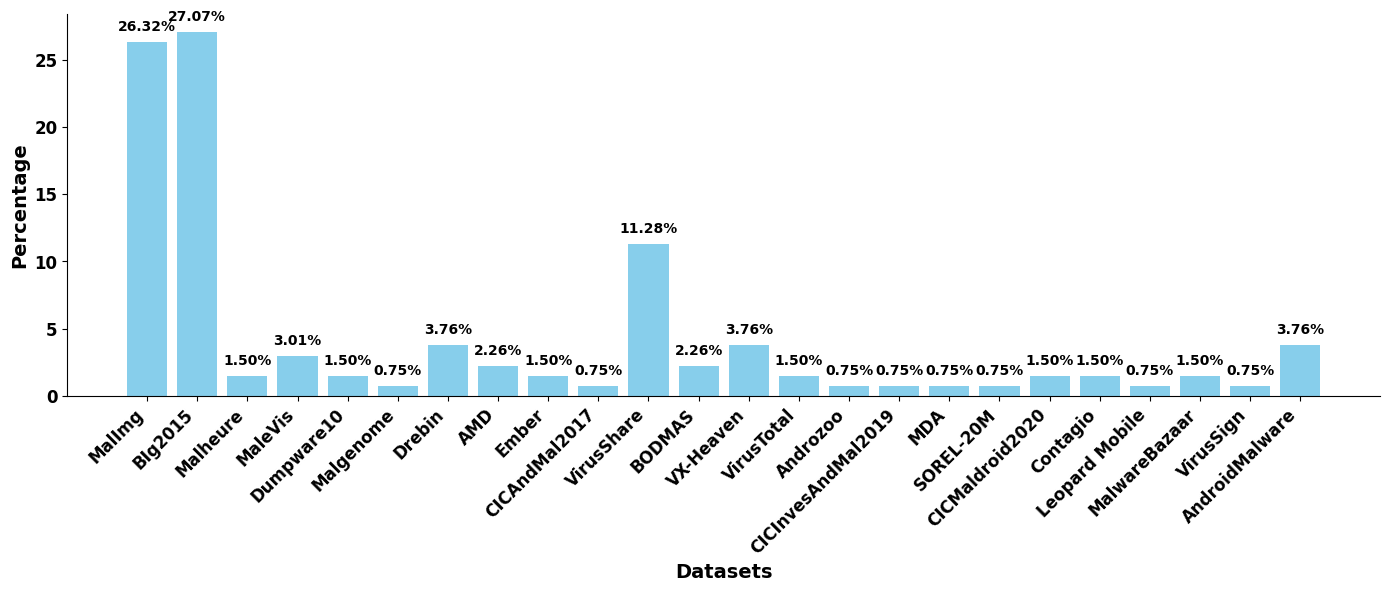}
    \caption{The percentage of datasets used in the considered papers during the year 2018-2025 \label{fig:percentageDataset}}
\end{figure*}

Figure \ref{fig:percentageDataset} illustrates the percentage of datasets used in the articles considered during the years 2018-2025. When researchers assess models, they face choices like how to partition datasets into training, validation, and test sets.  The different approaches vary substantially, even on the same dataset.   Often, clarification is needed on how the test set
\begin{table*}
{\scriptsize
\centering
\renewcommand{\arraystretch}{1.5}
\caption{The most used datasets for visualization-based malware detection \label{tab:datasets_table}}
\begin{tabular}{|p{1.8cm}|l|p{1.5cm}|p{1.5cm}|p{0.6cm}|l|p{1cm}|p{0.5cm}|l|p{4.4cm}|}

\hline
\textbf{Dataset Name} & \textbf{Year} & \textbf{\#Malicious Samples}  & \textbf{\#Benign Samples} & \textbf{\#Fam.} & \textbf{POT} & \textbf{TOF} & \textbf{Avail.} & \textbf{\#Cit.}  & \textbf{Ref.}  \\ \hline
VX-Heaven & 1990 & 0 & 0 & 0 & 1990-2010 & any & \cmark & $217$ & \cite{yakura2018malware,hashemi,li2021cnn,xiao2021image,hashemi2023ifmd} \\ \hline

VirusTotal & 2004 & 0 & 0 & 42 & 2004-2025 & any &  \cmark & 0 & \cite{oshau,jiang2023apyramid}   \\ 
\hline

Leopard Mobile & 2009 & 14,733 & 2,486 & 0 & - & apk & \cmark & $217$ & \cite{naeem2020malware} \\ \hline
MalImg & 2011 & 9,339 & 0 & 25  & - & Gray\-scale & \xmark & 1,089 & \cite{kumar2018malicious,cui2018detection,xiao2020malfcs,gibert2019using,liu2019anew,akarsh2019deep,venkatraman2019ahybrid,naeem2020malware,vasan2020imcec,bozkir2021catch,jain2020convo,vasan2020imcfn,roseline2020intelligent,go2020visualization,verma2020multiclass,pinhero2021malware,bensaoud2022deep,bagane2021classification,sudhakar2021mcft,anandhi2021malware,maldetect,zhong2022malware,kumar2022dtmic,mallik2022conrec,conti2022few,paardekooper2022designing,qiu2022malware,rustam2023malw,kim2023attention,shaukat2023anovel,guo2023mdenet,zhan2023amgmal,yang2025variant,vasan2024broad,ambekar2025fasnet,alam2025miracle,zhang2025imcmk,sharma2024migan,kumar2024imcnn,nataraj2011malware} \\ \hline 

Malheur & 2011 & 3,131 & 0 & 24  & 2006-2009 & .exe & \xmark & 890 & \cite{naeem2019identification,shiva2018windows}  \\ \hline 

VirusSign & 2011 & 0 & 0 & 0 & 2020-2025 & any & \cmark &  0 & \cite{lekssays2020anovel} \\ \hline

Contagio Mobile & 2011 & 0 & 0 & 0 & 2011-2025 & apk & \cmark & 0 & \cite{mercaldo2020deep,bensaoud2022deep}   \\ 
\hline 

Malgenome & 2012 & 1260 & 0 & 69 & 2010-2011 & apk & \xmark & 2,842 & \cite{d2020malware}\\ 
\hline
VirusShare & 2012 & 61,455,426 & 0 & 0 & 2012-2025 & any &  \cmark & 1,041 & \cite{hsiao2019malware,naeem2019identification,baptista2019anovel,hsiao2019malware,vasan2020imcec,ren2020end,hit4mal,bensaoud2022deep,li2021cnn,wang2021anovel,maldetect,zhu2022afew,shaukat2023anovel,yang2025variant,johny2025deep}   \\ 
\hline
Drebin& 2014 & 5,560 & 123,453 & 4 & 2010-2012 & apk & \cmark & 2,359 & \cite{singh2021classification,vinayakumar2022efficientnet,dhanya2023obfuscated,meng2025detecting,yang2025sac} \\ \hline
BIG2015 & 2015 & 10,860 & 0 & 9 & 2006-2015 & .exe & \cmark & 327 & \cite{yang2018aconvolutional,ni2018malware,sun2021deep,xiao2020malfcs,liu2019atmpa,gibert2019using,khormali2019copycat,qiao2019amulti,zhao2020amalware,venkatraman2019ahybrid,bozkir2021catch,dharmalaksana2022improved,roseline2020intelligent,tekerek2022anovel,pinhero2021malware,gibert2020hydra,darem2021visualization,acharya2021efficient,sudhakar2021mcft,anandhi2021malware,xiao2021image,bouchaib2021transfer,jian2021anovel,gibert2022fusing,kumar2022dtmic,mallik2022conrec,conti2022few,paardekooper2022designing,chaganti2022image,pratama2022malw,mctvd,kim2023attention,shaukat2023anovel,guo2023mdenet,yang2025variant,vasan2024broad,alam2025miracle,zhang2025imcmk,yu2025semantic,dong2024image} \\ 
\hline 
AndroZoo & 2017 & 22,707,999 & 0 & 42 & 2011-2025 & apk &  \cmark & 0 & \cite{keyes2021entroplyzer}  \\ 
\hline
AMD & 2017 & $24,553$ & 0 & $71$ & 2010-2016 & apk & \xmark          & $440$ & \cite{xiao2019animage,mercaldo2020deep,iadarola2021towards} \\ 
\hline
CICAndMal 2017 & 2018 & 426 & $5,065$ & $42$ & 2015-2017 & apk & \cmark & $217$ & \cite{lekssays2020anovel}   \\ 
\hline
Ember2018 & 2018 & $300,000$ & $300,000$ & 0 & 2012-2018 & .exe & \cmark & $338$ & \cite{vinayakumar2019robust,liu2019anew}   \\ \hline
MaleVis & 2019 & $8,750$ & $5,476$ & 25 & 2017-2018 & RGB-image & \cmark & $24$ & \cite{maldetect,roseline2020intelligent,yang2025variant,ambekar2025fasnet,alam2025miracle} \\ \hline

CICInvesAnd Mal2019 & 2019 & $426$ & $5,065$ & $42$ & 2017-2019 & apk &  \cmark & $143$ & \cite{dhanya2023obfuscated} \\ \hline

MDA & 2019 & $500$ & $500$ & $10$ & 2019-2025 & csv & \cmark & $23$ & \cite{yang2025variant} \\ \hline
SOREL-20M & 2020 & $10,000,000$ & $10,000,000$ & 0 & 2020-2020 & .exe & \cmark & $122$ & \cite{zhan2023amgmal} \\ 
\hline
CICMalDroid 2020 & 2020 & $17,341$ & 0 & 0 & 2017-2018 & apk &  \cmark & $94$ & \cite{vinayakumar2022efficientnet,yang2025sac}   \\ \hline

MalwareBazaar & 2020 & 0 & 0 & 0 & 2020-2025 & any &  \cmark & $217$ & \cite{wang2022image,conti2022few} \\ \hline

Dumpware10 & 2021 & $3,686$ & $608$ & $10$ & - & RGB image & \cmark & $44$ & \cite{bozkir2021catch,tekerek2022anovel}  \\ \hline

BODMAS & 2021 & $57,293$ & $7,142$ & 0 & 2019-2020 & .exe &  \cmark & $45$ &\cite{chai2022from,kim2023attention,meng2025detecting}   \\ \hline

Android Malware & 2025 & $14,285$ & $2,583$ & 2 & 2025 & Grayscale & \cmark & 2 & \cite{alam2025miracle} \\ \hline

\end{tabular}
}
\end{table*}
\noindent
\onecolumn

  integrates with training. Some authors repurpose the test set for validation during training, reporting its metrics as final results. Others follow a similar process but assess the model across the entire dataset.

Few reserve the test set exclusively for model evaluation. Moreover, specific datasets pose unique challenges. For instance, Big2015 comprises $10,000$ samples in both its training and test sets. Many articles only utilize the $10,000$ samples allocated for training in their experiments.
 Another problem is related to the issue of imbalance that we have discussed before. Different ways of addressing this issue may cause some models to outperform others simply because they oversampled or undersampled their dataset more effectively. Researchers must consider these problems because they limit the ability to confront and explain models and might generate results that are not representative of the model's real power. In the last few years, the accuracy of the best models reached almost $100\%$ \cite{mctvd,vasan2020imcec}.

\subsection{Image Generation}
\label{sec:visualization}
This section defines the methods used in the literature to visualize malware files in image form. First, we introduce which characteristics, dynamic or static, are usually employed to generate images. After this, we explore the techniques used to translate the data into a graphical form. 
  \label{vmethods}

\subsubsection{Image Visualization Techniques}
Our observations on the types of features utilized in visualization have led us to identify three distinct approaches: static visualization, dynamic visualization, and hybrid visualization. 

Static visualization of malware entails representing the content of malicious software, either in the form of hexadecimal binary code or source code written in assembly format, as images without executing them. The hexadecimal view displays the machine code as a series of hexadecimal values. Analysts can examine and graphically represent various pieces of information within the machine code, such as instruction code data structures, starting at a memory address. Additionally, we can visualize the entropy information derived from the hexadecimal code. Conversely, the assembly language source code encompasses not only the symbolic machine code instructions but also details about memory allocation, function calls, and variables.
\par Dynamic-based visualization techniques involve the execution of files in a sandbox environment, such as the Cuckoo sandbox, to extract relevant features representing malware behavior. These features are then processed to generate grayscale or color images based on the experimental requirements. However, compared to static visualization, we have observed a relatively limited amount of effort dedicated to dynamic visualization approaches (see Table \ref{tab:image_generation_fe}). Static visualizations lack real-time and interactive features and cannot depict temporal changes. Dynamic visualization techniques are essential to capture the evolving patterns and trends that unfold over time in malware utilizing polymorphic and obfuscation techniques. Dynamic features such as API calls \cite{wang2021anovel}, memory dumps \cite{bozkir2021catch}, network traffic \cite{xu2021hybrid} \cite{shen2023self}, system calls~\cite{casolare2021dynamic}, are widely used for malware visualization.

\par Hybrid visualization aims to harness the advantages of static and dynamic techniques. Static features are adept at capturing the structural organization and composition of the underlying binary, thereby uncovering commonalities among different malware variants. Conversely, researchers specifically consider dynamic features to model the behavioral aspects of malware. Within the framework of hybrid visualization, these two feature types can be amalgamated and consolidated into a unified image \cite{huang2020amethod} \cite{oshau} \cite{chaganti2023amulti} \cite{karbab2023swiftr}. In particular, Huang et al. \cite{huang2020amethod} used static analysis to generate an RGB representation of the static code using the byte sequence. Then, they used dynamic analysis to extract the API call sequence of the samples and generated malware images by assigning RGB values to each call. Finally, one can produce a hybrid image for classification by placing the dynamic and static representations one below the other.

We can visualize malware files as images using various techniques, including {\em(i)} raw bytes, {\em(ii)} meta-data information, or {\em(iii)} runtime data. In the following, we will dive into the details of these visualization approaches.

\textbf{Raw Bytecode Images}. Malware often appears as executable files\cite{vtstat}. Raw byte visualization converts these files into images by mapping their binary content, enabling analysis of structure, obfuscation patterns, and hidden data. This technique helps reveal embedded functionality and highlights similarities or differences between malware samples. Several studies have investigated the conversion of raw bytes from executable formats (e.g., EXE, APK, ELF) to grayscale or RGB images for visualization-based malware detection \cite{nataraj2011malware,xiao2019animage,ravi2023vit4mal}. Nataraj et al. \cite{nataraj2011malware} proposed a technique known as grayscale image conversion for visualization-based malware detection. This approach has gained widespread adoption in academic research for malware detection~\cite{makandar2015malware,choi2017malware,liu2020anovel}. They generate grayscale images by utilizing the byte values from the binary representation of executables or applications, mapping them to a range of 0 to 255, representing various levels of gray. These images consist exclusively of shades of gray, with black representing the lowest intensity and white representing the highest intensity. Many researchers divide binary executables into 8-bit substrings and convert each substring into a 1-D vector of decimal numbers, regardless of the image type or color map, before generating the grayscale image. However, some researchers have chosen different substring lengths, such as 16-bit~\cite{mohammed2021malware}, or calculated fixed lengths~\cite{kumar2016machine}, based on their specific requirements. Since the authors might not have access to the samples in binary form, they need to adapt the algorithm to transform the malware into an image to suit the specific format of the samples. The B2IMG conversion algorithm offers a variant of the standard bytecode image technique, suitable for cases where we use the sample's hex dump \cite{tekerek2022anovel}.
We can apply a similar procedure with opcodes instead of directly on the binaries by interpreting opcodes as integers and \textit{selecting only the most common ones} \cite{jian2021anovel} \cite{gibert2020hydra} \cite{li2021cnn} \cite{sun2021deep}. The values at each position can then be interpreted as bytes and transformed into grayscale images. 
 Qiao et al.~\cite{qiao2019amulti} presented another method that utilizes the Word2Vec model to compute a fixed-length vector representation for each assembly instruction in every malware sample. The word vectors are combined, generating a matrix of values for each sample, which can be interpreted as a grayscale image. Although researchers widely use and find it relatively easy to implement the bytecode method, which makes it the standard technique in most research papers. However, they often limit its utilization due to the potential violation of the \textit{locality principle} during the translation from binary to image. Directly converting non-image data to pixel values may fail to preserve the spatial relationships or visual coherence, leading to inaccuracies in representing the original data distribution within the resulting images. Advanced methods employ techniques such as hashing\cite{ni2018malware} and space-filling curves \cite{baptista2019anovel} to tackle this issue. By large, the bytecode method excels in its versatility, as it can process different types of information from a malware file, regardless of whether it is dynamic or static. It processes the information as a sequence of bytes, transforming it into a visual image for effective visualization.

\textbf{Metadata Information Images}. The metadata of malware files encompasses supplementary information linked to the file, which extends beyond the binary code of the malware. Metadata can help analyze and understand malware because it provides additional context and information that can help identify and classify the malware. Various types of metadata information can be used in different ways to create visual elements that aid malware analysis. For example, metadata from the \textit{Portable Executable (PE) header}, such as the address of entry points, sections, and optional header details, can be transformed into visual elements to help understand the structural aspects of the files. The PE header contains valuable content, including the file structure description and other information that researchers can exploit to identify and classify malware. Moreira et al. \cite{moreira2023improving} used this technique to classify 25 ransomware families. In their paper, the authors used the raw PE header without assuming the length of the fields or requiring expert knowledge to extract features. The PE header has a fixed length of 1024 bytes, which researchers can easily interpret as a $32 \times 32$ matrix with the application of an SFC. Researchers have also experimented with the use of boundaries on executables sections. In particular, Xiao et al. \cite{xiao2021image} used the section seizes extracted from the malware PE header to mark the limits of each section with a different color to facilitate visual distinction between each part. 
Metadata related to \textit{imported and exported functions or modules}, including their relationships and dependencies, can also be visualized to provide insights into the code's functionality and the interactions between different components. Metadata pertaining to \textit{digital signatures} can be overlaid on the images of the binaries to visually depict the trustworthiness or authenticity of the file, aiding in the identification of potentially malicious executables. In addition, metadata like \textit{file size}, \textit{creation date}, and \textit{file extensions} can be encoded as visual features and incorporated into the visualization to provide additional context and information about malware samples. The technique of utilizing metadata for malware visualization is particularly prevalent in analyzing Android malware, owing to the abundance of useful metadata information available in the APK files. Application signatures, certificates, and information in the \texttt{AndroidManifest.xml} file serve as valuable metadata sources for the analysis and comprehension of Android malware. In \cite{bakir2023droidencoder}, Bakir et al. use the \texttt{classes.dex} file as the basis for their analysis. This file can be read as an integer array and resized to form a 2-D matrix. Singh et al. \cite{singh2021classification} use the \texttt{classes.dex}, \texttt{resources.arsc}, \texttt{AndroidManifest.xml} files, and the content of the \texttt{META-INF} folder in order to generate the visualization representation of the malware. These files are interpreted as an array of bytes, resized in a matrix, and combined to represent the final 2-D grayscale image. Metadata information provides crucial details that aid researchers in understanding the characteristics and behaviors of Android malware.

\textbf{Run-time Information for Image Generation}. Another approach for generating an image from malicious binaries involves gathering features obtained through behavioral or runtime analysis. This method executes the malware within a secure environment or sandbox and uses the collected data to create images. This data includes network communications logs, file access, API calls, return values, call times, call process numbers, system calls, resource consumption, registry keys, memory artifacts, and more. The most commonly adopted technique is to monitor system calls ~\cite{wang2021anovel}. Different approaches exist in terms of interpreting the extracted information. For instance,~\cite{shiva2018windows} assigns a hexadecimal code to each extracted API call, resulting in a byte sequence that can interpreted as a grayscale image. Conversely, \cite{gibert2020hydra} models the image as a matrix, where each value corresponds to an API call, and it assigns a value of 0 if the malware does not use the API call or 1 if it does. Another method to interpret dynamically extracted information as images is achieved by substituting the dynamic analysis report with the malware sample's original binary and generating a grayscale image from that~\cite{ma2022visual}. In this case, the report file itself is used instead of the malware file to represent the sample in further analysis steps. The dynamic analysis data is gathered from \texttt{.json} files and converted into images using the standard grayscale method. This way, the techniques usually applied on the binary file for extracting static features can be applied on a \texttt{.json} file that contains the representation of dynamic information.
\par In their study, Bozkir et al.~\cite{bozkir2021catch} employed memory forensics techniques to extract temporary data from the computer's physical and virtual memory. They utilized tools such as ``Procdump'' (reference missing) to dump this data while executing the malware sample in a virtual environment. Due to the substantial file sizes produced by this method, researchers adopted RGB images. In this approach, each image pixel represents three sequential bytes, allowing for the inclusion of color, in contrast to the traditional grayscale method that relies on bytecode. The approach used by Xu et al.~\cite{xu2021hybrid} involves the use of network-captured packets as the foundation for image generation. Concatenating the captured PCAP files for each sample, researchers interpret their bytes as a grayscale image. Agrafiotis et al. \cite{agrafiotis2022image} have more recently explored using PCAP files in conjunction with transformers.

In summary, dynamic information visualization has excellent potential for malware analysts \cite{tang2019dynamic}. However, most of the research focuses on the static feature-based image generation technique (see Table \ref{tab:image_generation_fe}. The emphasis on static feature-based image generation arises from the challenges and diverse approaches associated with dynamic analysis, which become more complex with larger datasets, making testing challenging. Additional experimentation is needed to explore the real potential of dynamic analysis and compare it with more established static analysis techniques.

\subsubsection{Image Representation}
This section elucidates the methodologies employed in existing literature to convert extracted dataset information into visual images. The utmost importance lies in choosing an optimal technique for image representation because it defines how to organize the data visually. These approaches represent data in a visual format, maintaining essential features such as relational structures and interconnections. Researchers must balance creating an effective image layout for feature extractors and classifiers while minimizing information loss during data transformation.

\textbf{Space Filling Curves}. The conventional bytecode method for translating a binary file into an image leads to information loss, including a partial degradation of locality. This implies that arranging bytes in rows will cause line breaks to separate bytes originally adjacent in the code. In an attempt to address this issue, researchers have conducted experiments on the byte layout of the image, which refers to the order in which the pixels are generated and arranged within the image. Several studies, including those by Rustam et al.\cite{rustam2023malw}, Tekerek et al.\cite{tekerek2022anovel}, Qiu et al.\cite{qiu2022malware}, Pratama et al.\cite{pratama2022malw}, and Chaganti et al.~\cite{chaganti2022image}, discuss the most commonly used and straightforward approach, known as the ``line-by-line'' or ``carriage return'' method. In this method, pixels are arranged in a specific order, placing each pixel after the previous one until reaching the preferred line width. Subsequently, the next pixel moves to the start of the following line. However, this method overlooks the issue of locality.

In addition, researchers have explored the utilization of various Space-Filling Curves (SFC) to generate grayscale images. Space-filling curves are mathematical constructs that traverse and fill a given space continuously. In malware visualization, one can use these curves to fill a matrix of dimensions $N \times M$, where each pixel represents a value derived from the original malware sample. This approach allows for the generation of the corresponding image of the sample.

\begin{figure*}
     \centering
     \begin{subfigure}[b]{0.19\textwidth}
         \centering
         \includegraphics[width=.8\linewidth, height=2.3cm]{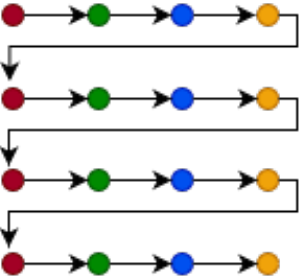}
         \caption{Line-by-line}
     \end{subfigure}
     \hfill
     \begin{subfigure}[b]{0.19\textwidth}
         \centering
         \includegraphics[width=.8\linewidth, height=2.3cm]{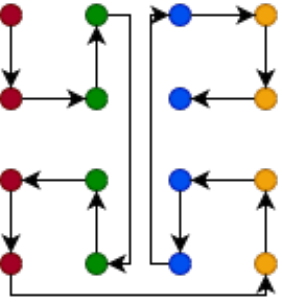}
         \caption{Gray-code}
     \end{subfigure}
     \hfill
     \begin{subfigure}[b]{0.19\textwidth}
         \centering
         \includegraphics[width=.8\linewidth, height=2.3cm]{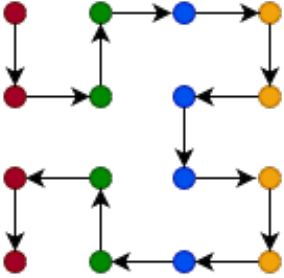}
         \caption{Hilbert}
     \end{subfigure}
     \hfill
     \begin{subfigure}[b]{0.19\textwidth}
         \centering
         \includegraphics[width=.8\linewidth,height=2.3cm]{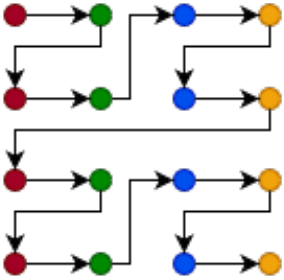}
         \caption{Z-order}
     \end{subfigure}
     \hfill
     \begin{subfigure}[b]{0.19\textwidth}
         \centering
         \includegraphics[width=.8\linewidth,height=2.3cm]{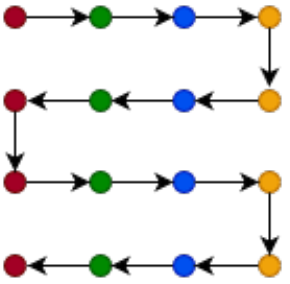}
         \caption{Wrap around}
     \end{subfigure}
     
        \caption{The figure shows a representation of the most common SFC used in the literature. The line-by-line method is the most common way grayscale bytecode images are generated.}
        \label{fig:sfc}
\end{figure*}

In~\cite{oshau}, O’Shaughnessy et al. use three different SFC techniques: Z-order, Gray-code, and Hilbert curves. Similarly, in the study conducted by the authors in~\cite{hit4mal}, the wrap-around and H-curve approaches are utilized in addition to the standard line-by-line method. Hashemi et al.~\cite{hashemi} also adopt the wrap-around method, among others, to generate the images. Another study by authors in~\cite{baptista2019anovel} leverages the website binvis.io ~\cite{binvis} to generate the images. This website offers various ways to represent ASCII printable characters by assigning each a distinct color and arranging the pixels using an SFC of choice, including line-by-line, Z-order, and Hilbert. The Figure\ref{fig:sfc} illustrates the most commonly used SFC in the literature.

\textbf{Markov Images}. In various studies,~\cite{mctvd,chai2022from,pinhero2021malware,hashemi} researchers explored the assumption that the order of instructions in an executable file is not random but based on the locality principle. Their experiments employed Markov chains to generate information-rich images, recognizing that the preceding attribute influences each attribute (bytes, mnemonics, opcodes, etc.) in the file.
To represent files visually, one approach is to treat each byte in the byte stream as a state, with 256 possible states for each element. By assuming that the next state depends only on the current state, we can conceptualize the byte sequence as a Markov chain, such that $P(s_{i + 1}|s_{0},...,s_{i})=P(s_{i + 1}|s_{i})$ if $s_{i}$ represents the attribute in position $i$. The formula in Equation \ref{P} expresses the probability of transitioning from state 
$m$ to the subsequent state $n$.
\begin{equation}
\label{P}
  P_{m,n}=P(n|m)=\frac{f(m,n)}{\sum^{256}_{n=0}f(m,n)}  
\end{equation}

Where $f(m,n)$ is the frequency of the transition from state $m$ to state $n$. In contrast, the denominator $\sum^{256}_{n=0}f(m,n)$ signifies the sum of frequencies for all possible transitions from state $m$. This approach allows for the extraction of a matrix of dimensions $256\times256$. The resulting matrix, often called a Markov image, encapsulates the transition probabilities and can visually represent the underlying malware characteristics.

\textbf{Hashing Schemes}. Researchers have used different hashing algorithms to generate images, allowing them to extract locality-sensitive information and combine local and visualization data in feature images. One particular algorithm, ``Minhash'', can create a concise visual summary or fingerprint of malware samples. Minhash \cite{broder1997on} is a prominent hashing algorithm employed for estimating the similarity between sets. It finds extensive utilization across diverse domains, including document similarity analysis, recommendation systems, and clustering. In the context of malware visualization, we represent each sample using a set of opcodes or opcode groups, applying the minhash hashing function to them. This method produces a fixed-size image that can be used in the following steps of the pipeline. In malware visualization, researchers extract relevant attributes or characteristics from the samples, such as opcode sequences, API calls, byte-level n-grams, or other suitable representations of the malware file. Next, researchers extract specific elements(i.e., shingles) of a fixed length contiguous subsequences and apply the hashing algorithm to the resulting set. This procedure culminates in the creation of grayscale images~\cite{sun2021deep}. 
Charikar et al.~\cite{charikar2002simhash} presented \textit{Simhash} as a locality-sensitive hashing scheme primarily used for recognizing similar text and identifying duplicated web pages. The idea behind \textit{Simhash} is the use of rounding algorithms to maintain the locality principle while hashing.
A rounding algorithm converts numerical values to a specified level of precision or rounding interval. This procedure maintains locality in its output by ensuring that nearby values in the input remain close to each other in the rounded result, preserving the proximity of data.
Common hashing algorithms have the desirable property of low collision rate, \textit{Simhash}, similar to \textit{Minhash}, instead aiming to make the hashes of two similar inputs similar themselves. In~\cite{ni2018malware}, \cite{darem2021visualization} and ~\cite{mctvd}. The technique of \textit{Simhash} involves representing a document as a vector with a specified number of bits. Each bit in the vector corresponds to a hash value calculated from the document's keywords. To be more specific, the value of the $n$-th bit of the vector is determined by computing the hash value of all the keywords present in the document. 
We can determine the value of the $n$-th bit as 1 or 0 based on whether the count of hash values with the $n$-th bit (for all keywords) set to 1 is greater than the count of hash values with the $n$-th bit set to 0 or not.

\textbf{Entropy Images}. Entropy is a concept derived from information theory and statistics~\cite{shannon}. It measures the randomness or uncertainty of a system or data. Experts in malware analysis primarily utilize it to study the techniques employed by malware authors to obfuscate or conceal their malicious intentions. Calculating entropy values involves dividing malware binaries into segments~\cite{fu2018malware} and then determining the entropy value for each segment. Subsequently, these entropy values are converted into pixels to create an image representation. Another approach, proposed by~\cite{zhu2022afew}, \cite{xiao2020malfcs} and~\cite{chai2022from}, involves using entropy graphs where the entropy of the entire binary serves as the image itself. This method utilizes entropy time series to capture the content of local discriminative features in each file. Employing this approach proves beneficial for the classifier in accurately distinguishing the samples into their respective malware classes.

\textbf{Colored Images}. Researchers have recently embraced the use of RGB and other colored images in malware detection due to their ability to convey more texture information compared to grayscale images, resulting in improved detection rates \cite{naeem2020malware,jian2021anovel,mctvd}. 
However, the true advantage of RGB images lies in their capability to incorporate different representations within the three color channels. There are two primary approaches to generating RGB images. The first approach involves utilizing the three RGB channels to represent three distinct characteristics \cite{xuan2024bitcn}, while the second approach involves directly extracting the three channels from the attributes of malware binaries \cite{vasan2020imcfn}. In the first case, we create three grayscale images using different techniques, and we use their output as each of the three channels of RGB~\cite{mctvd,pinhero2021malware}. For example, in~\cite{conti2022few}, the authors introduce a novel image variant referred to as the GEM image. It is a 3-channel image combining the gray-level matrix, Markov, and entropy graph images. 

Authors can utilize a combination of dynamic and static data to create more accurate representations of malware and its unique characteristics in generated images. Another approach involves reducing image dimensions and overlaying various binary sections, as described in~\cite{maldetect}. Although this approach does not provide additional information to feature extractors and classifiers using RGB images, it can be useful for resizing images to a fixed dimension. In \cite{vasan2020imcfn}, researchers utilized a color map to convert grayscale images into colored images, which improved classifier accuracy. Additionally, researchers have used RGB images to incorporate specific information extracted from executable headers. For instance, Xiao et al.~\cite{xiao2021image} used different colors to highlight the PE sections, improving classifier performance.

\subsection{Feature Processing Methods}
\label{sec:featureExtraction}
This section describes the data preprocessing phase utilized in the literature. This phase entails crafting a more manageable and informative dataset suitable for training and evaluating the models. Here, we examine the techniques for {\em(i)} extracting, {\em(ii)} selecting, and {\em(ii)} reducing features' dimensionality. Table \ref{tab:image_generation_fe} describes for each analyzed paper,

\subsubsection{Feature Extractors}
This section addresses the use of different feature extractors usually employed in combination with Machine Learning techniques. The algorithms presented here provide an explainable methodology to extract valuable characteristics from images and extrapolate discriminating features. In the following section we will analyze the literature related to Gabor-based Image Segmentation Technique (GIST), Local Binary Pattern method (LBP), Scale-Invariant Feature Transform (SIFT), Gray-Level Co-occurrence Matrix (GLCM), Histogram, Neural Network (NN), and the integration of more than one techniques.

\textbf{Gabor-based Image Segmentation Technique (GIST)}. The Gabor-based Image Segmentation Technique~(GIST) \cite{grigorescu2002comparison} was explicitly designed to segment and represent images by analyzing the spatial distribution of local spatial frequencies and orientations. Gabor filters imitate the functioning of the human visual cortex and consist of sinusoidal signals that detect texture variations in an image based on their frequency and orientation. These filters act as local bandpass filters, capturing localization properties in both the spatial and frequency domains. A set of Gabor filters with various orientations and frequencies ensures spatial localization, extracting precise features by applying them to the image. Gabor filters possess the capability to identify distinctive textural characteristics that differentiate malware from benign files or distinguish among different families of malware as state in~\cite{nataraj2011malware,anandhi2021malware,pinhero2021malware,sharma2022windows,naeem2019identification}. Tuning the parameters of these filters requires careful attention, and selecting suitable values can be challenging. Moreover, since GIST features depend on the global structure of an image, an adversary with knowledge of the method could potentially evade detection by rearranging different sections of the code, as mentioned in \cite{gibert2019using,liu2020anovel}. 
\par In~\cite{corum2019robust} the authors effectively distinguished between benign and malicious PDF samples by transforming the files into images by utilizing byte and Markov plots. Subsequently, distinctive visual attributes of the images were extracted using Keypoint descriptors such as SIFT (Scale-Invariant Feature Transform), Oriented FAST, and Rotated BRIEF, along with texture features like LBP (Local Binary Patterns), Gabor Filter, and Local Entropy. Remarkably, they achieved an impressive F1-score of $99.48\%$ in Random Forest (RF) classification by employing a byte plot with Gabor Filter.\\
\textbf{Local Binary Pattern method (LBP)}. Researchers widely employ the Local Binary Pattern method (LBP)~\cite{ojala1994performance,luo2017binary,hashemi,wu2017android,naeem2022android,wang2022image} as a feature extractor in malware visualization approaches. This method generates a new LBP image by comparing neighboring pixels and assigning them binary values (0 or 1), resulting in an 8-bit binary array. Afterward, these arrays are converted into decimal values by traversing either clockwise or anti-clockwise. Finally, store these decimal values in the LBP image at their respective pixel locations. To calculate the LBP value for each center pixel ($LBP_c$), researchers utilize Equation \ref{lbp}.

\begin{equation}
\label{lbp}
   LBP_{c}=\sum_{n=0}^{n-1}L(g_{n}-g_{c}))2^{n}
\end{equation}
Where $n$ is the number of neighborhood pixels, $g_{n}$ and $g_{c}$ are the gray levels of the neighborhood pixel and center pixel, respectively. We compute $L(x)$ by using Equation \ref{l} and then combine the resulting $LBP_{c}$ values to generate the LBP image.
\begin{equation}
\label{l}
   L(x) =
    \begin{cases}
      0 & \text{if $x<0$} \\
      1 & \text{if $x>=0$} \\
    \end{cases}   
\end{equation}

In~\cite{tuncer2020automated}, the authors propose a novel method for malware recognition based on byte code. The method incorporates several steps, including feature extraction using local neighborhood binary pattern (LNBP), concatenation of features, feature selection utilizing Neighborhood Component Analysis~(NCA), feature reduction through Principal Component Analysis~(PCA), and classification using Linear Discriminant Analysis. The LBP feature is known for its stability and noise resistance but is limited to focusing solely on local features\cite{gao2021malware}. In~\cite{wu2017android}, researchers have introduced an approach that utilizes Local Binary Patterns and Principal Component Analysis to detect Android malware. Notably, their method achieved an impressive accuracy rate of $90\%.$

\textbf{Scale-Invariant Feature Transform (SIFT)}. SIFT descriptors can increase detection accuracy~\cite{lowe1999object}. This method extracts features by computing gradient magnitude histograms in eight orientations of bins. Splitting the image into small sections generates these bins. A sliding window executes the computations, computing the gradient histograms of each image locality. Cascaded connection functions obtain the final image feature descriptors\cite{liu2019anew,naeem2019identification} \cite{naeem2020malware}.
To classify Android malware images, DeepVisDroid~\cite{bakour2021deepvisdroid} utilizes four distinct image-based local features and three diverse image-based global features. These features are employed to train a lightweight 1d-convolutional Neural Network model. Recently, the author in~\cite{kumar2022identification} converted the malicious file into grayscale images to extract local and global textural features. They employed SIFT and other descriptors for local feature extraction. GIST and other methods for global feature extraction. The author also developed a BoVW algorithm to construct a local feature map by selecting low-dimensional features from the grayscale images. These feature maps were combined to train K-NN, SVM, RF, NB, and ExtraTree classifiers.

\textbf{Gray-Level Co-occurrence Matrix (GLCM)}. Gray-Level Co-occurrence Matrix-based feature extraction (GLCM)~\cite{glcm} is a popular method to extract texture information from images. Extracting GLCM features involves considering the spatial relationship of distance and orientation between pixels. A GLCM matrix is created by estimating the probability density function of gray-level pairs with a predetermined spatial relationship, usually a distance of one in the four cardinal directions. From that, we calculate relevant statistics to represent the initial image. These statistics can vary from one research work to another. 
Table \ref{tab:image_generation_fe} contains, for each analyzed paper, the year of publication; the information used to create the image (i.e., [S] for static information, [M] for metadata, [D] for dynamic information); the technique used for the binary representation of the image (i.e, [LbyL] if the paper uses the classic line by line binary representation, [Others] if the paper uses other Space-Filling Curve (SFC) techniques to lay the pixel in the image other than LbyL); the image generation technique (i.e., [Markov] if the Markov technique is used to generate the images; [Hashing] if hashing schemes have been used to generate the images; [Entropy] if Entropy Analysis have been used to generate images; [Original] if the image generation technique employed are original); the information about the image color (i.e., [A] for colored images that use the color to add information, [B] for the colored images that do not add information); the used feature extraction technique (i.e, [G/GIST] for GIST or Gabor filters, [LBP] for Local Binary Pattern; [SIFT] for Scale-Invariant Feature Transform, [H/GLCM] for Gray-Level Co-occurrence Matrix or Haralick features, [Hist] if the paper uses histrograms of any type to extract features, [NN] - if the paper uses a Neural Network to extract feature. The specific architecture of the NN will be specified if possible and a Greek letter will indicate if [$\alpha$] the model is trained from scratch, [$\beta$] transfer learning has been applied, or [$\gamma$] if fine-tuning has been applied, [Original] if the feature extraction technique is original); if the model uses a fusion technique. Karanja et al. \par \cite{karanja2020analysis} also utilized GLCM to extract features from a grayscale image. In particular, researchers used only five of the proposed statistical measures of texture features to extract the relevant data. The selected features are Entropy, Contrast, Correlation, Angular Second Moment (ASM), and Inverse Differential Moment (IDM).
The authors in~\cite{verma2020multiclass} combines first-order and GLCM-based second-order statistical texture features with ensemble learning techniques to classify malware images. Recently, in~\cite{singh2021classification}, the authors selected nineteen statistics that measured with the previously described method, resulting in 76 features.

\textbf{Haralick textures}~\cite{haralick1973textural} evaluate image texture properties like coarseness, smoothness, and regularity using a co-occurrence matrix and equations for contrast, correlation, and angular moment. The authors in~\cite{ahmadi2016novel} conducted experiments that focused on global properties using Haralick features, but modern techniques have now surpassed them. Unver et al.\cite{unver2020android} transformed Android apps into grayscale images and used Haralick texture and Hu Moments to achieve over $98\%$ accuracy on malware classification. Karanja et al.~\cite{karanja2020analysis} proposed an approach to classify IoT malware using Haralick image features and classifiers like naïve Bayes, K-Nearest Neighbor, and Random Forest. The method involves converting a binary file into a grayscale image, computing GLCM for each image, and deriving five Haralick features.
\par \textbf{Histogram}. The histogram captures the color distribution within an image by comprising multiple bins, each representing the frequency of pixels at specific intensity levels~\cite{bakour2021visdroid}. For grayscale images, histogram construction involves 255 bins, which correspond to intensity values ranging from 0 to 255.
One can employ contrast-limited adaptive histogram equalization algorithms to enhance the local contrast of each region in a malware image and shift the focus from identifying similar codes to identifying similar image regions within a malware family. These algorithms adjust the pixel intensities within individual regions of the image, effectively improving the distinction between different regions~\cite{zhong2022malware}. The evaluation of VisMal demonstrates an impressive accuracy rate of $96\%$. Kumar et al.~\cite{kumar2019texture} demonstrate the effectiveness of the HOG feature descriptor in extracting texture statistics from grayscale images. The experiments highlighted the HOG descriptor's efficacy, achieving remarkable accuracy and outperforming the Gabor filter and LBP analysis techniques. The researchers in~\cite{bozkir2021catch} endeavored to generate RGB images from memory dumps to capture visual patterns. They create feature vectors using GIST and HOG descriptors, which were subsequently classified using various classifiers. Notably, the SMO algorithm achieved an impressive accuracy of $96.39\%$ when utilizing GIST and HOG feature vectors. Another similar approach, as discussed in~\cite{dai2018malware}, involves extracting a memory dump file and converting it into a grayscale image. Subsequently, features are extracted from the image using the histogram of gradients technique. 
 
\textbf{Neural Networks (NN)}. The use of Neural Networks as feature extractors in visualization-based malware detection is gaining popularity due to its ability to automate the analysis and categorization of malware images~\cite{vasan2020imcfn, davuluru2019convolutional,xiao2020malfcs}. By training NN on a large dataset of malware images, they can automatically identify crucial characteristics indicative of malicious behavior or traits. These features encompass shapes, textures, visual patterns, and other visual properties that distinguish malware images. Convolutional Neural Networks, in particular, are capable of detecting code modifications resulting from code reallocation by learning spatially invariant patterns~\cite{gibert2019using}. Hence, researchers often prefer the simplicity and utility of using NN as feature extractors. In the study~\cite{xiao2021image}, a fine-tuned VGG16 model, specifically the first five convolution blocks, was employed as the feature extractor for images. Binary texture analysis has shown improved accuracy and reduced time consumption, making it effective in addressing obfuscation issues. However, the computational cost of extracting complex texture features from malware images remains high~\cite{vasan2020imcfn}. In contrast, the authors in~\cite{johny2025deep} utilized the full set of convolutional layers from the pre-trained VGG16 (prior to fully connected layers). This choice allows the model to utilize rich features without the loss of spatial resolution that would occur in fully connected layers. The study\cite{yang2025variant}  utilizes a deep residual network (ResNet) as the primary model for feature extraction. ResNet is known for its capability to learn complex patterns in data due to its deep architecture and residual learning mechanism. The network can effectively capture both local and global features of malware images after the enhancement process, improving the overall accuracy of the detection model. One significant advantage of feature extraction through Neural Networks is that it does not require human intervention or description during the extraction process.
 
\textbf{Feature Integration}. Various feature extraction methodologies can ensure a more faithful representation of the original data. Authors have experimented with using local and global feature extractors to get information about the general features of the malware without losing focus on local peculiarities. Naeem et al. in \cite{naeem2019identification} designed an original feature extractor, LGMP, by fusing D-SIFT as the local feature extractor and GIST as the global feature extractor. Their test demonstrated a significant improvement in the accuracy of the classifier when using LGMP instead of just local ($5.4\%$ increase) or just global ($9.2\%$ increase) feature extractors. Feature integration, which involves synthesizing local and global features, achieves a balance in the influence of local and global features using Gaussian weights.
Several authors have also experimented with the fusion of features extracted from different CNNs \cite{vasan2020imcec} \cite{gibert2020hydra}\cite{dib2021multi}, showing sensible improvements from the best results of a single CNN.
The most used technique in feature integration among different CNNs is the simple concatenation of the first flattened layer, usually with previous dimensionality reduction, to make the ensemble flattened layer manageable. Gibert et al. \cite{gibert2020hydra} specifically use different generated images to extract features from the CNNs, all originating from the same binary file. Meng et al.~\cite{meng2025detecting} integrates features through a deep Neural Network model that combines local and global information from these distinct perspectives, allowing for a more comprehensive analysis of app behaviors. By leveraging multi-view vectorial fusion, LensDroid enhances the detection capabilities by capturing high-level semantics that are not inherently linked. The authors of \cite{yang2025sac} proposed the SAC framework that integrates malware features from DEX files by collaboratively modeling image and graph data. It uses a task-oriented CNN (IFNeXt) for local image features and a dual-channel GCN for global bytecode structure, capturing both content and structural malware characteristics. Yu et al. \cite{yu2025semantic} proposed a lossless feature extraction and integration method that transforms malicious code into semantically intact images. This technique preserves bytecode information and code text correlations through strategic pixel arrangement and interleaved encoding, mitigating semantic truncation. A multi-scale feature extraction module ensures uniform embedding of variable-length samples into fixed-size feature maps, capturing both local and global contextual features as long-text sequences within a matrix. By integrating these features, this approach overcomes information loss inherent in traditional resizing and cropping, enhancing malware classification accuracy. 
\par Recent advances in malware detection have explored integrating multiple visual modalities to enhance classification performance. One prominent approach is presented in~\cite{johny2025deep}, they propose a multimodal deep learning fusion framework that leverages grayscale images, entropy graphs, and simhash images generated from malware binaries. Using separate VGG16 models trained on each modality, the authors explored feature-level fusion using common operators such as addition, average, maximum, and concatenation. This study highlights the effectiveness of fusion in malware detection and the critical role of selecting suitable fusion operators for robust multimodal learning.

\subsubsection{Feature selection}
An encoding-based technique is mentioned in~\cite{naeem2019visual} to reduce the dimensions of D-SIFT and GIST features. During the local feature reduction phase, they employed D-SIFT to identify interest points in the malware image and extract 128-dimensional features from each patch obtained from a dense grid of patches. Subsequently, they applied the Bag of Features~(BOF) model to perform dimensionality reduction. The BOF model consisted of multiple steps, including identifying salient local points and creating a dictionary of local features using Fisher vector encoding. 

\begin{table*}
\caption{Techniques used to extract features from the starting malware.  }
\label{tab:image_generation_fe}
\begin{center}
{\scriptsize
\begin{tabular}{|p{2.3cm}|c|p{0.6cm}|p{1.5cm}|p{2.5cm}|p{0.9cm}|p{3.5cm}|p{1.6cm}|} \hline
\textbf{Ref.} & \textbf{Year} &\textbf{Info} & \textbf{Binary Repr.} & \textbf{Image Gen. Technique} & \textbf{Color} & \textbf{Feature Extraction Technique} & \textbf{Fusion Technique} \\ \hline


Fu et al. \cite{fu2018malware} & 2018 & S & LbyL & SimHash & \cmark & H/GLCM, Others & \xmark \\ \hline

Yakura et al. \cite{yakura2018malware} & 2018 & S & LbyL & - & \cmark & NN, Others & \xmark \\ \hline

Yang et al. \cite{yang2018aconvolutional} & 2018 & S & LbyL & - & \xmark & NN & \xmark \\ \hline

Darus et al.\cite{darus2018android} & 2018 & S & LbyL & - & \xmark & G/GIST & \xmark \\ \hline
 
Kumar et al. \cite{kumar2018malicious} & 2018 & S & LbyL & - & \xmark & NN & \xmark \\ \hline
  
Su et al. \cite{su2018lightweight} & 2018 & S & LbyL & - & \xmark & NN & \xmark \\ \hline
  
Cui et al. \cite{cui2018detection} & 2018 & S & LbyL & - & \xmark & Original & \xmark \\ \hline

Ni et al. \cite{ni2018malware} & 2018 & S & \xmark & Simhash & \xmark & NN & \xmark \\ \hline

Shiva et al. \cite{shiva2018windows} & 2018 & D & & Original & & Original & \xmark \\ \hline

Xiao et al. \cite{xiao2019animage} & 2019 & S & LbyL & Original & \cmark & NN & \xmark \\ \hline
  
Vinayakumar~et~al.~\cite{vinayakumar2019robust} & 2019 & S & LbyL & - & \xmark & NN, Others & \xmark \\ \hline
 
Liu et al. \cite{liu2019atmpa} & 2019 & S & LbyL & - & \xmark & G/GIST, NN & \xmark \\ \hline
  
  Hsiao et al. \cite{hsiao2019malware} & 2019 & S & LbyL & Average Hash & \xmark & NN & \xmark \\ \hline
  
  Gibert et al. \cite{gibert2019using} & 2019 & S & LbyL & - & \xmark & NN &\xmark \\ \hline
  
  Khormali et al. \cite{khormali2019copycat} & 2019 & S & LbyL & - & \xmark & NN & \xmark \\ \hline
 
  Naeem et al. \cite{naeem2019identification} & 2019 & S & LbyL & - & \xmark & G/GIST, SIFT & \xmark \\ \hline
 
  Liu et al. \cite{liu2019anew} & 2019 & S & LbyL & - & \xmark & LBP, SIFT & \xmark \\ \hline
  
  Akarsh et al. \cite{akarsh2019deep} & 2019 & S & LbyL & - & \xmark & NN & \xmark \\ \hline

  Venkatraman~et~al.~\cite{venkatraman2019ahybrid}& 2019 & S & LbyL & Entropy & \xmark & NN & \xmark \\ \hline
  
  Hashemi et al. \cite{hashemi}& 2019 & S & Others & Markov & \xmark & LBP & \xmark \\ \hline
  
  Baptista et al. \cite{baptista2019anovel} & 2019 & S & Others & - & \cmark & Original & \xmark \\ \hline

  Yang et al. \cite{yang2019anovel} & 2019 & M,D & \xmark & Original & \xmark & Original & \xmark \\ \hline
  
  Chen et al. \cite{chen2019applying} & 2019 & S & & Original & & NN & \xmark \\ \hline

    Qiao et al. \cite{qiao2019amulti} & 2019 & S & & Original & & NN & \xmark \\ \hline    
   
  Mercaldo et al. \cite{mercaldo2020deep} & 2020 & S & LbyL & - & & NN(FFNN) & \xmark \\ \hline

  Lekssays et al. \cite{lekssays2020anovel} & 2020 & S & LbyL & G/GIST & & NN & \xmark \\ \hline

  Naeem et al. \cite{naeem2020malware} & 2020 & S & LbyL & - & \cmark & NN & \xmark \\ \hline
  
  Vasan et al. \cite{vasan2020imcec} & 2020 & S & LbyL & - & & NN & \xmark \\ \hline
  
  Liu et al. \cite{liu2020anovel} & 2020 & S & LbyL & - & \xmark & NN & \xmark \\ \hline
  
  Karanja et al. \cite{karanja2020analysis} & 2020 & S & LbyL & - & \xmark & H/GLCM & \xmark \\ \hline

    Xiao et al. \cite{xiao2020malfcs} & 2020 & S & \xmark &  Entropy & \cmark & NN &\xmark \\ \hline

     Jain et al. \cite{jain2020convo} & 2020 & S & LbyL & Original & & NN & \xmark \\ \hline  

  Vasan et al. \cite{vasan2020imcfn} & 2020 & S & LbyL & - & \cmark & NN &\\ \hline

  Zhao et al. \cite{zhao2020amalware} & 2020 & S & LbyL & - & \cmark & NN & \xmark \\ \hline

    Roseline et al. \cite{roseline2020intelligent} & 2020 & S & LbyL & - & & Original & \xmark \\ \hline

  Go et al. \cite{go2020visualization} & 2020 & S & LbyL & - & & NN &  \\ \hline

  Verma et al. \cite{verma2020multiclass} & 2020 & S & LbyL & - & &  H/GLCM, Hist & \xmark \\ \hline

  Ren et al. \cite{ren2020end} & 2020 & S & LbyL & Original & &   NN & \xmark \\ \hline

  Vu et al. \cite{hit4mal} & 2020 & S & Others & Entropy & \cmark & H/GLCM, NN & \xmark \\ \hline

  Girbert et al. \cite{gibert2020hydra} & 2020 & S & LbyL & - & & NN, Others & \xmark \\ \hline

  Iadarola et al. \cite{iadarola2021towards} & 2021 & S & LbyL & - & & NN & \xmark \\ \hline

  Singh et al. \cite{singh2021classification} & 2021 & S & LbyL & - & & G/GIST, LBP, H/GLCM, NN & \xmark \\ \hline
 Dib et al. \cite{dib2021multi} & 2021 & S & LbyL & - & & NN (CNN$\alpha$) & \cmark \\ \hline

  Darem et al. \cite{darem2021visualization} & 2021 & S & LbyL & Original & & NN, Others & \xmark \\ \hline

  Sun et al. \cite{sun2021deep} & 2021 & S & \xmark & MinHash & \xmark & \xmark & \xmark \\ \hline

    Bozkir et al. \cite{bozkir2021catch} & 2021 & D & LbyL & - & \cmark & G/GIST, Hist & \xmark \\ \hline

      Pinhero et al. \cite{pinhero2021malware} & 2021 & S & LbyL & Markov, Entropy & \cmark & G/GIST, NN & \xmark \\ \hline

  Bagane et al. \cite{bagane2021classification} & 2021 & S & LbyL & - & & NN & \xmark \\ \hline

  Li et al. \cite{li2021cnn} & 2021 & S & LbyL & - & \cmark & NN (CNN$\alpha$) & \xmark \\ \hline

  Acharya et al. \cite{acharya2021efficient} & 2021 & S & LbyL & - & & NN (EfficientNet- B1$\beta$) & \xmark \\ \hline

  Sudhakar et al. \cite{sudhakar2021mcft} & 2021 & S & LbyL & - & &  NN (MCFT-CNN$\gamma$) & \xmark \\ \hline

    Anandhi et al. \cite{anandhi2021malware} & 2021 & S & & Markov & B & G/GIST, NN (VGG3$\gamma$, DenseNet$\gamma$) & \xmark \\ \hline

  Xiao et al. \cite{xiao2021image} & 2021 & S,M & LbyL & - & A & NN (VGG16$\gamma$) & \xmark \\ \hline

  Wang et al. \cite{wang2021anovel} & 2021 & D & LbyL & - & & N (BiLSTM) & \xmark \\ \hline
  
  Jian et al. \cite{jian2021anovel} & 2021 & S & LbyL & - & A & NN (SEResNet50$\alpha$,~BiLSTM,~Attention) & \cmark \\ \hline

  Bouchaib et al. \cite{bouchaib2021transfer} & 2021 & S & LbyL & - & & NN (VGG16$\gamma$) & \xmark \\ \hline

  Zhu et al. \cite{zhu2022afew} & 2022 & S & \xmark & Entropy & \xmark & \xmark & \xmark \\ \hline
 
  Chai et al. \cite{chai2022from} & 2022 & S & \xmark & Entropy & \xmark & \xmark & \xmark \\ \hline

  Dharmalaksana~et~al.~\cite{dharmalaksana2022improved} & 2022 & S & LbyL & - & \cmark & NN & \xmark \\ \hline

 Tekerek et al. \cite{tekerek2022anovel} & 2022 & S & LbyL & - & \cmark & NN & \xmark \\ \hline

  Bensaoud et al. \cite{bensaoud2022deep} & 2022 & S & LbyL & - & \cmark & NN & \xmark \\ \hline

  Wang et al. \cite{wang2022image} & 2022 & S & LbyL & - & & G/GIST, LBP, NN (ResNet50$\beta$) & \cmark \\ \hline

  Olorunjube et al. \cite{maldetect} & 2022 & S & LbyL & Original & B & NN & \xmark \\ \hline

    O'Shaughnessy~et~al.~\cite{oshau} & 2022 & S,D & Others & - & \cmark & G/GIST, LBP, Hist & \xmark \\ \hline
    Zhong et al. \cite{zhong2022malware} & 2022 & S & LbyL & - & & Hist, NN (CNN$\alpha$), Other & \xmark \\ \hline

  Kumar et al. \cite{kumar2022dtmic} & 2022 & S & LbyL & - & & NN (VGG16$\beta$, VGG19$\beta$, Inception V3$\beta$, ResNet50$\beta$) & \xmark \\ \hline
 Mallik et al. \cite{mallik2022conrec} & 2022 & S & LbyL & - & & NN (VGG16, BiLSTM) & VGG16-BiLSTM \\ \hline

  Conti et al. \cite{conti2022few} & 2022 & S & LbyL & Markov, Entropy, GEM Image & & H/GLCM, NN (Shallow-CNN, Siamese) & GEM \\ \hline
 Paardekooper~et~al.~\cite{paardekooper2022designing} & 2022 & S & LbyL & - & & NN (Fast-CNN GA) & \xmark \\ \hline

   Girbert et al. \cite{gibert2022fusing} & 2022 & S,M & LbyL & Entropy & & LBP, H/GLCM, NN (CNN) & \xmark \\ \hline

\end{tabular}
}
\end{center}
\end{table*}
\clearpage

\begin{center}
{\scriptsize
\begin{tabular}{|p{2.3cm}|c|p{0.6cm}|p{1.5cm}|p{2.5cm}|p{0.9cm}|p{3.5cm}|p{1.6cm}|} \hline
\textbf{Ref.} & \textbf{Year} &\textbf{Info} & \textbf{Binary Repr.} & \textbf{Image Gen. Technique} & \textbf{Color} & \textbf{Feature Extraction Technique} & \textbf{Fusion Technique} \\ \hline
 
Zhu et al. \cite{zhu2022afew} & 2022 & S & LbyL & Entropy & &NN (Siamese, VGG16$\gamma$, VGG16$\beta$, Inception V3$\gamma$, Xception$\gamma$, DNN, RNN) & \xmark \\ \hline

  Vinayakumar~et~al.~\cite{vinayakumar2022efficientnet} & 2022 & S & LbyL & - & B & NN (EfficientNet$\gamma$, various pre-trained networks) & Features from EfficientNet B0 to B7 \\ \hline

  Chai et al. \cite{chai2022from} & 2022 & S & LbyL & Markov, Original & A & NN (CNN), Others & \xmark \\ \hline

  Chaganti et al. \cite{chaganti2022image} & 2022 & S & LbyL & - & & NN (EfficientNet, B1$\gamma$) & \xmark \\ \hline

  Pratama et al. \cite{pratama2022malw} & 2022 & S & LbyL & - & B & NN (EfficientNet, B0-B7$\gamma$) & \xmark \\ \hline

  Qui et al. \cite{qiu2022malware} & 2022 & S & LbyL & - & & NN (ShuffleNet$\alpha$) & \xmark \\ \hline

  Ma et al. \cite{ma2022visual} & 2022 & S,D & LbyL & - & & NN (CNN$\alpha$) & \xmark \\ \hline

    Huaxin et al. \cite{mctvd} & 2023 & S & LbyL, Others & Simhash & A & NN (custom-CNN$\alpha$) & \xmark \\ \hline

  Hashemi et al. \cite{hashemi2023ifmd} & 2023 & S & Others & Original & A & NN (AlexNet$\beta$), Others & \xmark \\ \hline

  Rustam et al. \cite{rustam2023malw} & 2023 & S & LbyL & - & & NN (VGG16$\beta$, ResNet50$\beta$) & VGG16 + ResNet50 \\ \hline

  Kim et al. \cite{kim2023attention} & 2023 & S & LbyL & Entropy, Original & & NN (CNN $\alpha$) & \cmark\\ \hline

  Lee et al. \cite{lee2023robust} & 2023 & S & & Original & & Original & \xmark \\ \hline

  Karbab et al. \cite{karbab2023swiftr} & 2023 & S,D & LbyL & Original & - & NN (HNN), Others & \xmark \\ \hline

  Chaganti et al. \cite{chaganti2023amulti} & 2023 & S,D & & - & & NN (CNN$\alpha$), Others & \cmark \\ \hline
 
  Guo et al. \cite{guo2023mdenet} & 2023 & S,D & & - & & NN (CNN) & \xmark \\ \hline 

  Jiang et al. \cite{jiang2023apyramid} & 2023 & S & LbyL & - & &   NN (Stripe Pooling-CNN$\alpha$, Pyramid Stripe Pooling-CNN$\alpha$) & \xmark \\ \hline
 
  Shaukat et al. \cite{shaukat2023anovel} & 2023 & S & LbyL & - & B & NN (15 CNNs$\gamma$) & \xmark \\ \hline
 
  Moreira et al. \cite{moreira2023improving} & 2023 & S,M & LbyL, Others & - & B & NN (Xception$\alpha$) & \xmark \\ \hline

  Zhan et al. \cite{zhan2023amgmal} & 2023  & S & LbyL & - & \xmark & LeNet, VGGNet, ResNet, DenseNet, SqueezeNet   & \xmark \\ \hline
 
  Dhanya et al. \cite{dhanya2023obfuscated} & 2023  & S & others & Markov & \xmark &  NN & \xmark \\ \hline

  Xuan et al. \cite{xuan2024bitcn} & 2024 & S & LbyL &   Original & A & NN & Multi-Feature Fusion \\ \hline 
  
  Vasan et al. \cite{vasan2024broad} & 2024 & S & Others &   Original & \xmark & Original & \xmark  \\ \hline 
  
  Sharma et al. \cite{sharma2024migan} & 2024 & S & Lbyl & Original & A & NN & \xmark  \\ \hline 
    
  Kumar et al. \cite{kumar2024imcnn} & 2024 & S & Others & Original & B & NN(CNN) & \cmark  \\ \hline

  Yang et al. \cite{yang2025variant} & 2025 & S & LbyL &   Original & A & NN(ResNet) & \xmark \\ \hline 
  
  Johnny et al.\cite{johny2025deep} & 2025 & S & Others &   Original & A & NN(VGG16) & \cmark \\ \hline
  
  Meng et al. \cite{meng2025detecting} & 2025 & S,D,M & Others & Original & A & NN(CNN) & \cmark \\ \hline  
  
  Yang et al.\cite{yang2025sac} & 2025 & S & Others & Original & A & NN(CNN))& \cmark \\ \hline
  
  Ambekar et al.\cite{ambekar2025fasnet} & 2025 & S & Others & Original & A & NN(Siamese~Network)& \xmark \\ \hline
  
  Alam et al.\cite{alam2025miracle} & 2025 & S & LbyL & Original & A & NN(Sp-CNN)& \xmark \\ \hline  
  
  Zhang et al.\cite{zhang2025imcmk} & 2025 & S & Others & Original & A & GIST,NN(CNN)& \xmark \\ \hline 
 
  Yu et al.\cite{yu2025semantic} & 2025 & S & Others & Original & B & Original & \cmark \\ \hline
  \end{tabular}
  }
  \end{center}
  \FloatBarrier
They accomplished Fisher vector encoding by training a Gaussian mixture model~(GMM) to estimate the Fisher vector of dimension 2DK for D-SIFT features. Additionally, they employed PCA to further optimize the results by reducing the dimensionality of the feature vector to 100. In the second stage, they extracted global features from the malware image using the GIST feature description and reduced the dimension of these global features to 256.
\subsubsection{Feature Dimensionality Reduction}
Feature reduction plays a crucial role in image-based approaches, specifically in the context of image-based malware detection. It encompasses reducing the dimensionality of image data while retaining pertinent information. This section will explore the feature-reduction and selection techniques in image-based malware detection approaches.
\par Principal Component Analysis (PCA)~\cite{hotelling1933analysis} is a statistical technique that can extract texture from an image through dimensionality reduction. When applying PCA to an image, it aims to identify a set of orthogonal axes, termed principal components, that effectively represent the highest variation in the image data. Consider these fundamental components as representative patterns of texture. The first principal component captures the direction of maximum variance, while subsequent components record decreasing variance. \cite{vasan2020imcec} provides another example of dimensionality reduction. The authors apply PCA to reduce the features considered in their ensemble model. Once the features are extracted from the fine-tuned ResNet50 and VGG16 models, they were further reduced from 2048 to 205 and 6144 to 615, respectively. The resulting feature vectors from both models retained the component that explains $90\%$  of the data variation. Consequently, by employing this technique, the feature vector of the ensemble model was condensed to a mere 815 features.
\par PCA is utilized in the study by Naeem et al.~\cite{naeem2019identification} to extract the most pertinent features obtained through GIST. Similar to the approach in Vasan et al.~\cite{vasan2020imcec}, the authors employ PCA to merge the remaining features with others extracted using alternative methods. This capability of reducing the feature count without sacrificing representation power facilitates researchers in calibrating the number of features required for their experiments, thereby simplifying the combination of different techniques. Additionally, in the work by Karbab et al.\cite{karbab2023swiftr}, PCA is directly applied to the generated image to reduce dimensionality prior to the feature extraction step. The authors in \cite{vasan2024broad} introduce IMCBL, a malware classification method that transforms malware binaries into grayscale images and applies Truncated Singular Value Decomposition (SVD) to reduce dimensionality, improving computational efficiency while preserving essential features. The article~\cite{naeem2019visual} introduces an encoding-based approach to decrease the dimensions of D-SIFT and GIST features. Firstly, they used D-SIFT to identify interest points in the malware image. The Bag of Features (BOF) model was used for dimensionality reduction, followed by PCA to reduce the feature vector to 100 dimensions. In the second stage, global features are extracted from the malware image using GIST.
\subsection{Classification}
\label{sec:classification}
In this section, we explore the classification models presented by researchers for visualization-based techniques in malware detection and classification. The analysis primarily focuses on two key aspects of modeling techniques found in current literature: {\em(i)} Conventional ML-based approaches and {\em(ii)} DL-based approaches. Furthermore, we discuss work related to transformers and attention models, model fusion, data augmentation, and Few-shot Learning.

Table \ref{tab:classification_table} presents a summary of the state-of-the-art literature about classification models and explored methodologies during the classification phase. Each entry represents the primary contribution of the paper. The columns denote the following information: the paper reference; if the classification type is malware detection [D] or classification into families [C]); the type of classifier (i.e., [DT] for Decision Tree, [RF] for Random Forest, [KNN] for K-Nearest Neighbors, [SVM] for Support Vector Machine, [CNN] for Convolutional Neural Network classifier and specific architecture if applicable, [RNN] for Recurrent Neural Network classifier and specific architecture if available, [Other ML] for other Machine Learning techniques, [Other DL] other Deep Learning techniques;
the split notation, that is the proportion of data allocated to training, validation, and testing, or the indication of k-fold cross-validation or of leave-one-out; if there are some consideration of explainability and interpretability issues (identified by ``Explain.''); if there is the use of data augmentation and specified architecture (identified by ``Aug''); if there is an exploration of adversarial attacks, obfuscation, or packing (identified by ``Adv'');
if there is an exploration of few-shot learning;
if the paper considers sustainability and concept drift issues (identified by ``Sust.''); if the model used is an ensemble of classifiers(identified by "Ensemble"). \subsubsection{Conventional Machine Learning Classifiers}
Machine Learning algorithms play a vital role in malware image visualization by classifying and labeling various categories of malware based on their visual characteristics. These algorithms utilize image recognition techniques to identify distinct patterns and features specific to each malware type, achieving high accuracy by being paired with proper feature extraction techniques and training on extensive datasets of malware images. Below, we briefly introduce some of the algorithms widely employed in a substantial body of literature. One important advantage of classic ML approaches is that they are easy to implement and inherently explainable.  

The most commonly used Machine Learning classifiers in the literature are KNN, SVM, and Random Forests.
The K-nearest neighbor (KNN) model uses proximity to predict sample classes. The algorithm identifies the k-labeled data points in the sample space that are closest to the example to be classified. If the majority of these k-nearest instances belong to a specific category, then the data sample to be classified is also assigned to that category. The algorithm possesses specific advantageous characteristics. Firstly, it does not assume any specific distribution of the underlying data, making it highly flexible and adaptable to various datasets. Secondly, KNN does not require a dedicated training phase. Instead, it compares new instances with the labeled instances in the training dataset during the prediction phase, enabling quick and efficient decision-making. Another key strength of KNN is its robustness to outliers. Unlike linear models and other algorithms, KNN's decision boundaries are less affected by outliers. This resilience arises from the fact that class assignment in KNN is determined by the majority of neighbors, minimizing the impact of individual instances and preserving the overall integrity of the predictions. KNN models have been the classifier of choice for numerous models and serve as a benchmark technique against more complex classifiers~\cite{fu2018malware,yang2019anovel,naeem2019identification}.
Machine Learning algorithms like SVMs can perform regression or classification tasks. These algorithms utilize an $n$-dimensional space to classify samples, representing each data object as a point with $n$ attribute features. The categorization process involves identifying the hyperplane that effectively separates the class labels, enabling accurate classification. SVM performs well in high-dimensional spaces, making it suitable for problems with many features preventing overfitting. A negative characteristic of SVM, compared to other classifiers, like KNNs, is that they are computationally expensive, especially with large datasets. The use of SVMs has found some success in the past but has been largely overshadowed by better-performing machine-learning algorithms. Despite this, it has been used in recent papers as a benchmark but, more importantly, as an integral part of hybrid models based mainly on Deep Learning and CNNs \cite{vasan2020imcec} with great success, achieving better results than the softmax classifiers of the standard CNNs. 
Decision Trees~(DT) and Random Forests~(RF) have been among the best-performing Machine Learning algorithms across different feature extractors and datasets \cite{fu2018malware,yang2019anovel}. Roseline et al. \cite{roseline2020intelligent} demonstrated that a variation of the Random Forest model, known as Completely Random Forest (CRF), outperforms the classic RF model. Random Forest and Random Trees are known for their strong predictive power because they harness the combined knowledge of multiple decision trees, resulting in more accurate predictions than individual trees. Their ability to handle high-dimensional data stems from their feature subsampling technique, which allows them to focus on relevant attributes while mitigating the curse of dimensionality. 
However, the computational expense arises from training multiple trees, and the interpretability trade-off occurs in Random Forests as the ensemble nature makes it harder to attribute decisions to individual features. 
Traditional machine-learning classifiers have been dominating the field of malware image visualization. However, a shift in preference increasingly relegates them to a supporting role in favor of Deep Learning models.
Deep learning models typically outperform Machine Learning models when resource constraints are not a limiting factor. In particular, the advancements in hardware have been a big help in launching Deep Learning models among researchers. Another advantage is that Deep Learning allows for end-to-end learning, where the model learns to extract relevant features and make predictions without relying on manual feature engineering. This eliminates the need for domain-specific knowledge and simplifies the overall workflow. With a focus on visualization, the domain often involves working with large-scale datasets, such as image collections. Deep learning excels at processing and learning from massive amounts of data, leveraging techniques like mini-batch training and parallel computing to handle such datasets efficiently. 
\subsubsection{Deep Learning}
Recently, interest in Deep Learning and malware image-based classification techniques has grown significantly. This is because these approaches can alleviate the necessity for extensive feature engineering tasks, typically required in Machine Learning-based malware detection methods. This streamlines the classification process~\cite{yuan2020byte,vinayakumar2018scalable}. In this section, we provide a summary of the significance of various designs, including convolutional Neural Networks (CNN), recurrent Neural Networks (RNN), long short-term memory (LSTM), and others, in addressing visualization-based malware detection or classification challenges.

\textbf{Convolutional Neural Networks (CNNs)}. Convolutional Neural Networks (CNNs) are the most widely used architecture for malware image classification and general image-based analysis due to their effectiveness in processing spatially structured data. Introduced by LeCun et al.~\cite{cnn}, CNNs utilize local connectivity and weight sharing to efficiently capture spatial correlations. Through convolutional and pooling layers, they learn hierarchical and discriminative feature representations, enabling robust performance in image recognition and classification tasks.
A CNN typically consists of several layers, each serving a specific purpose in the learning process: 
\begin{itemize}
    \item \textbf{Convolutional Layer}: This layer applies a set of learnable filters to the input data, performing convolutional operations. It extracts local features and learns spatial hierarchies;
    \item \textbf{Activation Layer}: Typically following convolutional layers, activation functions (e.g., ReLU, sigmoid) introduce non-linearity, enabling the network to learn complex patterns and relationships;
    \item \textbf{Pooling Layer}: This layer downsamples the input, reducing spatial dimensions and summarizing information using operations like max or average pooling;
    \item \textbf{Fully Connected Layer}: Also called a dense layer, it connects all neurons between layers to learn global patterns in the extracted features;
    \item \textbf{Dropout Layer}: Dropout prevents overfitting by randomly setting a fraction of input units to zero during training, promoting robust feature learning;
    \item \textbf{Batch Normalization Layer}: This layer normalizes the activations of the previous layer, making the network more stable and accelerating convergence.
\end{itemize}
Researchers must tune hyperparameters, such as learning rate, batch size, epochs, optimizer, weight initialization, kernel size, filter count, and stride, to optimize CNN performance without overfitting or underfitting for a specific dataset.
\begin{table*}
\caption{Summary of classification models and explored methodologies. } 
\label{tab:classification_table}
\begin{center}
{\scriptsize
\begin{tabular}
{|p{1.5cm}|p{0.8cm}|p{1cm}|p{2.2cm}|p{1.2cm}|p{1cm}|p{1cm}|p{1cm}|p{1.1cm}|p{0.5cm}|p{1.3cm}|} \hline
\textbf{Ref.} & \textbf{Year} & \textbf{Classif. Type} & \textbf{Classifier} & \textbf{Split} & \textbf{Explain.} & \textbf{Aug} & \textbf{Adv} & \textbf{Few Shot} & \textbf{Sust.} & \textbf{Ensemble} \\ \hline


Fu et al. \cite{fu2018malware} & 2018 & C & RF, KNN, SVM & 90:10 & \xmark & \cmark & \xmark & \xmark & \xmark & \xmark \\ \hline

Yakura et al. \cite{yakura2018malware} & 2018 & C & CNN & 5f & \cmark  & \cmark & \cmark & \xmark & \xmark & \xmark\\ \hline

Yang et al. \cite{yang2018aconvolutional} & 2018 & C & CNN & 50/-/50 & \xmark & \xmark & \xmark & \xmark & \xmark & \xmark \\ \hline

Darus et al. \cite{darus2018android} & 2018 & D & KNN, SVM & 70:30 & \xmark & \xmark & \xmark & \xmark & \xmark & \xmark \\ \hline

Kurmar et al. \cite{kumar2018malicious} & 2018 & D & CNN & 90:10 & \xmark & \xmark & \xmark & \xmark & \xmark & \xmark \\ \hline

Su et al. \cite{su2018lightweight} & 2018 & C & CNN & - & \xmark & \xmark & \xmark & \xmark & \xmark & \xmark \\ \hline

Cui et al. \cite{cui2018detection} & 2018 & C & CNN & -& \xmark & \cmark & \xmark & \xmark & \xmark & \xmark \\ \hline

Ni et al. \cite{ni2018malware} & 2018 & C & KNN, SVM, CNN & 80:20 & \xmark & \xmark & \xmark & \xmark & \xmark & \cmark \\ \hline

Shiva et al. \cite{shiva2018windows} & 2018 & D & CNN & 10f & \xmark & \xmark & \xmark & \xmark & \xmark & \xmark \\ \hline

Xiao et al. \cite{xiao2019animage} & 2019 & D & CNN & 68:12:20 & \xmark & \xmark & \xmark & \xmark & \xmark & \xmark \\ \hline

Vinayakumar et al. \cite{vinayakumar2019robust} & 2019 & C & RF, DT, KNN, SVM, CNN, RNN (LSTM), Other ML & 10f & \xmark & \xmark & \cmark & \xmark & \xmark & \cmark \\ \hline

Liu et al. \cite{liu2019atmpa} & 2019 & D, C & RF, SVM, CNN & - & \xmark & \xmark & \cmark & \xmark & \xmark & \xmark \\ \hline

Hsiao et al. \cite{hsiao2019malware} & 2019 & C & CNN(Siamese), KNN & - & \xmark & \xmark & \xmark & Siamese & \xmark & \xmark \\ \hline

Girbert et al. \cite{gibert2019using} & 2019 & C & CNN & 10f & \xmark & \xmark & \xmark & \xmark & \xmark & \xmark \\ \hline

Khormali et al. \cite{khormali2019copycat} & 2019 & D & CNN & - & \xmark & \xmark & \cmark & \xmark & \xmark & \xmark \\ \hline

Naeem et al. \cite{naeem2019identification} & 2019 & C & KNN, SVM & 10:90,20:80, 30:70,40:60, 50:50,60:40, 70:30,80:20 & \xmark & \xmark & \xmark & \xmark & \xmark & \xmark \\ \hline

Liu et al. \cite{liu2019anew} & 2019 & C & RF, KNN & 10f & \xmark & \xmark & \xmark & \xmark & \xmark & \xmark \\ \hline

Akarsh et al. \cite{akarsh2019deep} & 2019 & C & CNN, RNN (LSTM) & 70:30 & \xmark & \xmark & \xmark & \xmark & \xmark & \cmark \\ \hline

Venkatraman et al. \cite{venkatraman2019ahybrid} & 2019 & D, C & CNN, RNN (LSTM) & 90:10 & \xmark & \xmark & \xmark & \xmark & \xmark & \cmark \\ \hline

Hashemi et al. \cite{hashemi} & 2019 & D & KNN & 10f & \xmark & \xmark & \xmark & \xmark & \xmark & \xmark \\ \hline

Baptista et al. \cite{baptista2019anovel} & 2019 & D &  Other ML & - & \xmark & \xmark & \xmark & \xmark & \xmark & \xmark \\ \hline

Yang et al. \cite{yang2019anovel} & 2019 & C & RF, KNN, Other ML & 5f & \xmark & \xmark & \xmark & \xmark & \xmark & \xmark \\ \hline

Chen et al. \cite{chen2019applying} & 2019 & D & CNN (InceptionV3) & - & \xmark & \xmark & \cmark & \xmark & \xmark & \xmark \\ \hline

Qiao et al. \cite{qiao2019amulti} & 2019 & C & CNN (LeNet5) & 50:50 & \xmark & \cmark & & \xmark & \xmark & \xmark \\ \hline

Liu et al. \cite{liu2020anovel} & 2020 & D & RF, SVM, CNN & 10f, 5f & \xmark & \xmark & \cmark & \xmark & \cmark & \xmark \\ \hline

Karanja et al. \cite{karanja2020analysis} & 2020 & D, C & RF, KNN & - & \xmark & \xmark & \xmark & \xmark & \xmark & \xmark \\ \hline

Xiao et al. \cite{xiao2020malfcs} & 2020 & C & SVM, CNN & 50:50 & \xmark & \xmark & \xmark & \xmark & \xmark & \xmark \\ \hline

Mercaldo et al. \cite{mercaldo2020deep} & 2020 & D, C & RF, DT, Other DL & 50:50 & \xmark & \xmark & \xmark & \xmark & \cmark & \xmark \\ \hline

Lekssays et al. \cite{lekssays2020anovel} & 2020 & D, C & KNN, CNN (3 Conv+2 Dense) & - & \xmark & \xmark & \xmark & \xmark & \xmark & \xmark \\ \hline

Naeem et al. \cite{naeem2020malware} & 2020 & D, C & CNN & 70:30 & \xmark & \xmark & \xmark & \xmark & \xmark & \xmark \\ \hline

Vasan et al. \cite{vasan2020imcec} & 2020 & C & CNN & 70:30:- & \xmark & \xmark & \cmark & \xmark & \xmark & VGG16+ ResNet50 \\ \hline

Jain et al. \cite{jain2020convo} & 2020 & C & CNN, ELM & - & \xmark & \xmark & \xmark & \xmark & \xmark & 50 ELM models\\ \hline

Vasan et al. \cite{vasan2020imcfn} & 2020 & C & Fine-tuned CNN & 70:30 & - & \cmark & \xmark & \xmark & \xmark & \xmark \\ \hline

Zhao et al. \cite{zhao2020amalware}& 2020 & C & Region-CNN & 70:30 & \xmark & \xmark & \xmark & \xmark & \xmark & \xmark \\ \hline 

Roseline et al. \cite{roseline2020intelligent} & 2020 & C & Deep RF & 80:20:- & \xmark & \xmark & \xmark & \xmark & \xmark & Complete- RF \\ \hline 

Go et al. \cite{go2020visualization} & 2020 & C & Other DL (ResNeXt) & 10f & \xmark & \xmark & \xmark & \xmark & \xmark & \xmark \\ \hline

Verma et al. \cite{verma2020multiclass} & 2020 & C & RF & 10f/ leave/ one out & \xmark & \xmark & \xmark & \xmark & \xmark & \xmark \\ \hline

Ren et al. \cite{ren2020end} & 2020 & D & CNN (Dex-CNN), Other DL (DexCRNN) & 80:10:10 & \xmark & \xmark & \xmark & \xmark & \xmark & \xmark \\ \hline 

Vu et al. \cite{hit4mal} & 2020 & C & CNN, Other ML (XGBoost) & 80:10:- & \xmark & \xmark & \xmark & \xmark & \xmark & \xmark \\ \hline

Gibert et al. \cite{gibert2020hydra} & 2020 & D, C & CNN & 80:20 & \xmark & \xmark & \xmark & \xmark & \xmark & \xmark \\ \hline

Iadarola et al. \cite{iadarola2021towards} & 2021 & C & CNN & 80:20, 80:20:- & Grad-CAM & \xmark & \xmark & \xmark & \xmark & \xmark \\ \hline

\end{tabular}
}
\end{center}
\end{table*}

\begin{center}
{\scriptsize
\begin{tabular}{|p{1.5cm}|p{0.8cm}|p{1cm}|p{2.2cm}|p{1.2cm}|p{1cm}|p{1cm}|p{1cm}|p{1.1cm}|p{0.5cm}|p{1.3cm}|}\hline
\textbf{Ref.} & \textbf{Year} & \textbf{Classif. Type} & \textbf{Classifier} & \textbf{Split} & \textbf{Explain.} & \textbf{Aug} & \textbf{Adv} & \textbf{Few Shot} & \textbf{Sust.} & \textbf{Ensemble} \\ \hline
Singh et al. \cite{singh2021classification} & 2021 & C & Random Forest, KNN, SVM, CNN & - & \xmark & \xmark & \xmark & \xmark & \xmark & \xmark \\ \hline
 Dib et al. \cite{dib2021multi}& 2021 & C & CNN,LSTM &10f  & \xmark & \xmark & \xmark & \xmark & \xmark & \xmark \\ \hline

Bozkir et al. \cite{bozkir2021catch} & 2021 & D, C & RF, DT, SVM, Other ML & 80:20:-, 3f & \xmark & \xmark & \xmark & \xmark & \xmark & \xmark \\ \hline

Sun et al. \cite{sun2021deep} & 2021 & C & CNN, RNN & - & \xmark & \xmark & \xmark & \xmark & \xmark & \xmark \\ \hline

Darem et al. \cite{darem2021visualization} & 2021 & D & CNN, Other ML (XGBoost) & f & \xmark & \xmark & Obfusca\-tion Detection & \xmark & \xmark & CNN, XGBoost\\ \hline

Pinhero et al. \cite{pinhero2021malware} & 2021 & D, C & CNN (VGG3, ResNet50) & 70:30 & \xmark & \xmark & \xmark & \xmark & \xmark & \xmark \\ \hline

Bagane et al. \cite{bagane2021classification} & 2021 & C & CNN (LSTM, BiLSTM) & 80:20 & \xmark & \xmark & \xmark & \xmark & \xmark & \xmark \\ \hline
                             
Li et al. \cite{li2021cnn} & 2021 & C & CNN (Attention, SPP) & 80:20 & \xmark & \xmark & \xmark & \xmark & \xmark & \xmark \\ \hline

Acharya et al. \cite{acharya2021efficient} & 2021 & C & CNN (EfficientNet-B1) & 75:25 & \xmark & \xmark & \xmark & \xmark & \xmark & \xmark \\ \hline

Sudhakar et al. \cite{sudhakar2021mcft} & 2021 & C & CNN (ResNet50) & 75:25 & \xmark & \xmark & \xmark & \xmark & \xmark & \xmark \\ \hline

Anandhi et al. \cite{anandhi2021malware} & 2021 & D, C & CNN (DenseNet201) & 70:30 & \xmark & \xmark & Additive Noise & \xmark & \xmark & \xmark \\ \hline

Xiao et al. \cite{xiao2021image} & 2021 & C & Linear-SVM, CNN (VGG16) & 10f & \xmark & \xmark & \xmark & \xmark & \xmark & \xmark \\ \hline

Wang et al. \cite{wang2021anovel} & 2021 & C & RNN (BiLSTM) & 73:20:20, 62:20:20 & \cmark & \xmark & \xmark & \cmark & \xmark & \xmark \\ \hline

Jian et al. \cite{jian2021anovel} & 2021 & D &  CNN (SEResNet50), RNN (BiLSTM) & 80:10:10, 70:15:15, 60:20:20 & \cmark & \cmark & \xmark & \xmark & \xmark & SERes\-Net50, BiLSTM \\ \hline

Bouchaib et al. \cite{bouchaib2021transfer} & 2021 & C & CNN (VGG16) & 50:50 & \xmark & \cmark & \xmark & \xmark & \xmark & \xmark \\ \hline

Zhu et al. \cite{zhu2022afew} & 2022 & C & CNN (VGG16, Xception, InceptionV3) & 80:20 & \cmark & \xmark & \xmark & Siamese & \xmark & \xmark \\ \hline

Chai et al. \cite{chai2022from} & 2022 & C & CNN, Other ML, Other DL & - & \xmark & \xmark & \xmark & \cmark & \xmark & \xmark \\ \hline

Dharmalak\-sana et al. \cite{dharmalaksana2022improved} & 2022 & C & CNN & - & \xmark & \xmark & \xmark & \xmark & \xmark & \xmark \\ \hline 

Tekerek et al. \cite{tekerek2022anovel} & 2022 & C & CNN (DenseNet121) & 80:20, 10f & \xmark & Cycle GAN & \xmark & \xmark & \xmark & \xmark \\ \hline

Bensaoud et al. \cite{bensaoud2022deep} & 2022 & D, C & CNN & 5f & \xmark & Cycle GAN & Obfusca\-tion & \xmark & \xmark & \xmark \\ \hline

Wang et al. \cite{wang2022image} & 2022 & C & Random Forest, Other ML & Stratified KFold & \xmark & \xmark & \xmark & \xmark & \xmark & MLP, RF, XGBoost \\ \hline 

O'Shaugh\-nessy et al. \cite{oshau} & 2022 & C & RF, KNN, SVM & 5f & \xmark & \xmark & \xmark & \xmark & \xmark & \xmark \\ \hline

Zhong et al. \cite{zhong2022malware} & 2022 & C & CNN (3conv, 3FC) & 10f & \xmark & \xmark & \xmark & \xmark & \xmark & \xmark \\ \hline

Gibert et al. \cite{gibert2022fusing} & 2022 & D & Other ML (XGBoost, GBT) & 10f & \xmark & \xmark & \xmark & \xmark & \xmark & \xmark \\ \hline

Olorunjube et al. \cite{maldetect} & 2022 & D, C & RF, DT, KNN, SVM, CNN (ResNet50), Other ML (Extra Tree Classifier, XGBoost, Bagging, Na\"ive Bayes) & - & \xmark & \cmark & DGAN & \xmark & \xmark & \xmark \\ \hline

Kumar et al. \cite{kumar2022dtmic} & 2022 & C & Random Forest, KNN, SVM, CNN (Fine-tuned VGG16 + FC, VGG19, Inception V3, ResNet50) & 90:10 & \xmark & \xmark & \xmark & \xmark & \xmark & \xmark \\ \hline

Mallik et al. \cite{mallik2022conrec} & 2022 & C & Decision Tree, KNN, RNN (BiLSTM), Other DL & 80:20 & \xmark & Rotation shifting flipping & \xmark & \xmark & \xmark & DCNN, BiLSTM \\ \hline

Conti et al. \cite{conti2022few} & 2022 & C & Shallow CNN & 70:30 & \xmark & \xmark & \xmark & Siamese & \xmark & \xmark \\ \hline       

Chaganti et al. \cite{chaganti2022image} & 2022 & C & CNN (Efficient\-NetB1) & 70:30 & \cmark & \xmark & \xmark & \xmark & \xmark & \xmark \\ \hline

Pratama et al. \cite{pratama2022malw} & 2022 & C & CNN (Efficient\-NetB7) & 70:20 & \xmark & \xmark & \xmark & \xmark & \xmark & \xmark \\ \hline

Zhu~et~al.~\cite{zhu2022afew} & 2022 & C & CNN (InceptionV3, ResNet50, VGG16, Xception) & 80:20 & t-SNE & \cmark & \xmark & Siamese & \xmark & \xmark \\ \hline

Paardekooper et al. \cite{paardekooper2022designing} & 2022 & C & CNN (2Conv, 1 FC) & 90:10 & \xmark & \xmark & \xmark & \xmark & \xmark & \xmark \\ \hline

Chai et al. \cite{chai2022from} & 2022 & C & CNN & 65:20:20, few-shot & Ablation study & Random Rotation & \xmark & \cmark & \xmark & \xmark\\ \hline

Qiu et al. \cite{qiu2022malware} & 2022 & C & CNN (ShuffleNet) & 70:30 & \xmark & \xmark & \xmark & \xmark & \xmark & \xmark \\ \hline

Ma et al. \cite{ma2022visual} & 2022 & C & CNN & 80:20 & \xmark & \xmark & \xmark & \xmark & \xmark & \xmark \\ \hline
  \end{tabular}
  }
  \end{center}

\clearpage

\begin{center}
{\scriptsize
\begin{tabular}{|p{1.5cm}|p{0.8cm}|p{1cm}|p{2.2cm}|p{1.2cm}|p{1cm}|p{1cm}|p{1cm}|p{1.1cm}|p{0.5cm}|p{1.3cm}|} \hline
\textbf{Ref.} & \textbf{Year} & \textbf{Classif. Type} & \textbf{Classifier} & \textbf{Split} & \textbf{Explain.} & \textbf{Aug} & \textbf{Adv} & \textbf{Few Shot} & \textbf{Sust.} & \textbf{Ensemble} \\ \hline

Vinayakumar et al. \cite{vinayakumar2022efficientnet} & 2022 & D & CNN (Variants of DenseNet, EfficientNet, Inception, MobileNet, ResNet, VGG) & 75:25 & t-SNE & \xmark & \xmark & \xmark & \xmark & Stacked SVM, RF, LR \\ \hline 

Hashemi et al. \cite{hashemi2023ifmd} & 2023 & D &CNN (AlexNet) & 50:50 & \xmark & \xmark & \xmark & \xmark & \xmark & \xmark \\ \hline

Huaxin~et~al.~\cite{mctvd} & 2023 & C & CNN (MCTVD) & 10f & \xmark & \xmark & \xmark & \xmark & \xmark & \xmark \\ \hline

Rustam et al. \cite{rustam2023malw} & 2023 & C & RF, KNN, SVM, Other ML & 80:20 & \cmark & \xmark & \xmark & \xmark & \xmark & \xmark \\ \hline

Kim et al. \cite{kim2023attention} & 2023 & C & CNN & 10f & \xmark & \xmark & \xmark & \xmark & \xmark & \xmark \\ \hline

Jiang et al. \cite{jiang2023apyramid} & 2023 & D, C & CNN (VGG, Stripe Pooling-CNN, Pyramid Stripe Pooling-CNN) & 80:20 & \xmark & \xmark & \xmark & \xmark & \xmark & \xmark \\ \hline

Lee et al. \cite{lee2023robust} & 2023 & D, C & RF, DT, Other ML & - & \cmark & \xmark & \xmark & \xmark & \xmark & RF, MLP \\ \hline

Shaukat~et~al.~\cite{shaukat2023anovel} & 2023 & D, C & SVM & 60:20:20,~10f & \xmark & \cmark & \xmark & \xmark & \xmark & \xmark \\ \hline

Guo et al. \cite{guo2023mdenet} & 2023 & C & CNN & 80:20 & \xmark & \xmark & \xmark & \xmark & \xmark & \xmark \\ \hline

Chaganti et al. \cite{chaganti2023amulti} & 2023 & D & CNN (1conv+ 2Dense), RNN (LSTM), Other DL (4 DNN) & 73:27 & \cmark & \xmark & \xmark & \xmark & \xmark & \xmark \\ \hline

Moreira et al. \cite{moreira2023improving} & 2023 & D & RF, KNN, SVM, CNN (Xception, Inception-ResNetV2 EfficientNetV2S), Other ML (NB, LR, SGD) & 10f & \cmark & \xmark & \xmark & \xmark & \xmark & \xmark \\ \hline

Karbab et al. \cite{karbab2023swiftr} & 2023 & D, C & Other ML & 5f & Compres\-sed Byte Entropy Matrix Visualization & \xmark & \xmark & \xmark & \xmark & \xmark \\ \hline

Zhan et al. \cite{zhan2023amgmal} & 2023 & D & CNN & 80:10:10 &Grad- & \xmark & \cmark & \xmark & \xmark & \xmark \\ \hline
Dhanya et al. \cite{dhanya2023obfuscated} & 2023 & D & CNN & 50:50, 60:40, 70:30, 80:20, 90:10 & \xmark & \xmark & \xmark &\xmark  & \cmark&\xmark  \\ \hline

Xuan et al. \cite{xuan2024bitcn} & 2024  & C  &BiTCN-TA EfficientNet &  5f & \xmark & \xmark & \cmark &\xmark  & \xmark &\xmark  \\ \hline

Vasan et al. \cite{vasan2024broad} & 2024  & C  & BLS-RVFLNN &  10f & \xmark & \xmark & \cmark &\xmark  & \xmark &\xmark  \\ \hline

Sharma~et~al.~\cite{sharma2024migan} & 2024  & C & CNN &  80:10:10 & Grad\-CAM & GAN & \xmark &\xmark  & \xmark &\xmark  \\ \hline
Dong~et~al.~\cite{dong2024image} & 2024  & D & CNN-GRU, VAE, CAE, Vanilla AE &  70:30 & \xmark & \cmark & \xmark  &\xmark  & \xmark &\xmark  \\ \hline
Kumar~et~al.~\cite{kumar2024imcnn} & 2024  & C & CNN,~KNN,~SVM,~RF &  - & \xmark & \xmark & \cmark  &\xmark  & \xmark &Soft~voting  \\ \hline

Yang~et~al.~\cite{yang2025variant}& 2025 & D & CNN,~RNN & -  & \xmark & \cmark & \xmark & \xmark & \xmark & \xmark \\ \hline

Meng~et~al.~\cite{meng2025detecting}& 2025 & D & DNN & -  & GNN\- Explainer, GradCAM & \xmark & \xmark & \xmark &  \xmark  & \xmark \\ \hline

Yang~et~al.~\cite{yang2025sac}& 2025 & D & CNN,GCN & 10f  & \xmark & \xmark & \xmark & \xmark & \xmark & \xmark \\ \hline

Johnny~et~al.~\cite{johny2025deep}& 2025 & D,C & CNN(VGG-16) & -  & GradCAM, SHAP, t-SNE & \cmark & \cmark & \xmark & \xmark & \xmark \\ \hline

Ambekar~et~al.~\cite{ambekar2025fasnet}& 2025 & D & Other~DL (Siamese) & KFold & \xmark & \cmark & \cmark & Siamese & \xmark & \xmark \\ \hline

Alam~et al.~\cite{alam2025miracle}& 2025 & C & CNN(spatial) & - & \xmark & \cmark & \xmark & \xmark & \xmark & \xmark \\ \hline

Zhang~et~al.~\cite{zhang2025imcmk}& 2025 & C & CNN & 90:10 & \xmark & \cmark & \cmark & \xmark & \xmark & \xmark \\ \hline

Yu~et al. \cite{yu2025semantic}& 2025 & C & CNN & 60:40 & \xmark & \xmark & \cmark & \xmark & \xmark & \xmark \\ \hline
  \end{tabular}
  }
  \end{center}
\FloatBarrier

\par The authors experimented with the design of CNN architecture from scratch to better 
fit the problem of visual malware classification. With this method, researchers are not bound by the designers' choices of other commonly used CNNs and can better select the best parameters for their CNN. In~\cite{ni2018malware}, the authors adapted the original multi-layer CNN by LeCun et al. The Neural Network architecture consists of a sequence of two convolutional layers, two subsampling layers, and three fully connected layers. They applied 32 filters of size 2x2 for the convolution process, and for subsampling, they utilized max pooling with a size of 2x2. In their study, the authors employed the BIG2015 dataset and obtained a classification accuracy of $99.260\%$. Cui et al.\cite{cui2018detection} presented a Deep Learning-based model that utilized 2D malware images from the Malimg dataset. They applied a custom algorithm to balance the dataset and achieved an accuracy of $94.5\%$ using CNN. The authors of \cite{naeem2020malware} introduced an architecture that integrates malware visualization with a DCNN model, enabling the detection of malicious activities within the IIoT environment. They achieved a detection accuracy rate of $97.81\%$ on the Leopard Mobile Malware Dataset and $98.47\% $on the MalImg dataset. In~\cite{gibert2019using}, Gibert et al. implemented an original convolutional Neural Network~(CNN) comprising three convolutional layers, followed by a fully-connected layer. The model proficiently classified samples from the MalImg and BIG2015 datasets, achieving accuracies of $98.48\%$ and $97.49\%$, respectively. Recent studies have also been performed by integrating Bi-Directional Temporal Convolutional Networks (BiTCN), which process input sequences in both directions to capture contextual dependencies, transfer learning with EfficientNet for feature adaptation, and Atrous Spatial Pyramid Pooling(ASPP) for multi-scale feature extraction. The authors improved the classification accuracy. In \cite{vasan2024broad}, the authors integrate a Broad Learning System (BLS), based on the Random Vector Functional Neural Network (RVFLNN), allowing dynamic model expansion without retraining, distinguishing it from traditional CNN-based methods. 

\par Researchers have used 1D-CNN~\cite{hasegawa2018one}~\cite{gohari2021android}~\cite{daoudi2021dexray} as well as 2D-CNN for malware classification. However, some researchers argue that the large number of parameters in 2D CNN adversely affects training speed~\cite{zhang2021android}. Additionally, standard convolution methods need to be improved for extracting features from bytecode sequences, as it is essential to take into account the sequential nature of the bytecode, which determines the intensity of successive pixels. Representing this sequential data as a 2D image introduces distortions and disrupts the original sequence. Furthermore, 2D CNN is ineffective for analyzing sequentially structured data, as the convolutional operations consider neighboring pixels in a 2D space, even when they are not inherently related. The Researchers of \cite{alam2025miracle} proposed a novel Spatial Convolutional Neural Network (Sp-CNN). In contrast to conventional CNNs employing uniform symmetric kernels, Sp-CNN introduces a partitioning strategy along the height dimension. This enables the network to focus on distinct input slices for individualized feature extraction. The iterative and independent processing of these slices facilitates the preservation of unique characteristics across malware classes and effectively captures nuanced spatial relationships between pixels. Furthermore, Sp-CNN operates on high-dimensional features generated by preceding conventional CNN layers, ensuring the retention of critical spatial information throughout the classification pipeline. This architecture demonstrates a significant enhancement in the accurate identification and categorization of diverse malware types, particularly in the context of imbalanced datasets. 

\par In \cite{zhang2025imcmk}, the authors presented Image-based Malware Classification with Multi-scale Kernels (IMCMK), a variant Convolutional Neural Network. Unlike standard CNN architectures that leverage stacked layers of small convolutional kernels to increase the receptive field, IMCMK introduces a Multi-scale Kernel (MK) block. This block combines the benefits of both large and small kernels, enabling the model to extract both detailed and broad contextual information from malware binary images. Additionally, an improved Squeeze-and-Excitation (SE) block is integrated to capture channel dependencies, allowing for enhanced feature selection and representation. The IMCMK architecture also employs efficient multi-scale kernel fusion strategies to mitigate the parameter overhead associated with larger kernels, thereby improving computational efficiency while preserving high classification performance.
In \cite{yu2025semantic}, the authors employed Spatial Pyramid Pooling (SPP) within a CNN framework to mitigate the challenge of varying input image dimensions inherent in malicious code representations. SPP addresses this issue by performing pooling operations across multiple spatial scales, producing fixed-size outputs irrespective of input image dimensions. This process divides feature maps into progressively finer sub-regions, capturing essential visual information at different spatial hierarchies. Consequently, the model achieves enhanced robustness and preserves semantic coherence within feature representations, leading to more accurate malware family classification by ensuring uniform input dimensions for subsequent processing stages. This is particularly advantageous for malicious code analysis, where input image sizes exhibit significant variability due to code complexity. 

\par In most of the research, CNNs have shown high accuracy~(98-99.9\%) in detecting malware based on visual features extracted. They can learn to differentiate between benign and malicious visual patterns, leading to reliable detection results\cite{vasan2020imcfn,liu2020anovel,tekerek2022anovel}. Compared to traditional ML techniques, CNNs have the capability of hierarchical representation learning, where they can capture low-level features and more abstract and complex features in different layers, which allows the network to learn progressively and combine features at different levels of abstraction.

\textbf{Recurrent Neural Networks (RNN)}. Researchers have extensively explored the application of Recurrent Neural Networks (RNNs) for image-based malware detection, capitalizing on their remarkable ability to handle sequential data. 
RNN analyzes the images, pixel by pixel or via extracted feature vectors, to identify malware-indicative patterns. It remembers past inputs through internal states, enhancing pattern recognition when context matters. Precisely, RNN architectures like Long Short-Term Memory~LSTM) effectively capture spatial dependencies within images through sequential processing. By transforming image data into sequences of vectors, RNNs empower the network to discern and recognize patterns indicative of malware during the training process. The significant advantage of RNNs lies in their capability to capture essential sequential information, such as pixel arrangement, contributing to highly effective malware detection.

\textbf{Transfer Learning and Fine Tuning}. In the field of transfer learning, pre-trained Neural Networks are employed for tasks that differ from their original training domains. This approach harnesses knowledge gained from various tasks, enhancing the effectiveness of visualization-based malware detection. Rather than training powerful and deep Neural Networks from scratch on malware images, transfer learning allows researchers to utilize existing models, saving time and resources. Fine-tuning is an additional technique to improve Neural Network performance in the specific domain of malware image classification. This method leverages the pre-trained model's knowledge from the larger dataset and adapts it to the smaller, task-specific dataset. During fine-tuning, the earlier layers of the pre-trained model are often frozen, while the later layers are updated to accommodate the new dataset. Transfer learning is particularly beneficial for limited or imbalanced datasets, which is common in malware detection. Researchers have developed a middle-ground approach that combines creating a convolutional Neural Network (CNN) from scratch with a known architecture. This method involves starting with a general CNN structure and then adapting the Neural Network to achieve an optimal balance between power and simplicity. This technique is commonly employed to facilitate transfer-learning with a simplified CNN, as demonstrated in studies \cite{mctvd} and \cite{kumar2022dtmic}.

\par Kalash et al.~\cite{kalash2018malware} utilized a deep Convolutional Neural Network (D-CNN) model to categorize malware binaries in grayscale image format, where the images had dimensions 224x224. Their study introduced a CNN model architecture based on VGG-16. 
They achieved an accuracy of $98.52\%$ on the MalImg dataset and $99.97\%$ for the BIG2015 dataset.
 Vasan et al. proposed IMCFN~\cite{vasan2020imcfn}, a malware classification algorithm that utilized malware visualization with a pre-trained VGG-16 model. Subsequently, they compared the performance of their proposed solution with ResNet50 and Inception V3, achieving an accuracy of $98.82\%$ and $97.35\%$ for the Malimg malware dataset and IoT-android mobile dataset, respectively. The authors proposed an alternative approach and conducted experiments with ResNet models in \cite{rustam2023malw} \cite{sudhakar2021mcft}. The enhanced model obtained an accuracy of 99.18\% on the MalImg malware dataset and 98.63\% on the BIG2015 malware dataset by modifying the final layer of ResNet50 in~\cite{sudhakar2021mcft}. Another commonly used Neural Network archetype is the EfficientNet family of CNNs developed in 2019 \cite{efficientnet}. These networks have demonstrated superiority over older counterparts in both the ImageNet challenge and the realm of malware analysis, as evident in studies by \cite{pratama2022malw} \cite{chaganti2022image} \cite{acharya2021efficient}.
 
Researchers in~\cite{tekerek2022anovel} opted for the family of DenseNet CNNs~\cite{densenet}. DenseNet-based CNNs exploit their dense connectivity, which allows each layer to access and utilize information from all preceding layers. DenseNet promotes feature reuse, reduces vanishing gradient issues, and enhances overall network performance. The authors in~\cite{ahmed2023inception} employed InceptionV3 to obtain a classification accuracy of $98.76\%$ on the BIG15 dataset.  In addition, the author in~\cite{chen2019applying} also experimented using InceptionV3 and obtained $90\%$ accuracy for a self-created dataset of $10,849$ malware. The research detailed in~\cite{qiu2022malware} delves into ShuffleNet, an alternative CNN-based model. ShuffleNet divides filters into groups, allowing for feature map shuffling within each layer. This approach promotes cross-channel interactions and facilitates streamlined information flow for improved efficiency. The customized ShuffleNet V2 model was employed in the Malimg dataset, achieving an accuracy of 99.03\%. The study \cite{kumar2024imcnn} introduced Intelligent Malware Classification using Deep Convolutional Neural Networks (IMCNN), a novel approach leveraging pre-trained CNN models (VGG16, VGG19, InceptionV3, and ResNet50) for efficient malware image feature extraction. Unlike traditional CNN methods, IMCNN customizes these models specifically for malware detection, enhancing feature relevance and reducing complexity to achieve effective and high-performing classification. This targeted application highlights the adaptability of pre-trained CNN architectures in malware analysis.

\subsubsection{Transformers and Attention}
Attention mechanisms, autoencoders, and transformers have emerged as pivotal advancements, drawing considerable interest due to their capacity to process and represent information proficiently. These cutting-edge techniques have sparked new avenues of research and application, particularly in domains such as natural language processing and computer vision. Attention \cite{bahdanau2016neural} is a mechanism that allows models to focus on relevant parts of input when processing an element. Transformers \cite{vaswani2023attention} is a Neural Network architecture mainly used in the field of NLP that extensively employs attention. It can be used as an explainability tool for black-box Neural Networks or exploited to generate novel malware classification models. Transformers in computer vision actively assign importance to various image regions through self-attention, enabling them to capture complex spatial structures and extended dependencies. By analyzing the entire input sequence at once, self-attention equips the model with access to global context. This contrasts with traditional sequential models like RNNs, where the fixed processing order limits information flow. 
\par Authors in \cite{guo2023mdenet} employ an encoder with a multi-scale CNN structure to generate malware images and organize tokenized malware features into sentences. It then leverages language models as textual encoders. This model obtained an accuracy of $99.32\%$ on the MalImg dataset, which is equal to the state of the art. Chen et al. \cite{chen2022malicious} use a combination of a transformer with a lambda layer which replaces the original attention mechanism to calculate the potential association information between different tokens by substituting it with a quicker single linear function. The outcomes achieved using this method showcase state-of-the-art performance, reaching an accuracy of $99.30\%$ on the Big2015 dataset. An alternative method employed by \cite{lee2021aclassification} involves employing distinct autoencoders, each fine-tuned to recognize a specific malware family. In this approach, every autoencoder undergoes exclusive training solely with samples from a particular class.  The fundamental concept lies in the premise that if an input image corresponds to the family a specific autoencoder is specialized in, the reconstruction error—indicating the distinction between the original and reconstructed images—will be minimal. Researchers leverage this reconstruction error observation to determine the family or class to which an input image pertains. Lately, researchers in~\cite{agrafiotis2022image} applied both CNN and Vision Transformer (ViT) to analyze images derived from data within PCAP files. 
\subsubsection{Model Fusion}
Enhancing malware visualization systems involves leveraging model fusion, integrating feature representations from multiple models, or combining predictions made by individual models, as explored in studies by Lad et al. \cite{lad2020malware}, Vasan et al. \cite{vasan2020imcec}, and Jian et al. \cite{jian2021anovel}. Ensembling strategies, such as weighted voting, majority voting, stacking, or boosting, stand out for their ability to generate robust and accurate outcomes. In a related contribution detailed in~\cite{roseline2019towards}, authors introduced a unique approach to ensembling known as the hybrid stacked sequential flow. This method intertwines the cascade process with representation learning using ensembled tree forests, achieving an accuracy of 98.91\% on the MalImg dataset and 96.84\% on the Big2015 dataset.
The approach, as proposed in~\cite{lin2021towards,ccayir2021random}, entails constructing an ensemble of multiple CNNs and subsequently combining the outputs of these networks to address the malware classification problem. In~\cite{lad2020malware}, the authors aimed to explore the feature extraction capabilities of CNNs and harness the classification power of SVMs. The proposed hybrid CNN + L2-SVM model achieved an accuracy of $99.59\%$. Quan Le et al.~\cite{le2018deep} employed Deep Learning on raw input images and transformed them into 1D vectors using a generic image scaling algorithm. The resulting CNN-BiLSTM classification model achieved an accuracy of $98\%$ on the BIG2015 dataset. However, a drawback of this approach was the considerable conversion time of the generic image scaling algorithm, which took approximately $191.2$ seconds. More recently, researchers in \cite{rustam2023malw} explored models created using the VGG family, and they introduced a bi-model architecture wherein they sequentially stacked the same models to achieve improved performance. They utilized the output of the first model to train the second one, resulting in $100\%$ accuracy for the MalImg dataset. 

\subsubsection{Augmentation}
The effectiveness of detection models heavily relies on the availability of a sufficient number of training samples. Unfortunately, this is a common issue with many datasets today, as there often aren't enough training samples. This presents a significant challenge for Deep Learning models, where the quality and quantity of data are critical to their effectiveness. For instance, the BIG2015 dataset, as shown in Table~\ref{tab:datasets_table}, has a high level of imbalance, with the \texttt{Kelihos\_ver3} malware family consisting of 2942 samples, while the \texttt{Simda} malware family has only 42 samples. This imbalance adversely affects both the training process and classification performance. A trivial solution is using oversampling or undersampling. Inevitably, this technique discards some of the initial information, and researchers will experience a trade-off between achieving a good classifier and utilizing all possible information. The problem of imbalanced datasets has pushed authors to consider alternative ways to use the information found in the samples. Data augmentation techniques are frequently employed to tackle the issue of imbalanced datasets in image-based malware detection. These techniques generate additional training data by applying a broader range of data variations to the original images. These variations can generate similar but different images from the available sample information. The new images will not represent actual malware found in the wild but can be helpful for the classifiers to generalize features and patterns found in the images. Different methods can achieve these variations, with the most straightforward involving spatial transformations such as rotation, scaling, flipping, color adjustments, and noise injection. MIGAN\cite{sharma2024migan} integrates image synthesis with classification using a generative adversarial network (GAN). The authors claim that they enhance malware detection by generating high-quality synthetic images, addressing class imbalance, and improving classification accuracy. Another two-phase framework uses autoencoders (vanilla, variational, and conditional) to reconstruct and enhance malware representations, then applies a CNN-GRU hybrid classifier to address class imbalance in greyscale and RGB IoT datasets\cite{dong2024image}.
According to the authors of \cite{vasan2020imcfn}, properly implementing data augmentation methods can mitigate overfitting problems and enhance the model's robustness. They utilized techniques such as rescaling and shearing, resulting in a modest 1.01\% improvement compared to the absence of data augmentation on the MalImg dataset. In their work, as discussed in \cite{bouchaib2021transfer}, the authors utilize the Synthetic Minority Oversampling Technique (SMOTE) algorithm, drawing from the principles of the $K$ minority class nearest neighbors model, as outlined in the source \cite{kevin2011smote}. This technique entails creating extra images for families with fewer samples while reducing the sample size of families with excess samples, addressing dataset balance issues.
\par Unsupervised learning becomes increasingly important when labeled data is scarce, and knowledge extraction is essential. Autoencoders use principles from data compression algorithms to capture the core features of the original data in a compressed feature set. Unlike traditional Neural Networks, autoencoders aim to approximate the input closely using fewer data without explicit input/output pairs or supervision. This autonomous approach effectively addresses the issue of dataset size in learning processes. D'Angelo et al.~\cite{d2020malware} converted the sequences of API calls invoked by apps into sparse image-like matrices called API images. They utilized autoencoders to extract the most informative features from these images. They subsequently presented the extracted features into an artificial Neural Network~(ANN)-based classifier for detection. They obtained an accuracy of $0.95$ on a customized dataset created using samples from Malgenome and contagio mobile datasets.

\par Another approach involves the utilization of the Variational Auto Encoders(VAE)~\cite{kingma2013auto}, which comprises an encoder and decoder structure. While an Autoencoder primarily focuses on data compression and encoding, the VAE emphasizes the creation of a latent vector by estimating the probability distribution of input data through its encoder. In particular, VAE functions as a statistical probability distribution model that learns the distribution of the dataset and facilitates the generation of novel data samples. This capability enables the generation of new data samples using the decoder. Similarly, the Conditional Variational Autoencoder~(CVAE)\cite{sohn2015learning} follows the same structure and objectives as the VAE. However, in contrast to the VAE, the CVAE incorporates condition information and input data during the learning process. In~\cite{van2022attention}, the authors introduce an attention mechanism to selectively assess weights in the VAE, facilitating the acquisition of crucial features in the latent space. Additionally, they integrate a lightweight CNN to capture lower-level features, ensuring a diverse representation of features. They obtained an accuracy of 99.40\% using RandomForest on the MalImg dataset.
\par Another solution is to leverage Generative Adversarial Networks (GANs)~\cite{goodfellow2014generative}, which consists of a generator and a discriminator Neural Network. Y. Lu and J. Li employed DCGAN~\cite{lu2019generative}, a GAN-based approach utilizing convolutional Neural Networks, which led to a $6\%$ increase in accuracy. Using GAN-generated samples, they trained an 18-layer deep residual network as the malware classifier. The training process included multiple convolutional transpose layers, generating 2,250 synthetic samples for each class. The deep residual network achieved an overall average testing accuracy of $84\%$, with improved performance in all other metrics for classes with larger sample sizes. Researchers utilize a Cycle GAN to generate novel training samples in their study detailed in~\cite{tekerek2022anovel}. This GAN learns the distinctive features between input and output images, employing these feature maps to execute image-to-image transformations and create new samples. This approach aims to produce images closely resembling realistic malware images. Augmenting datasets with such authentic-looking images could significantly benefit the development of an improved classifier. However, it remains crucial to rigorously test the model on unseen test samples to ensure it doesn't erroneously generalize by relying on features absent in real malware. In a related investigation by Burks et al. \cite{burks2019data}, researchers compare a VAE model and a GAN model using a ResNet18 classifier. In the MalImg dataset, using GAN led to a performance increase of $6\%$, while the integration of VAE-generated samples resulted in a 2\%  improvement. Based on these results, the researchers reported that generating artificial malware data using GANs is more effective than using VAEs for Deep Learning-based malware classification.

\subsubsection{Few-shot Learning}
\label{sec:fewshot}
\par Few-shot learning (FSL)~\cite{fei2006perona} is a Machine Learning technique that enables models to recognize and generalize to new classes with a limited number of labeled examples per class. Several Few-Shot Learning (FSL) models employ a meta-learning framework. This framework continually updates the model using mini-batches covering diverse tasks. These tasks are drawn from a latent task distribution based on the training set's characteristics. Consequently, the model acquires extensive and advanced knowledge that can be universally applied to all tasks, facilitating proficient classification across various mini-batches. The selection of the loss function holds paramount importance in Few-Shot Learning (FSL) since it steers the model in generalizing from a restricted set of examples.

\par The $N$-way-$K$-shot classification task is a common consideration, aiming to classify examples into $N$ distinct classes accurately. The challenge lies in having only $K$-labeled examples available for each class, as discussed in works by Hsiao et al. \cite{hsiao2019malware} and Barros et al. \cite{barros2022malware}. The imbalanced data distribution among different malware families significantly influences the effectiveness and generalization ability of Few-shot learning approaches~\cite{bai2020unsuccessful}. Researchers have proposed solutions using Siamese Neural Networks to tackle this challenge. A Siamese network, comprising two identical networks sharing weights, aims to embed malware instances into a continuous vector space. This network is trained on pairs of samples, treating instances from the same family as positive and those from different families as negative. The resulting feature extractor, trained by the Siamese network, can subsequently link to a classifier for predicting class labels. Hsiao et al. in ~\cite{hsiao2019malware} utilized Siamese Neural Networks to classify malware images, employing techniques such as average hashing and malware visualization during data preprocessing. Similarly, in\cite{bai2020unsuccessful}, the suggestion was made to use Siamese Neural Networks for Android malware classification, connecting the Siamese Neural Network to a multilayer perceptron for classifying generated embeddings. Another approach in ~\cite{rong2021umvd} involved generating grayscale images from network traffic data of malware and implementing a prototype-based few-shot learning model. The Siamese network utilized in the study \cite{ambekar2025fasnet} specifically employs a shared-weight architecture, where two identical sub-networks process pairs of images: one for the original malware image and the other for its adversarial counterpart created using the Fast Gradient Sign Method (FGSM). This design allows the network to learn a consistent feature embedding for both images, enabling it to compute a similarity score that quantifies the degree of similarity.

\subsection{Evaluation}
\label{sec:evaluation}
This section provides a comprehensive analysis of the evaluation methods used in the field with a particular focus on the comparability of results obtained by different research groups. We present the different software choices that researchers have to make to implement their model and how these have to be considered when presenting the results to the scientific community. Finally, we discuss the metrics commonly found in papers and their efficacy in judging a model.


\subsubsection{Software}

ML can uncover patterns and insights from large and complex datasets that may be difficult or impossible for humans to discern. Implementing ML algorithms from scratch can be a challenging and time-consuming task. Libraries provide pre-implemented algorithms, tools, and functionalities that simplify developing and deploying models. Selecting the suitable ML library holds significant importance during model implementation and when evaluating the results. Even when employing the same algorithm, the back-end implementation in various libraries may exhibit variations, leading to differences in the output that could significantly impact the overall model evaluation metrics \cite{stancin2019anoverview,gevorykan2019review}.
In the following paragraphs, we'll present the prevalent Python libraries frequently utilized by researchers within this field.

\par Several tools and libraries are crucial in visualization-oriented malware detection and machine learning research. OpenCV \cite{opencv} and the Python Imaging Library (PIL) \cite{pil} are powerful image-processing libraries that support essential tasks such as image loading, transformation, enhancement, feature extraction, visualization, and image comparison, while also integrating seamlessly with Machine Learning frameworks. Scikit-learn \cite{scikit} offers a wide range of ML algorithms focusing on simplicity and readability, enabling rapid prototyping and model validation. PyTorch \cite{pytorch}, known for its dynamic computational graph and modular design, provides flexibility for implementing cutting-edge deep learning models, while TensorFlow \cite{tensorflow}, with its scalable architecture and efficient computation graph, is suited for building and deploying complex ML models. Keras \cite{keras}, a high-level API running on TensorFlow, further simplifies neural network development with its user-friendly interface. For data visualization, Matplotlib \cite{matplotlib} is indispensable, supporting diverse plot styles from basic charts to intricate 3D visuals. Additionally, AVClass \cite{avclass} aids in automated malware classification by aggregating antivirus labels, offering consistent and meaningful insights into malware taxonomy.
\par Hardware setup is critical for evaluating ML and DL models, as computational resources directly impact training speed, model scalability, and reproducibility—yet over 40\% of studies omit detailed hardware specifications.
\subsubsection{Metrics}
The metrics commonly used to evaluate the models are:\\
    \textbf{Accuracy}. It is a metric used to evaluate the overall performance of a classification model. It represents the proportion of correctly classified instances relative to the total number of instances. Although widely used due to its simplicity, accuracy may not provide a reliable measure when dealing with imbalanced datasets. The formula for computing accuracy is given below:

        $$\text{Accuracy} = \frac{TP + TN}{TP + TN + FP + FN}$$
    
    \noindent here, TP (True Positives) denotes the number of actual positive instances that were correctly predicted, TN (True Negatives) indicates the number of correctly predicted negative instances, FP (False Positives) corresponds to negative cases incorrectly classified as positive, and FN (False Negatives) refers to positive instances that were wrongly predicted as negative.
   
    \textbf{Precision}. It evaluates how many of the instances predicted as positive are actually correct. It emphasizes the reliability of positive predictions, making it particularly important in scenarios where false positives carry significant consequences. Precision is calculated as the ratio of true positive predictions to the total number of predicted positive instances (i.e., the sum of true positives and false positives). The formula for precision is as follows:
   
        $$\text{Precision} = \frac{TP}{TP + FP}$$

    \noindent, here TP (True Positives) represents the correctly predicted positive instances, and FP (False Positives) identifies the incorrectly predicted positive instances.
   
    \textbf{Recall}. It is also referred to as sensitivity or the true positive rate, which measures the ability of a classifier to identify all relevant positive instances in the dataset. It reflects how well the model minimizes false negatives. Recall is defined as the proportion of true positive predictions out of the total actual positive cases, calculated using the following formula:

        $$\text{Recall} = \frac{TP}{TP + FN}$$
        
    \noindent, here TP (True Positives) represents the correctly predicted positive instances, and FN (False Negatives) represents the incorrectly predicted negative instances.
    \textbf{F1-score}. This metric combines precision and recall into a single value. It provides a balanced measure of a classifier's performance. F1-score is calculated as the harmonic mean of precision and recall, giving equal weight to both metrics. 
    
        $$\text{F1-Score} = 2 \times \frac{\text{Precision} \times \text{Recall}}{\text{Precision} + \text{Recall}}$$

    \textbf{Matthews Correlation Coefficient (MCC)}. MCC is A measure that takes into account all values in the confusion matrix, giving a more comprehensive assessment of the model's performance.

    $$MCC = \frac{TP \times TN - FP \times FN}{\sqrt{(TP + FP)(TP + FN)(TN + FP)(TN + FN)}}$$

    \noindent, here TP (True Positives) represents the correctly predicted positive instances, TN (True Negatives) refers to the correctly predicted negative instances, FP (False Positives) identifies the incorrectly predicted positive instances, and FN (False Negatives) represents the incorrectly predicted negative instances.

    \textbf{Jaccard Index (JI)}.It measures the similarity between predicted and actual classifications by calculating the ratio of their intersection to their union. It provides an indication of how closely the predicted labels match the actual labels.

    $$JI = \frac{TP}{TP + FP + FN}$$

    \noindent
    here TP (True Positives) represents the correctly predicted positive cases, FP (False Positives) identifies the incorrectly predicted positive instances, and FN (False Negatives) represents the incorrectly predicted negative instances.

    \textbf{Fowlkes-Mallows Index (FMI)}. It is defined as the geometric mean of precision and recall, offering a balanced evaluation of the performance of a classifier. It is often used as a complementary metric to assess the quality of the predictions.

    $$FMI = \sqrt{\text{Precision} \times \text{Recall}}$$
    
   \textbf{Confusion matrices}. A confusion matrix is a table used to evaluate the performance of a classification model by comparing predicted labels with actual labels. It includes four key components: true positives, true negatives, false positives, and false negatives. This matrix offers a comprehensive view of the model's prediction outcomes and serves as the foundation for deriving metrics like accuracy, precision, and recall.
    
    \textbf{ROC curve}. The Receiver Operating Characteristic (ROC) curve is a graphical representation of a classifier's performance across different discrimination thresholds. It plots the true positive rate (recall) against the false positive rate (1 - specificity) at various threshold settings. The curve shows the trade-off between true positive rate and false positive rate, and the area under the ROC curve (AUC) is the summary measure of a classifier's performance. A greater AUC value signifies improved classification performance. These assessments extend to multi-class classification challenges, such as those tackled through the One-vs-All or One-vs-One methods. Both methods enable the evaluation of a multi-class classifier through ROC curves but involve converting the multi-class issue into several binary classification tasks.
    
    \textbf{Resource Consumption Measurements}. Additionally, certain researchers prioritize the efficacy of their proposed models. They offer metrics such as \textit{training time}, \textit{classification time}, and \textit{CPU Usage} to highlight the time and resources expended by high-performing models, demonstrating their strengths and weaknesses. It is crucial to provide context to the quantitative outcomes by considering the hardware and software setup used for training and testing.

\begin{table*}
{\scriptsize
\caption{Metrics used to evaluate classification models and reference articles}
\label{metrics_table}
\centering
\begin{tabular}{|p{3cm}|p{12cm}|}
\hline
\textbf{Metric} & \textbf{Ref.} \\ \hline
 Accuracy &   \cite{vinayakumar2019robust} \cite{fu2018malware} \cite{ni2018malware} \cite{yang2018aconvolutional} \cite{hashemi} \cite{darus2018android} \cite{kumar2018malicious} \cite{su2018lightweight} \cite{cui2018detection} \cite{sun2021deep} \cite{xiao2019animage} \cite{xiao2020malfcs} \cite{hsiao2019malware} \cite{gibert2019using} \cite{khormali2019copycat} \cite{naeem2019identification} \cite{liu2019anew} \cite{qiao2019amulti} \cite{baptista2019anovel} \cite{yang2019anovel} \cite{chen2019applying} \cite{akarsh2019deep} \cite{karanja2020analysis} \cite{shiva2018windows} \cite{venkatraman2019ahybrid} \cite{mercaldo2020deep} \cite{lekssays2020anovel} \cite{naeem2020malware} \cite{vasan2020imcec} \cite{bozkir2021catch}  \cite{jain2020convo} \cite{vasan2020imcfn} \cite{zhao2020amalware}  \cite{dharmalaksana2022improved} \cite{roseline2020intelligent} \cite{go2020visualization} \cite{verma2020multiclass} \cite{ren2020end} \cite{hit4mal} \cite{gibert2020hydra} \cite{iadarola2021towards} \cite{singh2021classification} \cite{darem2021visualization}  \cite{bagane2021classification}\cite{tekerek2022anovel} \cite{pinhero2021malware} \cite{bensaoud2022deep} \cite{li2021cnn} \cite{acharya2021efficient} \cite{sudhakar2021mcft} \cite{wang2022image} \cite{anandhi2021malware} \cite{xiao2021image} \cite{wang2021anovel} \cite{jian2021anovel} \cite{bouchaib2021transfer} \cite{maldetect} \cite{hashemi2023ifmd} \cite{gibert2022fusing} \cite{oshau} \cite{zhong2022malware} \cite{kumar2022dtmic} \cite{mallik2022conrec} \cite{conti2022few} \cite{zhu2022afew} \cite{vinayakumar2022efficientnet} \cite{chai2022from} \cite{chaganti2022image} \cite{qiu2022malware} \cite{mctvd} \cite{ma2022visual} \cite{rustam2023malw} \cite{kim2023attention} \cite{jiang2023apyramid}\cite{lee2023robust} \cite{karbab2023swiftr} \cite{shaukat2023anovel} \cite{moreira2023improving} \cite{alam2025miracle} \cite{zhang2025imcmk} \cite{yu2025semantic} \cite{dong2024image} \cite{kumar2024imcnn}\\ \hline
 Precision &  \cite{fu2018malware} \cite{ni2018malware} \cite{hashemi} \cite{cui2018detection}  \cite{naeem2019identification} \cite{xiao2019animage} \cite{xiao2020malfcs} \cite{gibert2019using}  \cite{yang2019anovel} \cite{karanja2020analysis} \cite{shiva2018windows}  \cite{mercaldo2020deep} \cite{naeem2020malware} \cite{vasan2020imcec} \cite{bozkir2021catch} \cite{vasan2020imcfn} \cite{roseline2020intelligent} \cite{go2020visualization} \cite{verma2020multiclass} \cite{ren2020end} \cite{iadarola2021towards} \cite{singh2021classification} \cite{darem2021visualization} \cite{tekerek2022anovel} \cite{pinhero2021malware} \cite{bensaoud2022deep}  \cite{bagane2021classification}\cite{li2021cnn} \cite{wang2022image} \cite{anandhi2021malware} \cite{xiao2021image} \cite{wang2021anovel} \cite{jian2021anovel} \cite{bouchaib2021transfer} \cite{maldetect} \cite{oshau} \cite{zhong2022malware} \cite{kumar2022dtmic} \cite{mallik2022conrec}\cite{zhu2022afew} \cite{paardekooper2022designing}\cite{paardekooper2022designing} \cite{vinayakumar2022efficientnet}\cite{chaganti2022image} \cite{pratama2022malw} \cite{qiu2022malware} \cite{mctvd} \cite{ma2022visual} \cite{rustam2023malw} \cite{jiang2023apyramid} \cite{lee2023robust} \cite{karbab2023swiftr} \cite{shaukat2023anovel} \cite{moreira2023improving} \cite{alam2025miracle} \cite{zhang2025imcmk} \cite{yu2025semantic} \cite{dong2024image} \cite{kumar2024imcnn}\\ \hline
 Recall &  \cite{fu2018malware}  \cite{ni2018malware} \cite{hashemi} \cite{cui2018detection} \cite{xiao2019animage}  \cite{chaganti2022image} \cite{sudhakar2021mcft}\cite{acharya2021efficient} \cite{xiao2020malfcs} \cite{gibert2019using} \cite{naeem2019identification} \cite{yang2019anovel} \cite{karanja2020analysis} \cite{shiva2018windows} \cite{mercaldo2020deep} \cite{naeem2020malware} \cite{vasan2020imcec} \cite{bozkir2021catch} \cite{vasan2020imcfn} \cite{roseline2020intelligent} \cite{go2020visualization} \cite{verma2020multiclass} \cite{ren2020end} \cite{iadarola2021towards} \cite{singh2021classification} \cite{darem2021visualization}\cite{tekerek2022anovel} \cite{pinhero2021malware} \cite{bensaoud2022deep} \cite{bagane2021classification} \cite{li2021cnn} \cite{wang2022image} \cite{anandhi2021malware} \cite{xiao2021image} \cite{wang2021anovel} \cite{jian2021anovel} \cite{bouchaib2021transfer} \cite{maldetect} \cite{oshau} \cite{zhong2022malware} \cite{kumar2022dtmic} \cite{mallik2022conrec} \cite{zhu2022afew}\cite{paardekooper2022designing} \cite{vinayakumar2022efficientnet} \cite{pratama2022malw} \cite{qiu2022malware} \cite{mctvd} \cite{ma2022visual} \cite{rustam2023malw} \cite{jiang2023apyramid} \cite{lee2023robust} \cite{karbab2023swiftr} \cite{shaukat2023anovel} \cite{moreira2023improving} \cite{alam2025miracle} \cite{zhang2025imcmk} \cite{yu2025semantic} \cite{dong2024image} \cite{kumar2024imcnn}\\ \hline
 F1-score &  \cite{fu2018malware} \cite{ni2018malware}  \cite{hashemi} \cite{xiao2019animage}   \cite{xiao2020malfcs} \cite{gibert2019using} \cite{naeem2019identification} \cite{yang2019anovel} \cite{karanja2020analysis} \cite{shiva2018windows} \cite{mercaldo2020deep} \cite{bozkir2021catch} \cite{vasan2020imcfn}  \cite{dharmalaksana2022improved} \cite{roseline2020intelligent} \cite{go2020visualization} \cite{verma2020multiclass} \cite{ren2020end} \cite{gibert2020hydra} \cite{iadarola2021towards} \cite{darem2021visualization} \cite{tekerek2022anovel}  \cite{bagane2021classification} \cite{pinhero2021malware} \cite{bensaoud2022deep} \cite{li2021cnn} \cite{acharya2021efficient} \cite{sudhakar2021mcft} \cite{wang2022image} \cite{anandhi2021malware} \cite{xiao2021image} \cite{wang2021anovel} \cite{jian2021anovel} \cite{bouchaib2021transfer} \cite{maldetect} \cite{hashemi2023ifmd} \cite{zhong2022malware} \cite{mallik2022conrec}\cite{conti2022few} \cite{zhu2022afew} \cite{paardekooper2022designing}\cite{vinayakumar2022efficientnet} \cite{pratama2022malw} \cite{qiu2022malware} \cite{mctvd} \cite{rustam2023malw}\cite{kim2023attention} \cite{jiang2023apyramid} \cite{lee2023robust} \cite{karbab2023swiftr} \cite{shaukat2023anovel} \cite{moreira2023improving} \cite{alam2025miracle} \cite{zhang2025imcmk} \cite{yu2025semantic}\cite{dong2024image} \cite{kumar2024imcnn}\\ \hline
 Confusion Matrix & \cite{cui2018detection} \cite{gibert2019using} \cite{naeem2019identification} \cite{akarsh2019deep} \cite{naeem2020malware} \cite{vasan2020imcec} \cite{bozkir2021catch} \cite{vasan2020imcfn} \cite{roseline2020intelligent} \cite{verma2020multiclass} \cite{gibert2020hydra} \cite{iadarola2021towards} \cite{singh2021classification} \cite{darem2021visualization} \cite{tekerek2022anovel} \cite{pinhero2021malware} \cite{bensaoud2022deep} \cite{bagane2021classification} \cite{li2021cnn}  \cite{acharya2021efficient} \cite{sudhakar2021mcft} \cite{wang2022image} \cite{anandhi2021malware} \cite{xiao2021image} \cite{wang2021anovel} \cite{bouchaib2021transfer} \cite{maldetect} \cite{oshau} \cite{zhong2022malware} \cite{kumar2022dtmic} \cite{mallik2022conrec} \cite{zhu2022afew} \cite{vinayakumar2022efficientnet} \cite{chaganti2022image} \cite{pratama2022malw} \cite{mctvd} \cite{rustam2023malw}\cite{kim2023attention} \cite{karbab2023swiftr} \cite{dong2024image}\\ \hline
 ROC Curves &    \cite{vinayakumar2019robust} \cite{yang2018aconvolutional} \cite{akarsh2019deep}  \cite{karanja2020analysis} \cite{mercaldo2020deep} \cite{vasan2020imcec} \cite{roseline2020intelligent} \cite{ren2020end} \cite{bagane2021classification} \cite{maldetect} \cite{kumar2022dtmic} \cite{zhu2022afew} \cite{pratama2022malw} \cite{mctvd} \cite{karbab2023swiftr} \cite{yang2025sac} \cite{kumar2024imcnn}\\ \hline
 Time \& Resource & \cite{liu2020anovel} \cite{iadarola2021towards} \cite{bakir2023droidencoder} \cite{lachtar2023ransomshield} \cite{fu2018malware} \cite{ni2018malware} \cite{xiao2020malfcs} \cite{gibert2019using} \cite{naeem2019visual} \cite{naeem2019identification} \cite{baptista2019anovel} \cite{shiva2018windows} \cite{naeem2020malware} \cite{vasan2020imcec} \cite{bozkir2021catch} \cite{davuluru2019convolutional} \cite{vasan2020imcfn} \cite{zhao2020amalware} \cite{roseline2020intelligent} \cite{verma2020multiclass} \cite{hit4mal} \cite{darem2021visualization} \cite{pinhero2021malware} \cite{sudhakar2021mcft} \cite{anandhi2021malware} \cite{xiao2019animage} \cite{wang2021anovel} \cite{jian2021anovel} \cite{maldetect} \cite{hashemi2023ifmd} \cite{oshau} \cite{zhu2022afew} \cite{zhu2022afew} \cite{conti2022few}  \cite{dong2024image}\\ \hline
 Others (MCC, JI, FMI, etc.) & \cite{ambekar2025fasnet} \cite{yu2025semantic} \cite{alam2025miracle} \cite{yang2025sac} \cite{yang2025variant} \cite{meng2025detecting}\\ \hline

\end{tabular}
}
\end{table*}

Merely assessing accuracy does not offer a comprehensive evaluation of a model. Therefore, considering precision, recall, and the F1-score as metrics is crucial for a thorough evaluation. Moreover, a confusion matrix aids in pinpointing how a model misclassifies samples within a specific class, providing an immediate overview of the classifier's discriminatory behavior. Lastly, ROC curves serve in binary classification directly or are adapted for multi-class scenarios to showcase a model's capability to differentiate between different groups. Table \ref{metrics_table} showcases the chosen evaluation metrics for each paper concerning their models. Observe that, the ``Other'' row represents papers that employed extra pertinent metrics related to the visualization aspect of the problem.

\subsection{Model Robustness and Adaptation}
\label{sec:adversarial}
This section highlights adversarial attacks targeting visualization-based malware detectors.
Researchers have demonstrated that ML and DL are susceptible to Adversarial Examples (AEs) \cite{szegedy2013intriguing,yan2022survey}. Adversarial attacks aim to either misclassify input into a different class from the legitimate source class or intentionally misclassify samples from any source class into a specifically chosen target class. The adversarial domain has been structured using a taxonomy~\cite{papernot2016limitations} designed for multi-class Deep Learning classifiers. 
Adversaries can be classified according to their level of knowledge about the targeted model needed to carry out attacks: {\em (i)} white-box attack, {\em (ii)} black-box attack, and {\em (iii)} grey-box attack. A white-box attack requires complete access to the model. In a black-box attack, the attacker has limited or no visibility into the internal parameters or architecture of the targeted machine-learning model. In a grey-box attack, the attacker possesses limited information about the model, including architecture, training data, and gradients. An Adversarial Example (AE) pertains to an altered sample from the initial dataset, intentionally designed with slight modifications to trick Machine Learning-based malware detectors~\cite{szegedy2013intriguing}. The adversarial image $x_{adv}$ is generated using equation~\ref{adversarial example}. 

\begin{equation}
    \label{adversarial example}
    x_{adv}=x_{or} + \delta
\end{equation}

where a small perturbation called $\delta$ is applied to the original image $x_{org}$. Various approaches can be employed to create adversarial malware images. One method involves implementing adversarial perturbations~\cite{warde201611}, which entail carefully crafting imperceptible modifications to deceive image classification algorithms. Gradient-based techniques leverage gradients to generate adversarial examples, employing methods such as the Fast Gradient Sign Method (FGSM)~\cite{goodfellow2014explaining}. Generative Adversarial Networks~(GANs) can also train models that generate realistic, misleading images. Researchers generally classify malware adversarial attacks into two categories: Attacks in the problem domain and attacks in the feature domain \cite{ling2023adversarial,rosenberg2021adversarial}. In problem-domain attacks, adversarial attacks are designed to fool the model by manipulating the input data itself. This can be done by adding small, imperceptible perturbations to the input data that cause the model to make a different prediction. For example, an attacker could add a small amount of noise to an image. Conversely, feature-domain attacks, on the other hand, are designed to fool the model by manipulating the intermediate features that the model extracts from the input data. This can be done by adding or removing small amounts of activation to the model's neurons. These attacks frequently utilize gradient-based techniques in the image domain, such as DeepFool~\cite{moosavi2016deepfool} and FGSM~\cite{goodfellow2014explaining}. It's important to recognize that mapping from the feature space back to the original sample might not always succeed, possibly leading to creating samples that cannot be executed.

Targeted adversarial attacks aim to influence the model to predict a particular chosen target class. These adversarial examples are carefully crafted by subtly altering the data to guide the model's predictions toward the desired target class. Conversely, untargeted adversarial attacks pursue a broader goal of causing misclassification, not restricted to a specific target class. For example, they might create a function $F$ that modifies input $X'$ in a way that leads model $M$ to misclassify $X'$ as any class other than its original one. The challenge here is to ensure that the differences between $X$ and $X'$ remain imperceptible to human observers. In contrast, targeted attacks seek to bias model $M$ toward a particular target class $Y$. Attackers devise the function $F$ to produce a modified version of any normal input, prompting $M$ to predict class $Y$.

In~\cite{liu2019atmpa}, Liu et al. introduce a framework known as ATMPA, the first white-box adversarial attack method to undermine visualization-based malware detectors. ATMPA starts by transforming the malware sample into a grayscale image representing its binary texture, and then altering the associated adversarial example using subtle perturbations produced with the help of FGSM and C\&W techniques. However, a significant limitation of ATMPA is that the resulting adversarial grayscale image disrupts the original malware structure, rendering it unsuitable for practical PE malware detection in real-world scenarios. Khormali et al. introduce COPYCAT~\cite{khormali2019copycat}, an adversarial attack that makes use of existing generic adversarial attacks (e.g., FGSM, C\&W, DeepFool, PGD, etc.) similar to ATMPA  targeting visualization-based malware detectors that utilize CNNs. Subsequently, COPYCAT takes a different approach by attaching the adversarial image to the tail of the original malware image rather than integrating it directly into it.
In contrast, Park et al. in~\cite{park2019generation} proposed an alternative adversarial attack technique that generates an adversarial image of malware using existing off-the-shelf adversarial attacks. Subsequently, the adversarial malware alignment obfuscation (AMAO) algorithm injects minimal \texttt{NOP} instructions into the original executable malware. This process aims to align the executables with the previously generated adversarial image, ultimately evading visualization-based malware detectors. The authors introduce AMGmal \cite{zhan2023amgmal}, an Adaptive Mask-Guided adversarial attack designed to evade malware detection with minimal perturbation while preserving malware functionality. They achieve this by leveraging GradCAM++ for saliency detection to identify crucial bytes in malware binaries and modifying only slack areas in the PE format. The approach employs three adaptive mask transformation strategies—Dilation, Erosion, and Heuristic—to balance evasion effectiveness and perturbation minimization. Experimental results show AMGmal reduces perturbation from $3.34\%$ to $0.90\%$ on Malimg and from $2.32\%$ to $0.73\%$ on SOREL-20M, achieving evasion rates of $68.09\%$ and $64.75\%$. The method ensures functionality preservation and adaptability to other attack models but relies on slack area availability and accurate saliency detection. They suggest enhancing saliency precision, exploring black-box attacks, and developing robust dynamic and behavior-based detection techniques to counter adversarial malware evasion in the future. 
\par The authors of \cite{johny2025deep} utilized Generative Adversarial Networks (GANs) to improve the robustness of its malware detection model against adversarial attacks. This process involves generating adversarial examples that closely resemble original malware images by perturbing them with a generator network, while a discriminator network learns to distinguish between real and adversarial samples. These generated adversarial examples are then incorporated into the training dataset alongside the original images, leveraging a similarity metric (Structural Similarity Index Measure) to ensure their resemblance to the originals. This adversarial retraining enhances the model's ability to correctly classify both legitimate and perturbed images, ultimately improving its resilience against attacks, as evidenced by more concentrated activation patterns in heatmaps following retraining. In~\cite{asmitha2024deep}, the authors evaluated deep learning models under white-box adversarial attacks using noise perturbations like Gaussian and Salt \& Pepper. The models exhibited minor accuracy drops on MalImg (4.1\%) and BIG2015 (11.9\%), but performance on the Malhub data set dropped sharply (up to 98.5\%) due to data imbalance. These results highlight that even simple adversarial noise can significantly affect detection reliability, especially in imbalanced settings, emphasizing the importance of incorporating adversarial robustness into malware detection systems. The study \cite{ambekar2025fasnet} employs adversarial training using the FGSM to enhance the resilience of the FASNet model against malicious attacks. Adversarial images are produced through the addition of carefully calculated perturbations to the original malware images, making them difficult to classify correctly. As the model undergoes training, it learns to classify both the original and adversarial images, which enhance its capability to handle deliberately misleading inputs. This exposure ensures that the model can accurately distinguish between benign and adversarial samples, thereby bolstering its robustness against adversarial attacks in real-world environments.

\textbf{Attacks on models.} Machine learning (ML) has enhanced malware detection by enabling the identification of both known and novel threats, it simultaneously opens new attack surfaces. Adversaries increasingly exploit these vulnerabilities through sophisticated techniques to subvert and manipulate ML-based detection systems. Recent work in\cite{li2021backdoor} presents a novel backdoor attack targeting ML-based Android malware detectors, exposing critical vulnerabilities in these systems. Unlike traditional adversarial attacks, their approach operates without access to the training data, enhancing its realism and threat level. The authors inject adversarial Android applications containing stealthy, trigger-based perturbations selected using a Genetic Algorithm into the training pipeline. These samples are mislabeled as benign and modified at the APK level to preserve functionality while evading detection. Evaluated on four prominent detectors—Drebin, MaMaDroid, DroidCat, and DroidAPIMiner—the attack achieves evasion rates of up to 99\%, with minimal impact on overall model accuracy. The study identifies feature interdependencies as a limitation and underscores the need for robust defenses via improved feature selection and labeling.
\par Another effort in this direction is the Jigsaw Puzzle (JP) attack proposed in \cite{yang2023jigsaw}, which improves the stealth of backdoor strategies by selectively targeting specific malware families. Unlike traditional attacks that poison entire classes, JP activates only when a unique trigger aligns with the latent features of the attacker’s malware. The authors design the attack in a clean-label setting without controlling training or labels, enhancing realism. JP achieves high evasion on attacker-owned samples while preserving accuracy and maintaining low false positives. Evaluations on a large Android dataset show JP evades advanced defenses like MNTD by violating their broad-spectrum backdoor assumptions. While the study focuses solely on Android malware and uses fixed hyperparameters across families, it highlights the urgent need for adaptive defenses and calls for further research to generalize selective backdoor strategies to broader classification contexts. In \cite{zhan2025practical}, the authors present a framework that exploits third-party crowdsourced threat intelligence, allowing adversaries to inject poisoned data without altering labels. They embed universal adversarial triggers in unused PE header bytes, preserving functionality while manipulating predictions. Using heuristic search, they craft triggers that misclassify malware as benign. Experiments show high attack success and evasion of defenses with minimal impact on clean sample accuracy. Despite relying on soft labels and fixed regions, the method exposes weaknesses in static detection systems. The study emphasizes the need for research into sample-specific triggers, adaptability to dynamic detectors, and stronger defenses against clean-label backdoor attacks. 
\par In \cite{severi2021explanation}, the authors highlight that the widespread use of crowdsourced threat intelligence by security vendors creates a practical avenue for injecting poisoned data into training pipelines. They demonstrate the vulnerability of feature-based malware classifiers and propose a model-agnostic, explainable AI approach using SHAP to embed stealthy backdoors via influential features. The method achieves high attack success across diverse datasets and models while remaining difficult to detect due to benign sample diversity. The study highlights the dual-use risk of explainable AI in adversarial contexts, assuming knowledge of the victim model's feature space and showing variability across datasets. It calls for research into generalizable attacks and robust defenses, especially against stealthy clean-label threats in hybrid detection systems.
\section{Interpretability}
\label{sec:interpretability}

This section focuses on the vital realm of interpretable techniques for detecting image-based malware. Explainability holds immense significance in ML, addressing the need for clarity and responsibility in automated decision-making across diverse sectors like healthcare, finance, and autonomous systems \cite{mathews2019explainable}. Legislations like GDPR's Article 22 underscore the right for individuals to comprehend the logic and implications behind automated decisions, driving the call for transparency \cite{voigt2017theeu}. In cybersecurity, this transparency is non-negotiable, as ambiguity within security systems can lead to grave consequences. Understanding the 'why' behind these systems' decisions is crucial to prevent vulnerabilities and malicious activities. While some Machine Learning models naturally explain their outputs, Neural Networks—a prevalent model class—lack inherent transparency. This ambiguity has created a need for methods that make these models more understandable. This isn't just about trust and fairness but also about ethically using these systems.

\par The researchers in~\cite{yakura2019neural} applied a CNN with attention to identifying critical regions for classification, facilitating the extraction of distinctive byte sequences unique to the malware family. Even without prior knowledge, this method offers significant insights to human analysts. Moreover, examining the connections in attention maps between malware samples belonging to the same family assists analysts in pinpointing specific positions unique to targeted malware families. In their work \cite{jo2023malware}, the authors investigate how the Vision Transformer (ViT) exploits its strengths to offer a robust understanding of complex patterns in malware images. The attention map produced during detection identifies crucial elements such as class and method names. Real-world dataset validation demonstrates an $80.27\%$ accuracy in malware detection. Furthermore, they computed an ``interpretability score'' \cite{wu2021android}, surpassing other interpretable Machine Learning (ML) methods like Drebin, LIME, and XMal. This underscores its potential contribution to explainable artificial intelligence within cybersecurity.


The authors of~\cite{iadarola2021towards} presented an interpretable approach to detect malware in the Android environment. They employed the Grad-CAM algorithm to facilitate the visual debugging of models and gain insights into the specific areas that receive focus when predicting the maliciousness of Android application images. Hamad et al. in~\cite{naeem2022explainable} developed a pre-trained Inception-v3 transfer learning model to analyze malware in IoT devices. They used Grad-CAM to generate visual explanations through cumulative heatmaps, providing insights into the learned features of the CNN models. Additionally, t-SNE was employed to assess feature density within the proposed CNN models. 
In~\cite{chen2018deep}, the authors have utilized LIME to generate super-pixel~(patches of pixels) plots for malware images. An interpretable approach using an ensemble of eight CNN-based pre-trained models is proposed in~\cite{lin2021towards}. They have also conducted experiments to enhance accuracy by leveraging the resultant interpretations. 
Furthermore, the authors of~\cite{fidel2020explainability} suggested that explainable AI models demonstrate resilience against adversarial attacks. They highlighted the potential of these models in detecting adversarial inputs or samples by generating distinctive explanations using the Shapley Additive Explanations \ (SHAP) for perturbed samples. They generated SHAP Signatures for the internal layers of a Deep Neural Network classifier, which helped assess features' relevance and importance in distinguishing between normal and adversarial inputs. The authors of~\cite{johny2025deep} use SHAP to attribute individual feature contributions to model predictions, helping identify key influences on outcomes. Grad-CAM generates heatmaps to highlight important image regions, showing which areas most impact classifications. Additionally, t-SNE visualizes high-dimensional data in two dimensions, allowing the examination of clusters and patterns among different malware families.

\par The quantitative evaluation of interpretation quality is para\-mount. Metrics such as Fidelity~\cite{guo2018lemna}, stability\cite{fan2020can}, and Robust\-ness\cite{fan2020can} are three well-established measures used to assess the quality of explanations. Fidelity evaluates individual explanation accuracy, while robustness quantifies dissimilarity between explanations from different families. Stability assesses similarity among explanations from the same interpretation method on identical models. 
\section{Applications and Real-World Deployment Contexts}
Visualization-based malware detection has found applications across platforms including Windows, Android, and IoT environments such as smart factories and autonomous vehicles \cite{patel2022deep, prathiba2024blockchain, naeem2022explainable, ravi2023vit4mal}. These methods typically convert binary or behavioral data into grayscale images for classification using convolutional neural networks (CNNs). Recent work shows that embedding ML-based image analysis directly into IoT and vehicle platforms can enable low-latency malware detection. For example, Patel et al.\cite{patel2022deep} continuously monitor an autonomous vehicle’s network traffic, convert malware binaries into gray‑scale images, and apply a ResNet50V2 CNN to classify them. In a smart‑factory IIoT case, Kim and Lee deploy a three‑tier edge/cloud architecture (on-device, edge server, and cloud) that runs a CNN on visualization images of malware (from the Malimg dataset) and reaches 98.9\% accuracy. Prathiba et al.\cite{prathiba2024blockchain} take a complementary approach by using blockchain to label trustworthy AV nodes: onboard sensors detect malicious code and isolate infected vehicles, and only vetted vehicle IDs are recorded on-chain. This Blockchain-enabled scheme achieves 0.99 F1 for malware detection. ViT4Mal\cite{ravi2023vit4mal} demonstrates partial deployment of a lightweight vision transformer on a Xilinx FPGA, achieving real-time malware detection on edge devices with minimal latency and high accuracy, though decoder execution remains offloaded to a host CPU. In practice these systems span Windows executables, Android apps, and general IoT devices, leveraging CNNs on binary-to-image conversions and integrating with SOC or SIEM infrastructures. The blockchain logs and on-device inference bolster trust and privacy, but real-world deployment must still tackle issues of cross-device scalability, heterogeneous malware datasets, and the interpretability of deep models.
\section{Lesson Learned}
\label{sec:lesson}
This section presents a comprehensive discussion of the insights gained from the analysis conducted in this study. The main findings derived from our survey are outlined below:

\begin{itemize}
    \item Relying on outdated datasets risks poor model generalization to new malware variants, underscoring the necessity of using up-to-date datasets that reflect current threats for robust detection.
    \item Research largely focuses on static feature-based image generation, but combining it with dynamic visualization—despite its computational cost—offers more profound insights and improves the accuracy of malware analysis.
    \item The standard method found in the literature to generate images from malware is the bytecode gray scale method and its variants. This method interprets the bytes that form the malware file as grayscale values for a pixel. One can use the technique with the sample's opcodes with minor adaptation.
    \item Images can also be created by using dynamic features of the analyzed malware. Authors have experimented mainly with API calls, system calls, and network traffic. The methods used to transform these features in images vary a lot among authors, and the problem is often less explored;
    \item Researchers experimented with different methods to reorder and reinterpret the bytes to convey more useful information in an image. In particular, Markov images help classifiers to identify repeated patterns inside the code, SFC is used to maintain the vicinity of each byte of a sample, and hashing schemes are another way to maintain the locality of each byte. All these methods have proven to increase the accuracy of the analyzed classifiers. However, the literature needs a proper, well-defined benchmark dataset and environment with fixed parameters on which models can be objectively tested and compared.
    \item Entropy can also be used to represent a sample uniquely. The use of Entropy is interesting because it also conveys information about obfuscated malware samples.
    \item Various methods exist for extracting features from images, generally categorized into extracting global features and local features.
    \item The combination of local and global features has been proven more and more as the best-performing one
    \item The extraction of features through a Neural Network, particularly a CNN and its variants, has been proven better than any other feature extractors that do not use domain-specific information. Deep learning seems to be the way forward if we do not consider handcrafted features generated by domain experts.
    \item There is no way to objectively compare different feature extractors because different authors usually use different classifiers in their research, and so the final results are not comparable. At the feature level, few authors have pushed to interpret the extracted features, and the only explanations are related to inherently interpretable algorithms used to extract those features.
    \item Limited research has been conducted on interpretability in the context of DL malware image visualization, with CAM explanations emerging as the most promising avenues. Further experimentation in this domain is imperative, particularly in light of the growing demand for explanations driven by GDPR and similar laws requirements and other pressing security robustness concerns.
\end{itemize}

\section{Challenges and Future Directions}
\label{sec:challenges}
The cybersecurity research community has recently voiced concerns about the efficacy of a visualization-based approach in malware classification. Research on such approaches is gaining momentum. To contribute to the literature, we comprehensively investigated various steps and procedures in the visualization-based malware detection domain. This section delves into potential research challenges and future directions as visualization-based methods in malware detection become more prevalent. 

   \textbf{Obfuscated data}. By utilizing visualization techniques, analysts can better understand the behavior and structure of obfuscated code, enabling them to identify malicious patterns and enhance the detection process. Commonly employed obfuscation techniques include injecting dead code, reshuffling subroutines, rearranging code, etc.~\cite{you2010malware,shabtai2009detection}. Texture-based methods in malware detection enhance resilience against obfuscation techniques~\cite{nataraj2011malware}. Rather than focusing on code-level modifications, these methods analyze image texture patterns and visual characteristics, including color distribution, edges, and local patterns. They exhibit resilience to code obfuscation, packing, and encryption if they do not introduce significant visual alterations. These methods also demonstrate robustness to code-level modifications by effectively capturing stable high-level visual patterns. By performing statistical analysis on textures, such as histograms and co-occurrence matrices, these methods offer a robust representation that captures higher-level information. However, one significant challenge with many visualization approaches is their reliance on computing texture similarity. While these approaches effectively tackle code obfuscation issues, they demand substantial computational resources for extracting intricate texture features from malware images, including LBP, GIST, DSIFT, and GLCM. Analyzing specific malware obfuscations poses significant challenges due to their potential utilization of diverse packing techniques and resolutions. Consequently, examining them solely based on textual features becomes a difficult task\cite{venkatraman2019ahybrid}. In \cite{darem2021visualization}, the authors introduce a semi-supervised approach that uses grayscale images to visually analyze executables, aiming to detect and classify obfuscated malware. Meanwhile, Vasan et al.\cite{vasan2020imcfn} tackle the challenge of obfuscated malware classification by employing a fine-tuned CNN architecture. This architecture aims to extract resilient features by incorporating translation invariance, capturing distinct representations of obfuscated patterns, and adapting to evolving obfuscation techniques. Furthermore, in \cite{vasan2020imcec}, the same authors initially tested their model on MalImg and subsequently on packed and salted samples. The model achieved an accuracy of $98.11\%$ against packed malware, a common obfuscation technique, and $97.59\%$ against salted malware. Salting malware, observed in other studies like \cite{yajamanam2018deep}, involves inserting benign samples, commonly found on recent Windows PCs, into each malware family. Researchers in \cite{dhanya2023obfuscated} proposed a method to classify obfuscated malware and identify obfuscation types in Android IoT applications by converting bytecode into Markov images and applying a CNN. Although this approach improves robustness against obfuscation, its effectiveness depends on image quality, remains susceptible to concept drift, and may not address all obfuscation techniques. They recommend integrating dynamic analysis and implementing continuous learning to improve detection performance.
    
    \textbf{Sustainability}. Concept drift is a key challenge in sustainable malware detection, as the statistical properties of malware evolve over time, reducing the effectiveness of ML models. Traditional models trained on historical datasets become outdated and struggle to classify new malware accurately. While some researchers have addressed this issue, no systematic methodology has been established. A trade-off exists between using fixed benchmark datasets (e.g., Big2015, MalImg) and more current datasets. Solutions include testing models on benchmark datasets for comparability before evaluating them on newer data or periodically updating benchmark datasets with new samples. To mitigate concept drift, continuous model updates, adaptive learning techniques, ensemble methods, and the integration of new features and data sources are essential for maintaining effective malware classification. Unfortunately, the literature often overlooks a direct analysis of concept drift, as evident in Table \ref{tab:classification_table}. This oversight is rooted in the fact that benchmark datasets typically remain static over time. Consequently, evaluations of the presented models stand as singular snapshots, subject to judgment based on the future evolution of malware over an extended period. Malware visualization approaches have seen a limited exploration of concept drift, with only a few works addressing the issue in the context of malware detection using alternative features. Transcend~\cite{jordaney2017transcend} introduces a framework that innovatively identifies aging classification models and issues an early warning before the model consistently makes poor decisions due to outdated training. The framework achieves this by employing statistical comparisons of samples encountered during deployment with those utilized for model training, thereby establishing metrics for prediction quality. In\cite{barbero2022transcending}, the authors revisit the conformal evaluator and Transcend~\cite{jordaney2017transcend}, aiming to establish their internal workings on a solid theoretical foundation and identify optimal operational settings. They also introduce Transcendent, a framework for drift detection with computational efficiency. DroidSpan in\cite{cai2020assessing} introduces a behavior profile capturing the distribution of sensitive accesses in Android apps. This profile enables DroidSpan to demonstrate superior sustainability, consistently separating benign and malicious apps over eight years. The study also deals with the crucial role of features that endure over time in achieving sustainable learning-based malware detection. Another approach in\cite{guerra2022android} is to detect and mitigate concept drift in Android malware detection and demonstrate its effectiveness over a seven-year data set. The method also minimizes retraining with short-term datasets, evaluates the impact of timestamps, and characterizes concept drift in Android malware by analyzing essential features across time horizons. While sustainability is not explicitly addressed in the majority of papers, numerous authors examine the robustness of their models by testing them on diverse datasets. This approach, when combined with a reflection on the evolution of malware and model performance following the introduction of newer samples, can establish the foundation for a future study on sustainability \cite{liu2020anovel}. 
    
    \textbf{Outdated Datasets}. While many papers excel in malware detection or classification tasks using visualization-based approaches, these models are typically assessed using outdated datasets, posing a risk to the model's generalization ability for zero-day malware. While widely used datasets exist, they frequently encompass brief timeframes that might not sufficiently represent malware evolution or possess unevenly distributed samples across various families. Moreover, researchers frequently incorporate samples from various public repositories to either balance or augment the datasets. This practice, however, hampers the efficiency of method comparisons and reproducibility, as the samples vary across each research paper. These factors can significantly impact the performance of detection approaches. Hence, the creation of current and precise datasets that mirror ongoing trends becomes imperative, and we highlight this as a priority for future endeavors.
    
    \textbf{Computational Costs}. The majority of research focuses on techniques for generating static-feature-based images, particularly in greyscale. Although dynamic information visualization holds considerable potential for malware analysis, it comes with the drawback of being computationally intensive. Further efforts to apply hybrid approaches to images could enhance the robustness of these techniques, but this area still requires exploration.
    
    \textbf{Benchmark Dataset for Feature Fusion Strategies}. Though all feature extraction techniques enhance classifier accuracy, the existing literature lacks a well-defined benchmark dataset specifically tailored for applying these techniques. Moreover, a consistent environment with fixed parameters enables unbiased testing and model comparisons. Further exploration into combining local and global features is necessary. Additionally, investigating the fusion of various feature extractors and Deep Learning models warrants attention.
    
    \textbf{Resource-Constrained IoT devices}. In many instances, techniques for analyzing malware are crafted to enhance detection accuracy, often overlooking constraints such as memory footprint, power consumption, and network resources. Regrettably, these operational limitations give rise to notable performance bottlenecks in the context of IoT systems, resulting in a degradation of detection accuracy. For example, complex convolution and pooling operations slow down these techniques when used on edge devices. Extensive research is necessary for visualization-based approaches, which have proven effective for resource-constrained devices like IoT systems.

    \textbf{Interpretability}. Using deep models is limited by their lack of interpretability, as they essentially function as black boxes. However, understanding the reasoning behind predictive models is vital, especially in cybersecurity, where analysts must comprehend the security-related decisions made by algorithms. While explainability techniques currently exist to illuminate predictions from deep architectures, there is a need to explore more of these techniques to improve the interpretability of malware predictions through visualization-based malware detection approaches. This is crucial because very few works in this area involve creating methods that clearly explain why specific visual patterns indicate obfuscated behavior. Further research in this direction has the potential to benefit the fields of malware detection and analysis significantly.

    \textbf{Adversarial Attacks}. In a broad context, adversaries can target image classifiers by modifying pixel values to create adversarial images. Concerning malware, any alterations to a malware file must not compromise its functionality. The content within executable files is highly sensitive, and even a minor change can completely alter the malware functionality or render the file inoperable. Despite this constraint, adversaries in the visualization-based malware domain have mainly limited themselves to perturbation types, which struggle to maintain the functionality of modified files. Moreover, there are few available mechanisms to verify post-perturbation functionality. Consequently, future research should focus on developing tools that can automatically and efficiently verify the functionality of malware after undergoing perturbations.
  
    \textbf{Reproducibility}. The issue of result replicability is a well-known concern within the research community, and this study's domain is no exception. Daoudi et al. \cite{daoudi2021lessons} have addressed the critical challenge of reproducibility in Android image-based malware detection. The difficulty in replicating results, even with similar setups, highlights the need for researchers to consider this aspect carefully when aiming to make meaningful and testable contributions to the field. In this assessment, we're determining if the examined papers offer crucial details that can benefit fellow researchers, encompassing specifics on the model's structure, parameters, software components, and more. Guaranteeing thorough documentation of this data is vital to improve the ability to replicate research outcomes and drive progress in image-based malware detection.


\section{Conclusion}
\label{sec:conclusion}
The proliferation of malware, or malicious software, poses an ongoing and evolving challenge in cybersecurity. In response to the increasing threat landscape, cybersecurity defenses continually evolve to mitigate the impact of malware. Visualization-based detection approaches have emerged as a crucial component of these defense strategies. Adopting visualization-based approaches in malware detection enhances traditional methods by leveraging pictorial representations to uncover intricate relationships and anomalies in large datasets. This paper comprehensively reviews techniques that employ visualization for malware detection. We introduce fundamental knowledge essential for understanding visualization-based methods. Our review categorizes and provides insight into state-of-the-art works, outlining various steps adopted for visualization-based approaches and offering comprehensive descriptions of malware datasets represented as images.
This survey also underscores the vulnerability of image-based detection systems to effective adversarial attacks, as attackers aim to deceive these systems. In addition, our survey also discusses the obfuscation detection and sustainability of visualization-based approaches. We began by analyzing 248 works and ultimately included 102 papers published from 2018 to 2025 in our study to reveal the promising future of applying image-based methods to malware detection. As a result, we highlight the lessons learned and suggest future research directions based on current challenges in the field.


\begin{thebibliography}{100}

\bibitem{tensorflow}
Mart\'{i}n Abadi, Ashish Agarwal, Paul Barham, Eugene Brevdo, Zhifeng Chen, Craig Citro, Greg~S. Corrado, Andy Davis, Jeffrey Dean, Matthieu Devin, Sanjay Ghemawat, Ian Goodfellow, Andrew Harp, Geoffrey Irving, Michael Isard, Yangqing Jia, Rafal Jozefowicz, Lukasz Kaiser, Manjunath Kudlur, Josh Levenberg, Dandelion Man\'{e}, Rajat Monga, Sherry Moore, Derek Murray, Chris Olah, Mike Schuster, Jonathon Shlens, Benoit Steiner, Ilya Sutskever, Kunal Talwar, Paul Tucker, Vincent Vanhoucke, Vijay Vasudevan, Fernanda Vi\'{e}gas, Oriol Vinyals, Pete Warden, Martin Wattenberg, Martin Wicke, Yuan Yu, and Xiaoqiang Zheng.
\newblock {TensorFlow}: Large-scale machine learning on heterogeneous systems, 2015.
\newblock Software available from tensorflow.org.

\bibitem{acharya2021efficient}
Vasundhara Acharya, Vinayakumar Ravi, and Nazeeruddin Mohammad.
\newblock Efficientnet-based convolutional neural networks for malware classification.
\newblock In {\em 2021 12th International Conference on Computing Communication and Networking Technologies (ICCCNT)}, pages 1--6, 2021.

\bibitem{agrafiotis2022image}
Georgios Agrafiotis, Eftychia Makri, Ioannis Flionis, Antonios Lalas, Konstantinos Votis, and Dimitrios Tzovaras.
\newblock Image-based neural network models for malware traffic classification using pcap to picture conversion.
\newblock In {\em Proceedings of the 17th International Conference on Availability, Reliability and Security}, ARES '22, New York, NY, USA, 2022. Association for Computing Machinery.

\bibitem{ahmadi2016novel}
Mansour Ahmadi, Dmitry Ulyanov, Stanislav Semenov, Mikhail Trofimov, and Giorgio Giacinto.
\newblock Novel feature extraction, selection and fusion for effective malware family classification.
\newblock In {\em Proceedings of the sixth ACM conference on data and application security and privacy}, pages 183--194, 2016.

\bibitem{ahmed2023inception}
Mumtaz Ahmed, Neda Afreen, Muneeb Ahmed, Mustafa Sameer, and Jameel Ahamed.
\newblock An inception v3 approach for malware classification using machine learning and transfer learning.
\newblock {\em International Journal of Intelligent Networks}, 4:11--18, 2023.

\bibitem{akarsh2019deep}
S.~Akarsh, K.~Simran, Prabaharan Poornachandran, Vijay~Krishna Menon, and K.P. Soman.
\newblock Deep learning framework and visualization for malware classification.
\newblock In {\em 2019 5th International Conference on Advanced Computing \& Communication Systems (ICACCS)}, pages 1059--1063, 2019.

\bibitem{alam2025miracle}
Inzamamul Alam, Md~Samiullah, SM~Asaduzzaman, Upama Kabir, AM~Aahad, and Simon~S Woo.
\newblock Miracle: Malware image recognition and classification by layered extraction.
\newblock {\em Data Mining and Knowledge Discovery}, 39(1):10, 2025.

\bibitem{alsmadi2021asurvey}
Tibra Alsmadi and Nour Alqudah.
\newblock A survey on malware detection techniques.
\newblock In {\em 2021 International Conference on Information Technology (ICIT)}, pages 371--376, 2021.

\bibitem{ambekar2025fasnet}
Namrata~Govind Ambekar, Sonali Samal, N~Nandini Devi, and Surmila Thokchom.
\newblock Fasnet: Federated adversarial siamese networks for robust malware image classification.
\newblock {\em Journal of Parallel and Distributed Computing}, 198:105039, 2025.

\bibitem{hybrid}
Hybrid Analysis.
\newblock Hybrid sandbox.
\newblock \url{https://www.hybrid-analysis.com}, 2023.

\bibitem{anandhi2021malware}
V~Anandhi, P~Vinod, and Varun~G Menon.
\newblock Malware visualization and detection using densenets.
\newblock {\em Personal and Ubiquitous Computing}, pages 1--17, 2021.

\bibitem{ember}
H.~S. {Anderson} and P.~{Roth}.
\newblock {EMBER: An Open Dataset for Training Static PE Malware Machine Learning Models}.
\newblock {\em ArXiv e-prints}, April 2018.

\bibitem{aslan2020acomprehensive}
Omer~Aslan Aslan and Refik Samet.
\newblock A comprehensive review on malware detection approaches.
\newblock {\em IEEE Access}, 8:6249--6271, 2020.

\bibitem{asmitha2024deep}
KA~Asmitha, Vinod Puthuvath, KA~Rafidha~Rehiman, and SL~Ananth.
\newblock Deep learning vs. adversarial noise: a battle in malware image analysis.
\newblock {\em Cluster Computing}, 27(7):9191--9220, 2024.

\bibitem{bagane2021classification}
Pooja Bagane, Susheel~George Joseph, Abhishek Singh, Anurag Shrivastava, B.~Prabha, and Amit Shrivastava.
\newblock Classification of malware using deep learning techniques.
\newblock In {\em 2021 9th International Conference on Cyber and IT Service Management (CITSM)}, pages 1--7, 2021.

\bibitem{bahdanau2016neural}
Dzmitry Bahdanau, Kyunghyun Cho, and Yoshua Bengio.
\newblock Neural machine translation by jointly learning to align and translate, 2016.

\bibitem{bai2020unsuccessful}
Yude Bai, Zhenchang Xing, Xiaohong Li, Zhiyong Feng, and Duoyuan Ma.
\newblock Unsuccessful story about few shot malware family classification and siamese network to the rescue.
\newblock In {\em Proceedings of the ACM/IEEE 42nd International Conference on Software Engineering}, pages 1560--1571, 2020.

\bibitem{bakour2021deepvisdroid}
Khaled Bakour and Halil~Murat {\"U}nver.
\newblock Deepvisdroid: android malware detection by hybridizing image-based features with deep learning techniques.
\newblock {\em Neural Computing and Applications}, 33:11499--11516, 2021.

\bibitem{bakour2021visdroid}
Khaled Bakour and Halil~Murat {\"U}nver.
\newblock Visdroid: Android malware classification based on local and global image features, bag of visual words and machine learning techniques.
\newblock {\em Neural Computing and Applications}, 33:3133--3153, 2021.

\bibitem{bakir2023droidencoder}
Halit Bakır and Rezan Bakır.
\newblock Droidencoder: Malware detection using auto-encoder based feature extractor and machine learning algorithms.
\newblock {\em Computers and Electrical Engineering}, 110:108804, 2023.

\bibitem{baptista2019anovel}
Irina Baptista, Stavros Shiaeles, and Nicholas Kolokotronis.
\newblock A novel malware detection system based on machine learning and binary visualization.
\newblock In {\em 2019 IEEE International Conference on Communications Workshops (ICC Workshops)}, pages 1--6, 2019.

\bibitem{barbero2022transcending}
Federico Barbero, Feargus Pendlebury, Fabio Pierazzi, and Lorenzo Cavallaro.
\newblock Transcending transcend: Revisiting malware classification in the presence of concept drift.
\newblock In {\em 2022 IEEE Symposium on Security and Privacy (SP)}, pages 805--823. IEEE, 2022.

\bibitem{barros2022malware}
Pedro~H Barros, Eduarda~TC Chagas, Leonardo~B Oliveira, Fabiane Queiroz, and Heitor~S Ramos.
\newblock Malware-smell: A zero-shot learning strategy for detecting zero-day vulnerabilities.
\newblock {\em Computers \& Security}, 120:102785, 2022.

\bibitem{bensaoud2022deep}
Ahmed Bensaoud and Jugal Kalita.
\newblock Deep multi-task learning for malware image classification.
\newblock {\em Journal of Information Security and Applications}, 64:103057, 2022.

\bibitem{bhavan2024android}
Aaditya Vikram~Saravana Bhavan, Srujana Golla, Yuktha Poral, Alan~S Paul, Prasad~B Honnavalli, and S~Supreetha.
\newblock Android malware detection: A comprehensive review.
\newblock {\em Research Advances in Network Technologies}, pages 41--82, 2024.

\bibitem{bouchaib2021transfer}
Prima Bouchaib and Mohammed Bouhorma.
\newblock Transfer learning and smote algorithm for image-based malware classification.
\newblock In {\em Proceedings of the 4th International Conference on Networking, Information Systems \&; Security}, NISS2021, page 1–6, New York, NY, USA, 2021. Association for Computing Machinery.

\bibitem{kevin2011smote}
Kevin~W. Bowyer, Nitesh~V. Chawla, Lawrence~O. Hall, and W.~Philip Kegelmeyer.
\newblock {SMOTE:} synthetic minority over-sampling technique.
\newblock {\em CoRR}, abs/1106.1813, 2011.

\bibitem{bozkir2019utilization}
Ahmet~Selman Bozkir, Ahmet~Ogulcan Cankaya, and Murat Aydos.
\newblock Utilization and comparision of convolutional neural networks in malware recognition.
\newblock In {\em 2019 27th Signal Processing and Communications Applications Conference (SIU)}, pages 1--4. IEEE, 2019.

\bibitem{bozkir2021catch}
Ahmet~Selman Bozkir, Ersan Tahillioglu, Murat Aydos, and Ilker Kara.
\newblock Catch them alive: A malware detection approach through memory forensics, manifold learning and computer vision.
\newblock {\em Computers \& Security}, 103:102166, 2021.

\bibitem{opencv}
G.~Bradski.
\newblock {The OpenCV Library}.
\newblock {\em Dr. Dobb's Journal of Software Tools}, 2000.

\bibitem{broder1997on}
A.Z. Broder.
\newblock On the resemblance and containment of documents.
\newblock In {\em Proceedings. Compression and Complexity of SEQUENCES 1997 (Cat. No.97TB100171)}, pages 21--29, 1997.

\bibitem{brosolo2024sok}
Matteo Brosolo, Vinod Puthuvath, Asmitha Ka, Rafidha Rehiman, and Mauro Conti.
\newblock Sok: Visualization-based malware detection techniques.
\newblock In {\em Proceedings of the 19th International Conference on Availability, Reliability and Security}, pages 1--13, 2024.

\bibitem{burks2019data}
Roland Burks, Kazi~Aminul Islam, Yan Lu, and Jiang Li.
\newblock Data augmentation with generative models for improved malware detection: A comparative study.
\newblock In {\em 2019 IEEE 10th Annual Ubiquitous Computing, Electronics \& Mobile Communication Conference (UEMCON)}, pages 0660--0665. IEEE, 2019.

\bibitem{cai2020assessing}
Haipeng Cai.
\newblock Assessing and improving malware detection sustainability through app evolution studies.
\newblock {\em ACM Transactions on Software Engineering and Methodology (TOSEM)}, 29(2):1--28, 2020.

\bibitem{casolare2021dynamic}
Rosangela Casolare, Carlo De~Dominicis, Giacomo Iadarola, Fabio Martinelli, Francesco Mercaldo, and Antonella Santone.
\newblock Dynamic mobile malware detection through system call-based image representation.
\newblock {\em J. Wirel. Mob. Networks Ubiquitous Comput. Dependable Appl.}, 12(1):44--63, 2021.

\bibitem{ccayir2021random}
Aykut {\c{C}}ay{\i}r, U{\u{g}}ur {\"U}nal, and Hasan Da{\u{g}}.
\newblock Random capsnet forest model for imbalanced malware type classification task.
\newblock {\em Computers \& Security}, 102:102133, 2021.

\bibitem{chaganti2022image}
Rajasekhar Chaganti, Vinayakumar Ravi, and Tuan~D. Pham.
\newblock Image-based malware representation approach with efficientnet convolutional neural networks for effective malware classification.
\newblock {\em Journal of Information Security and Applications}, 69:103306, 2022.

\bibitem{chaganti2023amulti}
Rajasekhar Chaganti, Vinayakumar Ravi, and Tuan~D. Pham.
\newblock A multi-view feature fusion approach for effective malware classification using deep learning.
\newblock {\em Journal of Information Security and Applications}, 72:103402, 2023.

\bibitem{chai2022from}
Yuhan Chai, Jing Qiu, Lihua Yin, Lejun Zhang, Brij~B. Gupta, and Zhihong Tian.
\newblock From data and model levels: Improve the performance of few-shot malware classification.
\newblock {\em IEEE Transactions on Network and Service Management}, 19(4):4248--4261, 2022.

\bibitem{charikar2002simhash}
Moses~S. Charikar.
\newblock Similarity estimation techniques from rounding algorithms.
\newblock In {\em Proceedings of the Thiry-Fourth Annual ACM Symposium on Theory of Computing}, STOC '02, page 380–388, New York, NY, USA, 2002. Association for Computing Machinery.

\bibitem{chayal2024review}
Narendrakumar~Mangilal Chayal, Ankur Saxena, and Rijwan Khan.
\newblock A review on spreading and forensics analysis of windows-based ransomware.
\newblock {\em Annals of Data Science}, 11(5):1503--1524, 2024.

\bibitem{chen2019applying}
Chia-Mei Chen, Shi-Hao Wang, Dan-Wei Wen, Gu-Hsin Lai, and Ming-Kung Sun.
\newblock Applying convolutional neural network for malware detection.
\newblock In {\em 2019 IEEE 10th International Conference on Awareness Science and Technology (iCAST)}, pages 1--5, 2019.

\bibitem{chen2018deep}
Li~Chen.
\newblock Deep transfer learning for static malware classification.
\newblock {\em arXiv preprint arXiv:1812.07606}, 2018.

\bibitem{chen2022malicious}
Shi Chen, Ying Liu, Wei Hu, Jianyi Liu, Yating Gao, and Bingjie Lin.
\newblock Malicious code family classification method based on vision transformer.
\newblock In {\em 2022 IEEE 10th International Conference on Information, Communication and Networks (ICICN)}, pages 704--709, 2022.

\bibitem{chen2008towards}
Xu~Chen, Jon Andersen, Z.~Morley Mao, Michael Bailey, and Jose Nazario.
\newblock Towards an understanding of anti-virtualization and anti-debugging behavior in modern malware.
\newblock In {\em 2008 IEEE International Conference on Dependable Systems and Networks With FTCS and DCC (DSN)}, pages 177--186, 2008.

\bibitem{choi2017malware}
Sunoh Choi, Sungwook Jang, Youngsoo Kim, and Jonghyun Kim.
\newblock Malware detection using malware image and deep learning.
\newblock In {\em 2017 International Conference on Information and Communication Technology Convergence (ICTC)}, pages 1193--1195. IEEE, 2017.

\bibitem{keras}
Fran\c{c}ois Chollet et~al.
\newblock Keras.
\newblock \url{https://keras.io}, 2015.

\bibitem{conti2022few}
Mauro Conti, Shubham Khandhar, and P~Vinod.
\newblock A few-shot malware classification approach for unknown family recognition using malware feature visualization.
\newblock {\em Computers \& Security}, 122:102887, 2022.

\bibitem{binvis}
A.~Cortesi.
\newblock binvis.io: Visual analysis of binary files.
\newblock \url{ http://binvis.io/#/}, 2016.
\newblock Accessed: 2023-5-15.

\bibitem{corum2019robust}
Andrew Corum, Donovan Jenkins, and Jun Zheng.
\newblock Robust pdf malware detection with image visualization and processing techniques.
\newblock In {\em 2019 2nd International Conference on Data Intelligence and Security (ICDIS)}, pages 108--114. IEEE, 2019.

\bibitem{cuckoo}
cuckoosandbox.org.
\newblock Cuckoo sandbox.
\newblock \url{https://cuckoosandbox.org}, 2023.

\bibitem{cui2018detection}
Zhihua Cui, Fei Xue, Xingjuan Cai, Yang Cao, Gai-ge Wang, and Jinjun Chen.
\newblock Detection of malicious code variants based on deep learning.
\newblock {\em IEEE Transactions on Industrial Informatics}, 14(7):3187--3196, 2018.

\bibitem{cnn}
Y.~Le Cun, B.~Boser, J.~S. Denker, R.~E. Howard, W.~Habbard, L.~D. Jackel, and D.~Henderson.
\newblock {\em Handwritten Digit Recognition with a Back-Propagation Network}, pages 396–--404.
\newblock Morgan Kaufmann Publishers Inc., San Francisco, CA, USA, 1990.

\bibitem{dai2018malware}
Yusheng Dai, Hui Li, Yekui Qian, and Xidong Lu.
\newblock A malware classification method based on memory dump grayscale image.
\newblock {\em Digital Investigation}, 27:30--37, 2018.

\bibitem{daoudi2021lessons}
Kevin Daoudi, Nadia~andAllix, Tegawendé~F. Bissyandé, and Jacques Klein.
\newblock Lessons learnt on reproducibility in machine learning based android malware detection.
\newblock {\em Empirical Software Engineering}, 26:1573--7616, 2021.

\bibitem{daoudi2021dexray}
Nadia Daoudi, Jordan Samhi, Abdoul~Kader Kabore, Kevin Allix, Tegawend{\'e}~F Bissyand{\'e}, and Jacques Klein.
\newblock Dexray: a simple, yet effective deep learning approach to android malware detection based on image representation of bytecode.
\newblock In {\em Deployable Machine Learning for Security Defense: Second International Workshop, MLHat 2021, Virtual Event, August 15, 2021, Proceedings 2}, pages 81--106. Springer, 2021.

\bibitem{darem2021visualization}
Abdulbasit Darem, Jemal Abawajy, Aaisha Makkar, Asma Alhashmi, and Sultan Alanazi.
\newblock Visualization and deep-learning-based malware variant detection using opcode-level features.
\newblock {\em Future Generation Computer Systems}, 125:314--323, 2021.

\bibitem{darus2018android}
Fauzi~Mohd Darus, Noor Azurati~Ahmad Salleh, and Aswami~Fadillah Mohd~Ariffin.
\newblock Android malware detection using machine learning on image patterns.
\newblock In {\em 2018 Cyber Resilience Conference (CRC)}, pages 1--2, 2018.

\bibitem{davuluru2019convolutional}
Venkata Salini~Priyamvada Davuluru, Barath~Narayanan Narayanan, and Eric~J Balster.
\newblock Convolutional neural networks as classification tools and feature extractors for distinguishing malware programs.
\newblock In {\em 2019 IEEE National Aerospace and Electronics Conference (NAECON)}, pages 273--278. IEEE, 2019.

\bibitem{deldar2023deep}
Fatemeh Deldar and Mahdi Abadi.
\newblock Deep learning for zero-day malware detection and classification: A survey.
\newblock {\em ACM Comput. Surv.}, jun 2023.
\newblock Just Accepted.

\bibitem{mctvd}
Huaxin Deng, Chun Guo, Guowei Shen, Yunhe Cui, and Yuan Ping.
\newblock Mctvd: A malware classification method based on three-channel visualization and deep learning.
\newblock {\em Computers \& Security}, 126:103084, 2023.

\bibitem{dhanya2023obfuscated}
KA~Dhanya, P~Vinod, Suleiman~Y Yerima, Abul Bashar, Anwin David, T~Abhiram, Alan Antony, Ashil~K Shavanas, and Gireesh Kumar.
\newblock Obfuscated malware detection in iot android applications using markov images and cnn.
\newblock {\em IEEE Systems Journal}, 17(2):2756--2766, 2023.

\bibitem{dharmalaksana2022improved}
Putu~Sukma Dharmalaksana, Teddy Mantoro, Lutfil Khakim, and Muchlis Nurseno.
\newblock Improved malware detection results using visualization-based detection techniques ant convolutional neural network.
\newblock In {\em 2022 IEEE 8th International Conference on Computing, Engineering and Design (ICCED)}, pages 1--5, 2022.

\bibitem{dib2021multi}
Mirabelle Dib, Sadegh Torabi, Elias Bou-Harb, and Chadi Assi.
\newblock A multi-dimensional deep learning framework for iot malware classification and family attribution.
\newblock {\em IEEE Transactions on Network and Service Management}, 18(2):1165--1177, 2021.

\bibitem{dong2024image}
Huiyao Dong and Igor Kotenko.
\newblock Image-based malware analysis for enhanced iot security in smart cities.
\newblock {\em Internet of Things}, 27:101258, 2024.

\bibitem{d2020malware}
Gianni D’Angelo, Massimo Ficco, and Francesco Palmieri.
\newblock Malware detection in mobile environments based on autoencoders and api-images.
\newblock {\em Journal of Parallel and Distributed Computing}, 137:26--33, 2020.

\bibitem{maldetect}
Olorunjube~James Falana, Adesina~Simon Sodiya, Saidat~Adebukola Onashoga, and Biodun~Surajudeen Badmus.
\newblock Mal-detect: An intelligent visualization approach for malware detection.
\newblock {\em Journal of King Saud University - Computer and Information Sciences}, 34(5):1968--1983, 2022.

\bibitem{fan2020can}
Ming Fan, Wenying Wei, Xiaofei Xie, Yang Liu, Xiaohong Guan, and Ting Liu.
\newblock Can we trust your explanations? sanity checks for interpreters in android malware analysis.
\newblock {\em IEEE Transactions on Information Forensics and Security}, 16:838--853, 2020.

\bibitem{fei2006perona}
Fergus Fei-Fei.
\newblock Perona, 2006 fei-fei l., fergus r., perona p.
\newblock {\em One-shot learning of object categories, IEEE Trans. Pattern Anal. Mach. Intell}, 28(4):594--611, 2006.

\bibitem{felt2011mobile}
Adrienne~Porter Felt, Matthew Finifter, Erika Chin, Steve Hanna, and David Wagner.
\newblock A survey of mobile malware in the wild.
\newblock In {\em Proceedings of the 1st ACM Workshop on Security and Privacy in Smartphones and Mobile Devices}, SPSM '11, page 3–14, New York, NY, USA, 2011. Association for Computing Machinery.

\bibitem{fidel2020explainability}
Gil Fidel, Ron Bitton, and Asaf Shabtai.
\newblock When explainability meets adversarial learning: Detecting adversarial examples using shap signatures.
\newblock In {\em 2020 international joint conference on neural networks (IJCNN)}, pages 1--8. IEEE, 2020.

\bibitem{fu2018malware}
Jianwen Fu, Jingfeng Xue, Yong Wang, Zhenyan Liu, and Chun Shan.
\newblock Malware visualization for fine-grained classification.
\newblock {\em IEEE Access}, 6:14510--14523, 2018.

\bibitem{anyrun}
ANYRUN FZCO.
\newblock Any.run.
\newblock \url{https://any.run}, 2023.

\bibitem{gao2021malware}
Tan Gao, Lan Zhao, Xudong Li, and Wen Chen.
\newblock Malware detection based on semi-supervised learning with malware visualization.
\newblock {\em Math. Biosci. Eng}, 18(5):5995--6011, 2021.

\bibitem{gevorykan2019review}
Migran~N. Gevorkyan, Anastasia~V. Demidova, Tatiana~S. Demidova, and Anton~A. Sobolev.
\newblock Review and comparative analysis of machine learning libraries for machine learning.
\newblock {\em Discrete and Continuous Models and Applied Computational Science}, 27(4):305--315, 2019.

\bibitem{gibert2020hydra}
Daniel Gibert, Carles Mateu, and Jordi Planes.
\newblock Hydra: A multimodal deep learning framework for malware classification.
\newblock {\em Computers \& Security}, 95:101873, 2020.

\bibitem{gibertsurvey}
Daniel Gibert, Carles Mateu, and Jordi Planes.
\newblock The rise of machine learning for detection and classification of malware: Research developments, trends and challenges.
\newblock {\em Journal of Network and Computer Applications}, 153:102526, 2020.

\bibitem{gibert2019using}
Daniel Gibert, Carles Mateu, Jordi Planes, and Ramon Vicens.
\newblock Using convolutional neural networks for classification of malware represented as images.
\newblock {\em Journal of Computer Virology and Hacking Techniques}, 15:15--28, 2019.

\bibitem{gibert2022fusing}
Daniel Gibert, Jordi Planes, Carles Mateu, and Quan Le.
\newblock Fusing feature engineering and deep learning: A case study for malware classification.
\newblock {\em Expert Systems with Applications}, 207:117957, 2022.

\bibitem{go2020visualization}
Jin~Ho Go, Tony Jan, Manoranjan Mohanty, Om~Prakash Patel, Deepak Puthal, and Mukesh Prasad.
\newblock Visualization approach for malware classification with resnext.
\newblock In {\em 2020 IEEE Congress on Evolutionary Computation (CEC)}, pages 1--7, 2020.

\bibitem{gohari2021android}
Mahshid Gohari, Sattar Hashemi, and Lida Abdi.
\newblock Android malware detection and classification based on network traffic using deep learning.
\newblock In {\em 2021 7th International Conference on Web Research (ICWR)}, pages 71--77. IEEE, 2021.

\bibitem{goodfellow2014generative}
Ian~J Goodfellow, Jean Pouget-Abadie, Mehdi Mirza, Bing Xu, David Warde-Farley, Sherjil Ozair, Aaron Courville, and Yoshua Bengio.
\newblock Generative adversarial networks (2014).
\newblock {\em arXiv preprint arXiv:1406.2661}, 1406, 2014.

\bibitem{goodfellow2014explaining}
Ian~J Goodfellow, Jonathon Shlens, and Christian Szegedy.
\newblock Explaining and harnessing adversarial examples.
\newblock {\em arXiv preprint arXiv:1412.6572}, 2014.

\bibitem{grigorescu2002comparison}
Simona~E Grigorescu, Nicolai Petkov, and Peter Kruizinga.
\newblock Comparison of texture features based on gabor filters.
\newblock {\em IEEE Transactions on Image processing}, 11(10):1160--1167, 2002.

\bibitem{guerra2022android}
Alejandro Guerra-Manzanares, Marcin Luckner, and Hayretdin Bahsi.
\newblock Android malware concept drift using system calls: detection, characterization and challenges.
\newblock {\em Expert Systems with Applications}, 206:117200, 2022.

\bibitem{guo2023mdenet}
Jingcai Guo, Yuanyuan Xu, Wenchao Xu, Yufeng Zhan, Yuxia Sun, and Song Guo.
\newblock Mdenet: Multi-modal dual-embedding networks for malware open-set recognition, 2023.

\bibitem{guo2018lemna}
Wenbo Guo, Dongliang Mu, Jun Xu, Purui Su, Gang Wang, and Xinyu Xing.
\newblock Lemna: Explaining deep learning based security applications.
\newblock In {\em proceedings of the 2018 ACM SIGSAC conference on computer and communications security}, pages 364--379, 2018.

\bibitem{haralick1973textural}
Robert~M Haralick, Karthikeyan Shanmugam, and Its'~Hak Dinstein.
\newblock Textural features for image classification.
\newblock {\em IEEE Transactions on systems, man, and cybernetics}, pages 610--621, 1973.

\bibitem{hasegawa2018one}
Chihiro Hasegawa and Hitoshi Iyatomi.
\newblock One-dimensional convolutional neural networks for android malware detection.
\newblock In {\em 2018 IEEE 14th International Colloquium on Signal Processing \& Its Applications (CSPA)}, pages 99--102. IEEE, 2018.

\bibitem{hashemi}
Hashem Hashemi and Ali Hamzeh.
\newblock Visual malware detection using local malicious pattern.
\newblock {\em Journal of Computer Virology and Hacking Techniques}, 15:1--14, 2019.

\bibitem{hashemi2023ifmd}
Hashem Hashemi, Mohammad~Ebrahim Samie, and Ali Hamzeh.
\newblock Ifmd: image fusion for malware detection.
\newblock {\em Journal of Computer Virology and Hacking Techniques}, 19(2):271--286, 2023.

\bibitem{hotelling1933analysis}
Harold Hotelling.
\newblock Analysis of a complex of statistical variables into principal components.
\newblock {\em Journal of educational psychology}, 24(6):417, 1933.

\bibitem{hsiao2019malware}
Shou-Ching Hsiao, Da-Yu Kao, Zi-Yuan Liu, and Raylin Tso.
\newblock Malware image classification using one-shot learning with siamese networks.
\newblock {\em Procedia Computer Science}, 159:1863--1871, 2019.
\newblock Knowledge-Based and Intelligent Information \& Engineering Systems: Proceedings of the 23rd International Conference KES2019.

\bibitem{densenet}
Gao Huang, Zhuang Liu, and Kilian~Q. Weinberger.
\newblock Densely connected convolutional networks.
\newblock {\em CoRR}, abs/1608.06993, 2016.

\bibitem{huang2020amethod}
Xiang Huang, Li~Ma, Wenyin Yang, and Yong Zhong.
\newblock A method for windows malware detection based on deep learning.
\newblock {\em Journal of Signal Processing Systems}, 93:265--273, 2021.

\bibitem{matplotlib}
J.~D. Hunter.
\newblock Matplotlib: A 2d graphics environment.
\newblock {\em Computing in Science \& Engineering}, 9(3):90--95, 2007.

\bibitem{iadarola2021towards}
Giacomo Iadarola, Fabio Martinelli, Francesco Mercaldo, and Antonella Santone.
\newblock Towards an interpretable deep learning model for mobile malware detection and family identification.
\newblock {\em Computers \& Security}, 105:102198, 2021.

\bibitem{irolla2018duplication}
Paul Irolla and Alexandre Dey.
\newblock The duplication issue within the drebin dataset.
\newblock {\em Journal of Computer Virology and Hacking Techniques}, 14(3):245--249, 2018.

\bibitem{Microsoft}
Marc Jacob.
\newblock Microsoft digital defense report.
\newblock \url{https://news.microsoft.com/en-cee/2024/11/29/microsoft-digital-defense-report-600-million-cyberattacks-per-day-around-the-globe-2/}, 2025.
\newblock Accessed: 2025-5-3.

\bibitem{jain2020convo}
Mugdha Jain, William Andreopoulos, and Mark Stamp.
\newblock Convolutional neural networks and extreme learning machines for malware classification.
\newblock {\em Journal of Computer Virology and Hacking Techniques}, 16:pages 229–244, 2020.

\bibitem{jian2021anovel}
Yifei Jian, Hongbo Kuang, Chenglong Ren, Zicheng Ma, and Haizhou Wang.
\newblock A novel framework for image-based malware detection with a deep neural network.
\newblock {\em Computers \& Security}, 109:102400, 2021.

\bibitem{jiang2023apyramid}
Jiaqi Jiang and Yunchun Zhang.
\newblock A pyramid stripe pooling-based convolutional neural network for malware detection and classification.
\newblock {\em Journal of Ambient Intelligence and Humanized Computing}, 14(3):2785--2796, Mar 2023.

\bibitem{jo2023malware}
Jeonggeun Jo, Jaeik Cho, and Jongsub Moon.
\newblock A malware detection and extraction method for the related information using the vit attention mechanism on android operating system.
\newblock {\em Applied Sciences}, 13(11):6839, 2023.

\bibitem{joesandbox}
joesandbox.com.
\newblock Joe sandbox.
\newblock \url{https://www.joesandbox.com/#windows}, 2023.

\bibitem{johny2025deep}
Jahez~Abraham Johny, KA~Asmitha, P~Vinod, G~Radhamani, KA~Rafidha Rehiman, and Mauro Conti.
\newblock Deep learning fusion for effective malware detection: leveraging visual features.
\newblock {\em Cluster Computing}, 28(2):135, 2025.

\bibitem{jordaney2017transcend}
Roberto Jordaney, Kumar Sharad, Santanu~K Dash, Zhi Wang, Davide Papini, Ilia Nouretdinov, and Lorenzo Cavallaro.
\newblock Transcend: Detecting concept drift in malware classification models.
\newblock In {\em 26th USENIX security symposium (USENIX security 17)}, pages 625--642, 2017.

\bibitem{kalash2018malware}
Mahmoud Kalash, Mrigank Rochan, Noman Mohammed, Neil~DB Bruce, Yang Wang, and Farkhund Iqbal.
\newblock Malware classification with deep convolutional neural networks.
\newblock In {\em 2018 9th IFIP international conference on new technologies, mobility and security (NTMS)}, pages 1--5. IEEE, 2018.

\bibitem{karanja2020analysis}
Evanson~Mwangi Karanja, Shedden Masupe, and Mandu~Gasennelwe Jeffrey.
\newblock Analysis of internet of things malware using image texture features and machine learning techniques.
\newblock {\em Internet of Things}, 9:100153, 2020.

\bibitem{karbab2023swiftr}
ElMouatez~Billah Karbab, Mourad Debbabi, and Abdelouahid Derhab.
\newblock Swiftr: Cross-platform ransomware fingerprinting using hierarchical neural networks on hybrid features.
\newblock {\em Expert Systems with Applications}, 225:120017, 2023.

\bibitem{Kerr2025}
Kevin Kerr.
\newblock Trustwave's 2025 cybersecurity predictions: Ai-powered attacks, critical infrastructure risks, and regulatory challenges, 2025.
\newblock Accessed: 2025-05-03.

\bibitem{keyes2021entroplyzer}
David~Sean Keyes, Beiqi Li, Gurdip Kaur, Arash~Habibi Lashkari, Francois Gagnon, and Fr{\'e}d{\'e}ric Massicotte.
\newblock Entroplyzer: Android malware classification and characterization using entropy analysis of dynamic characteristics.
\newblock In {\em 2021 Reconciling Data Analytics, Automation, Privacy, and Security: A Big Data Challenge (RDAAPS)}, pages 1--12. IEEE, 2021.

\bibitem{khormali2019copycat}
Aminollah Khormali, Ahmed Abusnaina, Songqing Chen, DaeHun Nyang, and Aziz Mohaisen.
\newblock Copycat: practical adversarial attacks on visualization-based malware detection.
\newblock {\em arXiv preprint arXiv:1909.09735}, 2019.

\bibitem{kim2023attention}
Jeongwoo Kim, Joon-Young Paik, and Eun-Sun Cho.
\newblock Attention-based cross-modal cnn using non-disassembled files for malware classification.
\newblock {\em IEEE Access}, 11:22889--22903, 2023.

\bibitem{kingma2013auto}
Diederik~P Kingma and Max Welling.
\newblock Auto-encoding variational bayes.
\newblock {\em arXiv preprint arXiv:1312.6114}, 2013.

\bibitem{kumar2016machine}
Ajit Kumar, K~Pramod Sagar, KS~Kuppusamy, and G~Aghila.
\newblock Machine learning based malware classification for android applications using multimodal image representations.
\newblock In {\em 2016 10th international conference on intelligent systems and control (ISCO)}, pages 1--6. IEEE, 2016.

\bibitem{kumar2019texture}
Nitish Kumar and Toshanlal Meenpal.
\newblock Texture-based malware family classification.
\newblock In {\em 2019 10th International Conference on Computing, Communication and Networking Technologies (ICCCNT)}, pages 1--6. IEEE, 2019.

\bibitem{kumar2018malicious}
Rajesh Kumar, Zhang Xiaosong, Riaz~Ullah Khan, Ijaz Ahad, and Jay Kumar.
\newblock Malicious code detection based on image processing using deep learning.
\newblock In {\em Proceedings of the 2018 International Conference on Computing and Artificial Intelligence}, ICCAI 2018, page 81–85, New York, NY, USA, 2018. Association for Computing Machinery.

\bibitem{kumar2022dtmic}
Sanjeev Kumar and B.~Janet.
\newblock Dtmic: Deep transfer learning for malware image classification.
\newblock {\em Journal of Information Security and Applications}, 64:103063, 2022.

\bibitem{kumar2022identification}
Sanjeev Kumar, B~Janet, and Subramanian Neelakantan.
\newblock Identification of malware families using stacking of textural features and machine learning.
\newblock {\em Expert Systems with Applications}, 208:118073, 2022.

\bibitem{kumar2024imcnn}
Sanjeev Kumar, B~Janet, and Subramanian Neelakantan.
\newblock Imcnn: Intelligent malware classification using deep convolution neural networks as transfer learning and ensemble learning in honeypot enabled organizational network.
\newblock {\em Computer Communications}, 216:16--33, 2024.

\bibitem{avclass}
Malicia Lab.
\newblock Avclass.
\newblock \url{https://github.com/malicialab/avclass}, 2025.
\newblock Accessed: 2023-5-15.

\bibitem{lachtar2023ransomshield}
Nada Lachtar, Duha Ibdah, Hamza Khan, and Anys Bacha.
\newblock Ransomshield: A visualization approach to defending mobile systems against ransomware.
\newblock {\em ACM Trans. Priv. Secur.}, 26(3), mar 2023.

\bibitem{lad2020malware}
Sumit~S Lad, Amol~C Adamuthe, et~al.
\newblock Malware classification with improved convolutional neural network model.
\newblock {\em International Journal of Computer Network \& Information Security}, 12(6):30--43, 2020.

\bibitem{le2018deep}
Quan Le, Ois{\'\i}n Boydell, Brian Mac~Namee, and Mark Scanlon.
\newblock Deep learning at the shallow end: Malware classification for non-domain experts.
\newblock {\em Digital Investigation}, 26:S118--S126, 2018.

\bibitem{lee2023robust}
Hyunjong Lee, Sooin Kim, Dongheon Baek, Donghoon Kim, and Doosung Hwang.
\newblock Robust iot malware detection and classification using opcode category features on machine learning.
\newblock {\em IEEE Access}, 11:18855--18867, 2023.

\bibitem{lee2021aclassification}
Jongkwan Lee and Jongdeog Lee.
\newblock A classification system for visualized malware based on multiple autoencoder models.
\newblock {\em IEEE Access}, 9:144786--144795, 2021.

\bibitem{lekssays2020anovel}
Ahmed Lekssays, Bouchaib Falah, and Sameer Abufardeh.
\newblock A novel approach for android malware detection and classification using convolutional neural networks.
\newblock In {\em 15th International Conference on Software Technologies}, pages 606--614, 01 2020.

\bibitem{li2021backdoor}
Chaoran Li, Xiao Chen, Derui Wang, Sheng Wen, Muhammad~Ejaz Ahmed, Seyit Camtepe, and Yang Xiang.
\newblock Backdoor attack on machine learning based android malware detectors.
\newblock {\em IEEE Transactions on dependable and secure computing}, 19(5):3357--3370, 2021.

\bibitem{li2021cnn}
Qi~Li, Jiaxin Mi, Weishi Li, Junfeng Wang, and Mingyu Cheng.
\newblock Cnn-based malware variants detection method for internet of things.
\newblock {\em IEEE Internet of Things Journal}, 8(23):16946--16962, 2021.

\bibitem{lin2021towards}
Yuzhou Lin and Xiaolin Chang.
\newblock Towards interpretable ensemble learning for image-based malware detection.
\newblock {\em arXiv preprint arXiv:2101.04889}, 2021.

\bibitem{ling2023adversarial}
Xiang Ling, Lingfei Wu, Jiangyu Zhang, Zhenqing Qu, Wei Deng, Xiang Chen, Yaguan Qian, Chunming Wu, Shouling Ji, Tianyue Luo, et~al.
\newblock Adversarial attacks against windows pe malware detection: A survey of the state-of-the-art.
\newblock {\em Computers \& Security}, page 103134, 2023.

\bibitem{liu2020anovel}
Xinbo Liu, Yaping Lin, He~Li, and Jiliang Zhang.
\newblock A novel method for malware detection on ml-based visualization technique.
\newblock {\em Computers \& Security}, 89:101682, 2020.

\bibitem{liu2019atmpa}
Xinbo Liu, Jiliang Zhang, Yaping Lin, and He~Li.
\newblock Atmpa: attacking machine learning-based malware visualization detection methods via adversarial examples.
\newblock In {\em Proceedings of the International Symposium on Quality of Service}, pages 1--10, 2019.

\bibitem{liu2019anew}
Ya-shu Liu, Yu-Kun Lai, Zhi-Hai Wang, and Han-Bing Yan.
\newblock A new learning approach to malware classification using discriminative feature extraction.
\newblock {\em IEEE Access}, 7:13015--13023, 2019.

\bibitem{lowe1999object}
D.G. Lowe.
\newblock Object recognition from local scale-invariant features.
\newblock In {\em Proceedings of the Seventh IEEE International Conference on Computer vision}, volume~2, pages 1150--1157 vol.2, 1999.

\bibitem{lu2019generative}
Yan Lu and Jiang Li.
\newblock Generative adversarial network for improving deep learning based malware classification.
\newblock In {\em 2019 Winter Simulation Conference (WSC)}, pages 584--593. IEEE, 2019.

\bibitem{luo2017binary}
Jhu-Sin Luo and Dan Chia-Tien Lo.
\newblock Binary malware image classification using machine learning with local binary pattern.
\newblock In {\em 2017 IEEE International Conference on Big Data (Big Data)}, pages 4664--4667. IEEE, 2017.

\bibitem{survey2}
Ahmad M., Ahmed Ebada, and Aya Zoghby.
\newblock A survey on visualization-based malware detection.
\newblock {\em Journal of Cyber Security}, 4, 11 2022.

\bibitem{gopinath2023acomprehensive}
Gopinath M. and Sibi~Chakkaravarthy Sethuraman.
\newblock A comprehensive survey on deep learning based malware detection techniques.
\newblock {\em Computer Science Review}, 47:100529, 2023.

\bibitem{ma2022visual}
Zhidong Ma, Zhiwei Zhang, Chengliang Liu, Tianzhu Hu, Hongjun Li, and Baoquan Ren.
\newblock Visualizable malware detection based on multi-dimension dynamic behaviors.
\newblock In {\em 2022 International Conference on Networking and Network Applications (NaNA)}, pages 247--252, 2022.

\bibitem{mahoubi2020stochastic}
Arash Mahboubi, Seyit Camtepe, and Keyvan Ansari.
\newblock Stochastic modeling of iot botnet spread: A short survey on mobile malware spread modeling.
\newblock {\em IEEE Access}, 8:228818--228830, 2020.

\bibitem{makandar2015malware}
Aziz Makandar and Anita Patrot.
\newblock Malware analysis and classification using artificial neural network.
\newblock In {\em 2015 International conference on trends in automation, communications and computing technology (I-TACT-15)}, pages 1--6. IEEE, 2015.

\bibitem{mallik2022conrec}
Abhishek Mallik, Anavi Khetarpal, and Sanjay Kumar.
\newblock Conrec: malware classification using convolutional recurrence.
\newblock {\em Journal of Computer Virology and Hacking Techniques}, 18:297--313, 2022.

\bibitem{mathews2019explainable}
Sherin~Mary Mathews.
\newblock Explainable artificial intelligence applications in nlp, biomedical, and malware classification: A literature review.
\newblock In Kohei Arai, Rahul Bhatia, and Supriya Kapoor, editors, {\em Intelligent Computing}, pages 1269--1292, Cham, 2019. Springer International Publishing.

\bibitem{meng2025detecting}
Zhaoyi Meng, Jiale Zhang, Jiaqi Guo, Wansen Wang, Wenchao Huang, Jie Cui, Hong Zhong, and Yan Xiong.
\newblock Detecting android malware by visualizing app behaviors from multiple complementary views.
\newblock {\em IEEE Transactions on Information Forensics and Security}, 2025.

\bibitem{mercaldo2020deep}
Francesco Mercaldo and Antonella Santone.
\newblock Deep learning for image-based mobile malware detection.
\newblock {\em Journal of Computer Virology and Hacking Techniques}, 16:157 -- 171, 2020.

\bibitem{mohammed2021malware}
Tajuddin~Manhar Mohammed, Lakshmanan Nataraj, Satish Chikkagoudar, Shivkumar Chandrasekaran, and BS~Manjunath.
\newblock Malware detection using frequency domain-based image visualization and deep learning.
\newblock {\em arXiv preprint arXiv:2101.10578}, 2021.

\bibitem{moosavi2016deepfool}
Seyed-Mohsen Moosavi-Dezfooli, Alhussein Fawzi, and Pascal Frossard.
\newblock Deepfool: a simple and accurate method to fool deep neural networks.
\newblock In {\em Proceedings of the IEEE conference on computer vision and pattern recognition}, pages 2574--2582, 2016.

\bibitem{moreira2023improving}
Caio~C. Moreira, Davi~C. Moreira, and Claudomiro de~S.~de {Sales Jr.}
\newblock Improving ransomware detection based on portable executable header using xception convolutional neural network.
\newblock {\em Computers \& Security}, 130:103265, 2023.

\bibitem{naeem2022explainable}
Hamad Naeem, Bandar~M Alshammari, and Farhan Ullah.
\newblock Explainable artificial {Intelligence-Based} {IoT} device malware detection mechanism using image visualization and {Fine-Tuned} {CNN-Based} transfer learning model.
\newblock {\em Comput Intell Neurosci}, 2022:7671967, July 2022.

\bibitem{naeem2022android}
Hamad Naeem, Amjad Alsirhani, Mohammed~Mujib Alshahrani, and Abdullah Alomari.
\newblock Android device malware classification framework using multistep image feature extraction and multihead deep neural ensemble.
\newblock {\em Traitement du Signal}, 39(3), 2022.

\bibitem{naeem2019identification}
Hamad Naeem, Bing Guo, Muhammad~Rashid Naeem, Farhan Ullah, Hamza Aldabbas, and Muhammad~Sufyan Javed.
\newblock Identification of malicious code variants based on image visualization.
\newblock {\em Computers \& Electrical Engineering}, 76:225--237, 2019.

\bibitem{naeem2019visual}
Hamad Naeem, Bing Guo, Muhammad~Rashid Naeem, and Danish Vasan.
\newblock Visual malware classification using local and global malicious pattern.
\newblock {\em Journal of Computers}, 30(6):73--83, 2019.

\bibitem{naeem2020malware}
Hamad Naeem, Farhan Ullah, Muhammad~Rashid Naeem, Shehzad Khalid, Danish Vasan, Sohail Jabbar, and Saqib Saeed.
\newblock Malware detection in industrial internet of things based on hybrid image visualization and deep learning model.
\newblock {\em Ad Hoc Networks}, 105:102154, 2020.

\bibitem{survey3}
Shivarti Naik and Amita Dessai.
\newblock Malware classification approaches using machine learning techniques: A review.
\newblock In {\em 2021 5th International Conference on Electrical, Electronics, Communication, Computer Technologies and Optimization Techniques (ICEECCOT)}, pages 111--117, 2021.

\bibitem{malicia}
Antonio Nappa, M~Zubair Rafique, and Juan Caballero.
\newblock The malicia dataset: identification and analysis of drive-by download operations.
\newblock {\em International Journal of Information Security}, 14, 02 2014.

\bibitem{nataraj2011malware}
Lakshmanan Nataraj, Sreejith Karthikeyan, Gregoire Jacob, and Bangalore~S Manjunath.
\newblock Malware images: visualization and automatic classification.
\newblock In {\em Proceedings of the 8th international symposium on visualization for cyber security}, pages 1--7, 2011.

\bibitem{quoc2020asurvey}
Quoc-Dung Ngo, Huy-Trung Nguyen, Van-Hoang Le, and Doan-Hieu Nguyen.
\newblock A survey of iot malware and detection methods based on static features.
\newblock {\em ICT Express}, 6(4):280--286, 2020.

\bibitem{ni2018malware}
Sang Ni, Quan Qian, and Rui Zhang.
\newblock Malware identification using visualization images and deep learning.
\newblock {\em Computers \& Security}, 77:871--885, 2018.

\bibitem{ojala1994performance}
Timo Ojala, Matti Pietikainen, and David Harwood.
\newblock Performance evaluation of texture measures with classification based on kullback discrimination of distributions.
\newblock In {\em Proceedings of 12th international conference on pattern recognition}, volume~1, pages 582--585. IEEE, 1994.

\bibitem{oshau}
Stephen O'Shaughnessy and Stephen Sheridan.
\newblock Image-based malware classification hybrid framework based on space-filling curves.
\newblock {\em Computers \& Security}, 116:102660, 2022.

\bibitem{paardekooper2022designing}
Cornelius Paardekooper, Nasimul Noman, Raymond Chiong, and Vijay Varadharajan.
\newblock Designing deep convolutional neural networks using a genetic algorithm for image-based malware classification.
\newblock In {\em 2022 IEEE Congress on Evolutionary Computation (CEC)}, pages 1--8, 2022.

\bibitem{papernot2016limitations}
Nicolas Papernot, Patrick McDaniel, Somesh Jha, Matt Fredrikson, Z~Berkay Celik, and Ananthram Swami.
\newblock The limitations of deep learning in adversarial settings.
\newblock In {\em 2016 IEEE European symposium on security and privacy (EuroS\&P)}, pages 372--387. IEEE, 2016.

\bibitem{park2019generation}
Daniel Park, Haidar Khan, and B{\"u}lent Yener.
\newblock Generation \& evaluation of adversarial examples for malware obfuscation.
\newblock In {\em 2019 18th IEEE International Conference On Machine Learning And Applications (ICMLA)}, pages 1283--1290. IEEE, 2019.

\bibitem{pytorch}
Adam Paszke, Sam Gross, Francisco Massa, Adam Lerer, James Bradbury, Gregory Chanan, Trevor Killeen, Zeming Lin, Natalia Gimelshein, Luca Antiga, Alban Desmaison, Andreas Kopf, Edward Yang, Zachary DeVito, Martin Raison, Alykhan Tejani, Sasank Chilamkurthy, Benoit Steiner, Lu~Fang, Junjie Bai, and Soumith Chintala.
\newblock Pytorch: An imperative style, high-performance deep learning library.
\newblock In {\em Advances in Neural Information Processing Systems 32}, pages 8024--8035. Curran Associates, Inc., 2019.

\bibitem{patel2022deep}
Dev Patel, Dhairya Jadav, Rajesh Gupta, Nilesh~Kumar Jadav, Sudeep Tanwar, Bassem Ouni, and Mohsen Guizani.
\newblock Deep learning and blockchain-based framework to detect malware in autonomous vehicles.
\newblock In {\em 2022 International Wireless Communications and Mobile Computing (IWCMC)}, pages 278--283. IEEE, 2022.

\bibitem{scikit}
F.~Pedregosa, G.~Varoquaux, A.~Gramfort, V.~Michel, B.~Thirion, O.~Grisel, M.~Blondel, P.~Prettenhofer, R.~Weiss, V.~Dubourg, J.~Vanderplas, A.~Passos, D.~Cournapeau, M.~Brucher, M.~Perrot, and E.~Duchesnay.
\newblock Scikit-learn: Machine learning in {P}ython.
\newblock {\em Journal of Machine Learning Research}, 12:2825--2830, 2011.

\bibitem{pinhero2021malware}
Anson Pinhero, Anupama {M L}, Vinod P, C.A. Visaggio, Aneesh N, Abhijith S, and AnanthaKrishnan S.
\newblock Malware detection employed by visualization and deep neural network.
\newblock {\em Computers \& Security}, 105:102247, 2021.

\bibitem{pratama2022malw}
Handhika~Yanuar Pratama and Jeckson Sidabutar.
\newblock Malware classification and visualization using efficientnet and b2img algorithm.
\newblock In {\em 2022 International Conference on Advanced Computer Science and Information Systems (ICACSIS)}, pages 75--80, 2022.

\bibitem{prathiba2024blockchain}
Sahaya~Beni Prathiba, Pranav Murali, Rajalakshmi~Shenbaga Moorthy, Deepak~Kumar Anandhan, Arikumar~K Selvaraj, and Joel~JPC Rodrigues.
\newblock A blockchain-powered malicious node detection in internet of autonomous vehicles.
\newblock {\em IEEE Transactions on Intelligent Transportation Systems}, 2024.

\bibitem{qiao2019amulti}
Yanchen Qiao, Qingshan Jiang, Zhenchao Jiang, and Liang Gu.
\newblock A multi-channel visualization method for malware classification based on deep learning.
\newblock In {\em 2019 18th IEEE International Conference On Trust, Security And Privacy In Computing And Communications/13th IEEE International Conference On Big Data Science And Engineering (TrustCom/BigDataSE)}, pages 757--762, 2019.

\bibitem{qiu2022malware}
Lingfeng Qiu, Shuo Wang, Jian Wang, Yifei Wang, and Wei Huang.
\newblock Malware classification based on a light-weight architecture of cnn: Malshufflenet.
\newblock In {\em 2022 3rd International Conference on Computer Vision, Image and Deep Learning \& International Conference on Computer Engineering and Applications (CVIDL \& ICCEA)}, pages 1047--1050, 2022.

\bibitem{ravi2023vit4mal}
Akshara Ravi, Vivek Chaturvedi, and Muhammad Shafique.
\newblock Vit4mal: Lightweight vision transformer for malware detection on edge devices.
\newblock {\em ACM Transactions on Embedded Computing Systems}, 22(5s):1--26, 2023.

\bibitem{vinayakumar2022efficientnet}
Vinayakumar Ravi and Rajasekhar Chaganti.
\newblock Efficientnet deep learning meta-classifier approach for image-based android malware detection.
\newblock {\em Multimedia Tools and Applications}, 82:24891--24917, 2023.

\bibitem{upxwebsite}
John Reiser.
\newblock {MS Windows NT} kernel description.
\newblock \url{https://github.com/upx/upx}, 2025.

\bibitem{ren2020end}
Zhongru Ren, Haomin Wu, Qian Ning, Iftikhar Hussain, and Bingcai Chen.
\newblock End-to-end malware detection for android iot devices using deep learning.
\newblock {\em Ad Hoc Networks}, 101:102098, 2020.

\bibitem{rieck2011automatic}
Konrad Rieck, Philipp Trinius, Carsten Willems, and Thorsten Holz.
\newblock Automatic analysis of malware behavior using machine learning.
\newblock {\em J. Comput. Secur.}, 19(4):639–668, dec 2011.

\bibitem{ronen2018microsoft}
Royi Ronen, Marian Radu, Corina Feuerstein, Elad Yom-Tov, and Mansour Ahmadi.
\newblock Microsoft malware classification challenge.
\newblock {\em arXiv preprint arXiv:1802.10135}, 2018.

\bibitem{rong2021umvd}
Candong Rong, Gaopeng Gou, Chengshang Hou, Zhen Li, Gang Xiong, and Li~Guo.
\newblock Umvd-fsl: unseen malware variants detection using few-shot learning.
\newblock In {\em 2021 International Joint Conference on Neural Networks (IJCNN)}, pages 1--8. IEEE, 2021.

\bibitem{roseline2020intelligent}
S.~Abijah Roseline, S.~Geetha, Seifedine Kadry, and Yunyoung Nam.
\newblock Intelligent vision-based malware detection and classification using deep random forest paradigm.
\newblock {\em IEEE Access}, 8:206303--206324, 2020.

\bibitem{roseline2019towards}
S~Abijah Roseline, AD~Sasisri, S~Geetha, and C~Balasubramanian.
\newblock Towards efficient malware detection and classification using multilayered random forest ensemble technique.
\newblock In {\em 2019 International Carnahan Conference on Security Technology (ICCST)}, pages 1--6. IEEE, 2019.

\bibitem{rosenberg2021adversarial}
Ishai Rosenberg, Asaf Shabtai, Yuval Elovici, and Lior Rokach.
\newblock Adversarial machine learning attacks and defense methods in the cyber security domain.
\newblock {\em ACM Computing Surveys (CSUR)}, 54(5):1--36, 2021.

\bibitem{rustam2023malw}
Furqan Rustam, Imran Ashraf, Anca~Delia Jurcut, Ali~Kashif Bashir, and Yousaf~Bin Zikria.
\newblock Malware detection using image representation of malware data and transfer learning.
\newblock {\em Journal of Parallel and Distributed Computing}, 172:32--50, 2023.

\bibitem{severi2021explanation}
Giorgio Severi, Jim Meyer, Scott Coull, and Alina Oprea.
\newblock $\{$Explanation-Guided$\}$ backdoor poisoning attacks against malware classifiers.
\newblock In {\em 30th USENIX security symposium (USENIX security 21)}, pages 1487--1504, 2021.

\bibitem{shabtai2009detection}
Asaf Shabtai, Robert Moskovitch, Yuval Elovici, and Chanan Glezer.
\newblock Detection of malicious code by applying machine learning classifiers on static features: A state-of-the-art survey.
\newblock {\em information security technical report}, 14(1):16--29, 2009.

\bibitem{survey1}
Syed Shakir~Hameed Shah, Norziana Jamil, and Atta ur~Rehman Khan.
\newblock Performance comparison of visualization-based malware detection and classification techniques.
\newblock In {\em 2022 17th International Conference on Emerging Technologies (ICET)}, pages 200--205, 2022.

\bibitem{shannon}
Claude~Elwood Shannon.
\newblock A mathematical theory of communication.
\newblock {\em The Bell System Technical Journal}, 27:379--423, 1948.

\bibitem{sharma2022windows}
Osho Sharma, Akashdeep Sharma, and Arvind Kalia.
\newblock Windows and iot malware visualization and classification with deep cnn and xception cnn using markov images.
\newblock {\em Journal of Intelligent Information Systems}, pages 1--27, 2022.

\bibitem{sharma2024migan}
Osho Sharma, Akashdeep Sharma, and Arvind Kalia.
\newblock Migan: Gan for facilitating malware image synthesis with improved malware classification on novel dataset.
\newblock {\em Expert Systems with Applications}, 241:122678, 2024.

\bibitem{shaukat2023anovel}
Kamran Shaukat, Suhuai Luo, and Vijay Varadharajan.
\newblock A novel deep learning-based approach for malware detection.
\newblock {\em Engineering Applications of Artificial Intelligence}, 122:106030, 2023.

\bibitem{shen2023self}
Limin Shen, Jiayin Feng, Zhen Chen, Zhongkui Sun, Dongkui Liang, Hui Li, and Yuying Wang.
\newblock Self-attention based convolutional-lstm for android malware detection using network traffics grayscale image.
\newblock {\em Applied Intelligence}, 53(1):683--705, 2023.

\bibitem{singh2021classification}
Jaiteg Singh, Deepak Thakur, Tanya Gera, Babar Shah, Tamer Abuhmed, and Farman Ali.
\newblock Classification and analysis of android malware images using feature fusion technique.
\newblock {\em IEEE Access}, 9:90102--90117, 2021.

\bibitem{glcm}
Shruti Singh, Divya Srivastava, and Suneeta Agarwal.
\newblock Glcm and its application in pattern recognition.
\newblock In {\em 2017 5th International Symposium on Computational and Business Intelligence (ISCBI)}, pages 20--25, 2017.

\bibitem{virustotal}
Hispasec Sistemas.
\newblock Virus total.
\newblock \url{https://www.virustotal.com}, 2023.

\bibitem{shiva2018windows}
Shiva~Darshan S.L and Jaidhar C.D.
\newblock Windows malware detector using convolutional neural network based on visualization images.
\newblock {\em IEEE Transactions on Emerging Topics in Computing}, 9(2):1057--1069, 2021.

\bibitem{sohn2015learning}
Kihyuk Sohn, Honglak Lee, and Xinchen Yan.
\newblock Learning structured output representation using deep conditional generative models.
\newblock {\em Advances in neural information processing systems}, 28, 2015.

\bibitem{stancin2019anoverview}
I.~Stančin and A.~Jović.
\newblock An overview and comparison of free python libraries for data mining and big data analysis.
\newblock In {\em 2019 42nd International Convention on Information and Communication Technology, Electronics and Microelectronics (MIPRO)}, pages 977--982, 2019.

\bibitem{su2018lightweight}
Jiawei Su, Danilo~Vargas Vasconcellos, Sanjiva Prasad, Daniele Sgandurra, Yaokai Feng, and Kouichi Sakurai.
\newblock Lightweight classification of iot malware based on image recognition.
\newblock In {\em 2018 IEEE 42nd Annual Computer Software and Applications Conference (COMPSAC)}, volume~02, pages 664--669, 2018.

\bibitem{sudhakar2021mcft}
Sudhakar and Sushil Kumar.
\newblock Mcft-cnn: Malware classification with fine-tune convolution neural networks using traditional and transfer learning in internet of things.
\newblock {\em Future Generation Computer Systems}, 125:334--351, 2021.

\bibitem{sun2021deep}
Guosong Sun and Quan Qian.
\newblock Deep learning and visualization for identifying malware families.
\newblock {\em IEEE Transactions on Dependable and Secure Computing}, 18(1):283--295, 2021.

\bibitem{szegedy2013intriguing}
Christian Szegedy, Wojciech Zaremba, Ilya Sutskever, Joan Bruna, Dumitru Erhan, Ian Goodfellow, and Rob Fergus.
\newblock Intriguing properties of neural networks.
\newblock {\em arXiv preprint arXiv:1312.6199}, 2013.

\bibitem{efficientnet}
Mingxing Tan and Quoc~V. Le.
\newblock Efficientnet: Rethinking model scaling for convolutional neural networks.
\newblock {\em CoRR}, abs/1905.11946, 2019.

\bibitem{tang2019dynamic}
Mingdong Tang and Quan Qian.
\newblock Dynamic api call sequence visualization for malware classiﬁcation.
\newblock {\em IET Information Security}, 13, 07 2019.

\bibitem{tekerek2022anovel}
Adem Tekerek and Muhammed~Mutlu Yapici.
\newblock A novel malware classification and augmentation model based on convolutional neural network.
\newblock {\em Computers \& Security}, 112:102515, 2022.

\bibitem{vtstat}
Virus Total.
\newblock Virus total global statistics.
\newblock \url{https://www.virustotal.com/gui/stats}, 2023.

\bibitem{tuncer2020automated}
Turker Tuncer, Fatih Ertam, and Sengul Dogan.
\newblock Automated malware recognition method based on local neighborhood binary pattern.
\newblock {\em Multimedia Tools and Applications}, 79:27815--27832, 2020.

\bibitem{ucci2019survey}
Daniele Ucci, Leonardo Aniello, and Roberto Baldoni.
\newblock Survey of machine learning techniques for malware analysis.
\newblock {\em Computers \& Security}, 81:123--147, 2019.

\bibitem{pil}
P~Umesh.
\newblock Image processing in python.
\newblock {\em CSI Communications}, 23, 2012.

\bibitem{unver2020android}
Halil~Murat {\"U}nver and Khaled Bakour.
\newblock Android malware detection based on image-based features and machine learning techniques.
\newblock {\em SN Applied Sciences}, 2:1--15, 2020.

\bibitem{van2022attention}
Tuan Van~Dao, Hiroshi Sato, and Masao Kubo.
\newblock An attention mechanism for combination of cnn and vae for image-based malware classification.
\newblock {\em IEEE Access}, 10:85127--85136, 2022.

\bibitem{vasan2020imcfn}
Danish Vasan, Mamoun Alazab, Sobia Wassan, Hamad Naeem, Babak Safaei, and Qin Zheng.
\newblock Imcfn: Image-based malware classification using fine-tuned convolutional neural network architecture.
\newblock {\em Computer Networks}, 171:107138, 2020.

\bibitem{vasan2020imcec}
Danish Vasan, Mamoun Alazab, Sobia Wassan, Babak Safaei, and Qin Zheng.
\newblock Image-based malware classification using ensemble of cnn architectures (imcec).
\newblock {\em Computers \& Security}, 92:101748, 2020.

\bibitem{vasan2024broad}
Danish Vasan, Mohammad Hammoudeh, and Mamoun Alazab.
\newblock Broad learning: A gpu-free image-based malware classification.
\newblock {\em Applied Soft Computing}, 154:111401, 2024.

\bibitem{vaswani2023attention}
Ashish Vaswani, Noam Shazeer, Niki Parmar, Jakob Uszkoreit, Llion Jones, Aidan~N. Gomez, Lukasz Kaiser, and Illia Polosukhin.
\newblock Attention is all you need, 2023.

\bibitem{venkatraman2019ahybrid}
Sitalakshmi Venkatraman, Mamoun Alazab, and R.~Vinayakumar.
\newblock A hybrid deep learning image-based analysis for effective malware detection.
\newblock {\em Journal of Information Security and Applications}, 47:377--389, 2019.

\bibitem{verma2020multiclass}
Vinita Verma, Sunil~K Muttoo, and VB~Singh.
\newblock Multiclass malware classification via first-and second-order texture statistics.
\newblock {\em Computers \& Security}, 97:101895, 2020.

\bibitem{vinayakumar2019robust}
R.~Vinayakumar, Mamoun Alazab, K.~P. Soman, Prabaharan Poornachandran, and Sitalakshmi Venkatraman.
\newblock Robust intelligent malware detection using deep learning.
\newblock {\em IEEE Access}, 7:46717--46738, 2019.

\bibitem{vinayakumar2018scalable}
R~Vinayakumar, Prabaharan Poornachandran, and KP~Soman.
\newblock Scalable framework for cyber threat situational awareness based on domain name systems data analysis.
\newblock {\em Big data in engineering applications}, pages 113--142, 2018.

\bibitem{yara}
VirusTotal.
\newblock Yara rules.
\newblock \url{https://virustotal.github.io/yara}, 2025.

\bibitem{voigt2017theeu}
Paul Voigt and Axel Von~dem Bussche.
\newblock The eu general data protection regulation (gdpr).
\newblock {\em A Practical Guide, 1st Ed., Cham: Springer International Publishing}, 10(3152676):10--5555, 2017.

\bibitem{hit4mal}
Duc‐Ly Vu, Trong‐Kha Nguyen, Tam~V. Nguyen, Tu~N. Nguyen, Fabio Massacci, and Phu~H. Phung.
\newblock Hit4mal: Hybrid image transformation for malware classification.
\newblock {\em Trans. Emerg. Telecommun. Technol.}, 31(11), nov 2020.

\bibitem{wang2022image}
Bidong Wang, Hui Liu, Xinli Han, and Dongliang Xuan.
\newblock Image-based ransomware classification with classifier combination.
\newblock In {\em Proceedings of the 3rd International Conference on Advanced Information Science and System}, AISS '21, New York, NY, USA, 2022. Association for Computing Machinery.

\bibitem{wang2021anovel}
Peng Wang, Zhijie Tang, and Junfeng Wang.
\newblock A novel few-shot malware classification approach for unknown family recognition with multi-prototype modeling.
\newblock {\em Computers \& Security}, 106:102273, 2021.

\bibitem{warde201611}
David Warde-Farley and Ian Goodfellow.
\newblock 11 adversarial perturbations of deep neural networks.
\newblock {\em Perturbations, Optimization, and Statistics}, 311:5, 2016.

\bibitem{wu2021android}
Bozhi Wu, Sen Chen, Cuiyun Gao, Lingling Fan, Yang Liu, Weiping Wen, and Michael~R Lyu.
\newblock Why an android app is classified as malware: Toward malware classification interpretation.
\newblock {\em ACM Transactions on Software Engineering and Methodology (TOSEM)}, 30(2):1--29, 2021.

\bibitem{wu2017android}
Qixin Wu, Zheng Qin, Jinxin Zhang, Hui Yin, Guangyi Yang, and Kuangsheng Hu.
\newblock Android malware detection using local binary pattern and principal component analysis.
\newblock In {\em Data Science: Third International Conference of Pioneering Computer Scientists, Engineers and Educators, ICPCSEE 2017, Changsha, China, September 22--24, 2017, Proceedings, Part I}, pages 262--275. Springer, 2017.

\bibitem{xiao2020malfcs}
Guoqing Xiao, Jingning Li, Yuedan Chen, and Kenli Li.
\newblock Malfcs: An effective malware classification framework with automated feature extraction based on deep convolutional neural networks.
\newblock {\em Journal of Parallel and Distributed Computing}, 141:49--58, 2020.

\bibitem{xiao2021image}
Mao Xiao, Chun Guo, Guowei Shen, Yunhe Cui, and Chaohui Jiang.
\newblock Image-based malware classification using section distribution information.
\newblock {\em Computers \& Security}, 110:102420, 2021.

\bibitem{xiao2019animage}
Xusheng Xiao and Shao Yang.
\newblock An image-inspired and cnn-based android malware detection approach.
\newblock In {\em 2019 34th IEEE/ACM International Conference on Automated Software Engineering (ASE)}, pages 1259--1261, 2019.

\bibitem{xu2021hybrid}
Peng Xu, Claudia Eckert, and Apostolis Zarras.
\newblock hybrid-flacon: Hybrid pattern malware detection and categorization with network traffic and program code.
\newblock {\em arXiv preprint arXiv:2112.10035}, 2021.

\bibitem{xuan2024bitcn}
Bona Xuan, Jin Li, and Yafei Song.
\newblock Bitcn-taefficientnet malware classification approach based on sequence and rgb fusion.
\newblock {\em Computers \& Security}, 139:103734, 2024.

\bibitem{yajamanam2018deep}
Sravani Yajamanam, Vikash Raja~Samuel Selvin, Fabio Di~Troia, and Mark Stamp.
\newblock Deep learning versus gist descriptors for image-based malware classification.
\newblock In {\em Icissp}, pages 553--561, 2018.

\bibitem{yakura2018malware}
Hiromu Yakura, Shinnosuke Shinozaki, Reon Nishimura, Yoshihiro Oyama, and Jun Sakuma.
\newblock Malware analysis of imaged binary samples by convolutional neural network with attention mechanism.
\newblock In {\em Proceedings of the Eighth ACM Conference on Data and Application Security and Privacy}, CODASPY '18, page 127–134, New York, NY, USA, 2018. Association for Computing Machinery.

\bibitem{yakura2019neural}
Hiromu Yakura, Shinnosuke Shinozaki, Reon Nishimura, Yoshihiro Oyama, and Jun Sakuma.
\newblock Neural malware analysis with attention mechanism.
\newblock {\em Computers \& Security}, 87:101592, 2019.

\bibitem{yan2022survey}
Senming Yan, Jing Ren, Wei Wang, Limin Sun, Wei Zhang, and Quan Yu.
\newblock A survey of adversarial attack and defense methods for malware classification in cyber security.
\newblock {\em IEEE Communications Surveys \& Tutorials}, 25(1):467--496, 2022.

\bibitem{yang2018aconvolutional}
Chun Yang, Yu~Wen, Jianbin Guo, Haitao Song, Linfeng Li, Haoyang Che, and Dan Meng.
\newblock A convolutional neural network based classifier for uncompressed malware samples.
\newblock In {\em Proceedings of the 1st Workshop on Security-Oriented Designs of Computer Architectures and Processors}, SecArch'18, page 15–17, New York, NY, USA, 2018. Association for Computing Machinery.

\bibitem{yang2019anovel}
Hangfeng Yang, Shudong Li, Xiaobo Wu, Hui Lu, and Weihong Han.
\newblock A novel solutions for malicious code detection and family clustering based on machine learning.
\newblock {\em IEEE Access}, 7:148853--148860, 2019.

\bibitem{yang2025sac}
Jin Yang, Huijia Liang, Hang Ren, Dongqing Jia, and Xin Wang.
\newblock Sac: Collaborative learning of structure and content features for android malware detection framework.
\newblock {\em Neurocomputing}, page 130053, 2025.

\bibitem{yang2023jigsaw}
Limin Yang, Zhi Chen, Jacopo Cortellazzi, Feargus Pendlebury, Kevin Tu, Fabio Pierazzi, Lorenzo Cavallaro, and Gang Wang.
\newblock Jigsaw puzzle: Selective backdoor attack to subvert malware classifiers.
\newblock In {\em 2023 IEEE Symposium on Security and Privacy (SP)}, pages 719--736. IEEE, 2023.

\bibitem{bodmas}
Limin Yang, Arridhana Ciptadi, Ihar Laziuk, Ali Ahmadzadeh, and Gang Wang.
\newblock Bodmas: An open dataset for learning based temporal analysis of pe malware.
\newblock In {\em 4th Deep Learning and Security Workshop}, 2021.

\bibitem{yang2025variant}
Shumian Yang, Jiarui Hu, Xin Li, Dawei Zhao, Lijuan Xu, Fuqiang Yu, and Chunhui Wang.
\newblock A variant-sensitive malware detection method based on feature contrast enhancement.
\newblock {\em IEEE Transactions on Computational Social Systems}, 2025.

\bibitem{you2010malware}
Ilsun You and Kangbin Yim.
\newblock Malware obfuscation techniques: A brief survey.
\newblock In {\em 2010 International conference on broadband, wireless computing, communication and applications}, pages 297--300. IEEE, 2010.

\bibitem{yu2025semantic}
Yaoxiang Yu, Bo~Cai, Kamran Aziz, Xinyan Wang, Jian Luo, Muhammad~Shahid Iqbal, Prasun Chakrabarti, and Tulika Chakrabarti.
\newblock Semantic lossless encoded image representation for malware classification.
\newblock {\em Scientific Reports}, 15(1):7997, 2025.

\bibitem{yuan2020byte}
Baoguo Yuan, Junfeng Wang, Dong Liu, Wen Guo, Peng Wu, and Xuhua Bao.
\newblock Byte-level malware classification based on markov images and deep learning.
\newblock {\em Computers \& Security}, 92:101740, 2020.

\bibitem{zhan2023amgmal}
Dazhi Zhan, Yexin Duan, Yue Hu, Lujia Yin, Zhisong Pan, and Shize Guo.
\newblock Amgmal: Adaptive mask-guided adversarial attack against malware detection with minimal perturbation.
\newblock {\em Computers \& Security}, 127:103103, 2023.

\bibitem{zhan2025practical}
Dazhi Zhan, Kun Xu, Xin Liu, Tong Han, Zhisong Pan, and Shize Guo.
\newblock Practical clean-label backdoor attack against static malware detection.
\newblock {\em Computers \& Security}, 150:104280, 2025.

\bibitem{zhang2025imcmk}
Dandan Zhang, Yafei Song, Qian Xiang, and Yang Wang.
\newblock Imcmk-cnn: A lightweight convolutional neural network with multi-scale kernels for image-based malware classification.
\newblock {\em Alexandria Engineering Journal}, 111:203--220, 2025.

\bibitem{zhang2021android}
Wenhui Zhang, Nurbol Luktarhan, Chao Ding, and Bei Lu.
\newblock Android malware detection using tcn with bytecode image.
\newblock {\em Symmetry}, 13(7):1107, 2021.

\bibitem{zhao2021review}
Jiawei Zhao, Rahat Masood, and Suranga Seneviratne.
\newblock A review of computer vision methods in network security.
\newblock {\em IEEE Communications Surveys \& Tutorials}, 23(3):1838--1878, 2021.

\bibitem{zhao2020amalware}
Yuntao Zhao, Wenjie Cui, Shengnan Geng, Bo~Bo, Yongxin Feng, and Wenbo Zhang.
\newblock A malware detection method of code texture visualization based on an improved faster rcnn combining transfer learning.
\newblock {\em IEEE Access}, 8:166630--166641, 2020.

\bibitem{zhong2022malware}
Fangtian Zhong, Zekai Chen, Minghui Xu, Guoming Zhang, Dongxiao Yu, and Xiuzhen Cheng.
\newblock Malware-on-the-brain: Illuminating malware byte codes with images for malware classification.
\newblock {\em IEEE Transactions on Computers}, 72(2):438--451, 2022.

\bibitem{zhou2012dissecting}
Yajin Zhou and Xuxian Jiang.
\newblock Dissecting android malware: Characterization and evolution.
\newblock In {\em 2012 IEEE symposium on security and privacy}, pages 95--109. IEEE, 2012.

\bibitem{zhu2022afew}
Jinting Zhu, Julian Jang-Jaccard, Amardeep Singh, Ian Welch, Harith AI-Sahaf, and Seyit Camtepe.
\newblock A few-shot meta-learning based siamese neural network using entropy features for ransomware classification.
\newblock {\em Computers \& Security}, 117:102691, 2022.

\end{thebibliography}

\end{document}